\documentstyle[epsfig,12pt,here]{JHEP} 

\preprint{S 728-0799}

\title{Power corrections and resummation of radiative corrections 
in the single dressed gluon approximation -- 
the average thrust as a case study\footnote{Research 
supported in part by the EC 
program ``Training and Mobility of Researchers'', Network 
``QCD and Particle Structure'', contract ERBFMRXCT980194.}}

\author{{\bf Einan Gardi} \,\,\, and \,\,\,{\bf Georges Grunberg}\\
\vspace{.4in}
 Centre de Physique Th\'eorique de l'Ecole Polytechnique\footnote{CNRS
 UMR C7644}
\\ 91128 Palaiseau Cedex, France
\\email: {\tt gardi@cpht.polytechnique.fr,\, grunberg@cpht.polytechnique.fr}
}

\abstract{
Infrared power corrections for Minkowskian QCD observables are analyzed 
in the framework of renormalon resummation, motivated by analogy with 
the skeleton expansion in QED and the BLM approach. Performing the 
``massive gluon'' renormalon integral a renormalization scheme 
invariant result is obtained. Various regularizations of the integral 
are studied. In particular, we compare the infrared cutoff regularization 
with the standard principal value Borel sum and show that they yield equivalent
results once power terms are included. 
As an example the average thrust $\left< T \right>$ in 
$e^+e^-$ annihilation is analyzed. We find that a major part of the 
discrepancy between the known next-to-leading order calculation and 
experiment can be explained by resummation of higher order perturbative 
terms. This fact does not preclude the infrared finite coupling 
interpretation with a substantial $1/Q$ power term. Fitting 
the regularized perturbative sum plus a $1/Q$ term to experimental 
data yields 
$\alpha_s^{\hbox{$\overline{\hbox{\tiny MS}}\,$}}({\rm M_Z})=0.110\pm  0.002$.
}


\newcommand{\mycomm}[1]{\hfill\break
$\phantom{a}$\kern-3.5em{\tt===$>$ \bf #1}\hfill\break}
\newcommand{\mycommA}[1]{\hfill\break
$\phantom{a}$\kern-3.5em{\tt***$>$ \bf #1}\hfill\break}

\begin{document}
\catcode`\@=11 
\def\lsim{\mathrel{\mathpalette\@versim<}}
\def\gsim{\mathrel{\mathpalette\@versim>}}
\def\@versim#1#2{\vcenter{\offinterlineskip
        \ialign{$\m@th#1\hfil##\hfil$\crcr#2\crcr\sim\crcr } }}
\catcode`\@=12 
\def\beq{\begin{equation}}
\def\eeq{\end{equation}}
\def\MSbar {\hbox{$\overline{\hbox{\tiny MS}}\,$}}
\def\eff{\hbox{\tiny eff}}
\def\FP{\hbox{\tiny FP}}
\def\PV{\hbox{\tiny PV}}
\def\IR{\hbox{\tiny IR}}
\def\UV{\hbox{\tiny UV}}
\def\APT{\hbox{\tiny APT}}
\def\QCD{\hbox{\tiny QCD}}
\def\CMW{\hbox{\tiny CMW}}
\def\pinch{\hbox{\tiny pinch}}
\def\brem{\hbox{\tiny brem}}
\def\V{\hbox{\tiny V}}
\def\BLM{\hbox{\tiny BLM}}
\def\NLO{\hbox{\tiny NLO}}
\def\res{\hbox{\tiny res}}
\def\PT{\hbox{\tiny PT}}
\def\PA{\hbox{\tiny PA}}
\def\1loop{\hbox{\tiny 1-loop}}
\def\2loop{\hbox{\tiny 2-loop}}
\def\mysim{\kern -.1667em\lower0.8ex\hbox{$\tilde{\phantom{a}}$}}
\vskip 20pt

\section{Introduction}

Power corrections in QCD have been the subject of many interesting
theoretical developments in the recent years \cite{B1Z3}, especially concerning
observables that do not admit an operator product expansion.
Of particular interest in this respect are event shape variables in $e^+e^-$
annihilation. For these observables the available \cite{Ellis} next-to-leading 
order perturbative calculations in a standard choice of 
renormalization scheme and
scale are found not to agree with experimental data \cite{Moriond_Stenzel}. 
The discrepancy was originally bridged over by Monte-Carlo simulations
which account for non-perturbative ``hadronization corrections''. 
The data can also be
fitted by the next-to-leading order perturbative result plus a
power correction that typically (but not always) falls as $1/Q$, 
where $Q=\sqrt{s}$ is the center of mass energy.
Nevertheless, this situation is not satisfactory, especially because the
non-perturbative correction, which is not under control theoretically,
is numerically quite significant. Consequently, much theoretical effort
has been invested in the last five years in understanding the source 
of these power corrections. This effort turned out to be quite 
fruitful \cite{Manohar_Wise}--\cite{EEC_Broad}: 
a ``renormalon phenomenology'' has been developed \cite{B1Z3}, 
where information contained in perturbation theory is 
used to determine the form of the power correction, while its
normalization is left as a non-perturbative parameter, which is
determined by fitting experimental data.

On the other hand, the very same observables appear to have significant
sub-leading perturbative corrections and large renormalization
scheme and scale dependence when calculated up to the next-to-leading
order, as discussed in ref.~\cite{Beneke,CGM,Moriond_Stenzel}.
This observation raises concern about the reliability of the results
obtained in the usual procedure, where a perturbative expansion 
truncated at the next-to-leading order is 
used\footnote{For event shape {\em distributions} close to the 2-jet limit, 
large Sudakov logarithms related to the emission of soft and collinear
gluons appear explicitly in the perturbative coefficients. 
Due to multiple emission such contributions persist at high orders 
and make the fixed order calculation useless.
In ref.~\cite{CTTW} it was shown how these large logs can be 
systematically resummed to all orders. 
In this work we concentrate on {\em average} event shape observables, 
where these contributions are presumably not important.} as  
a starting point for the experimental fit \cite{Moriond_Stenzel}. 
Unfortunately, the full next-to-next-to-leading
order calculation for these observables is not yet available. 

In this paper, we adopt the view that the most important corrections are
related to the running of the coupling, in the spirit of the BLM
approach \cite{BLM}.
Following \cite{Grunberg-pow}, we assume that the perturbative series
can be reshuffled in a (yet hypothetical)  ``dressed skeleton expansion'' 
built in analogy with the Abelian theory, 
where each term is by itself {\em renormalization scheme invariant}. 
This expansion was the original motivation behind the BLM approach
(see ref.~\cite{Lu} for further discussion).
In the Abelian theory, the skeleton expansion is well defined.
The first term corresponds to an exchange of a
single photon, dressed by all possible vacuum polarization insertions which
build up the Gell-Mann Low effective-charge;
the second term corresponds to an exchange of two dressed photons and so on.
The expansion coincides with the standard expansion in $\alpha$ 
for a conformal theory where the coupling does not run. 
The leading skeleton term can be written compactly as a ``renormalon
integral'' -- an integral over all scales of the Gell-Mann Low 
effective-charge times an observable dependent
function which arises from the one-loop Feynman integrand and
represents the momentum distribution of the exchanged dressed photon.
When the leading skeleton term is expanded in terms of the coupling up
to large enough order, both large and small momentum regions 
give rise to factorially increasing perturbative
coefficients. These are the renormalons \cite{tHooft}--\cite{Z2} 
which are believed to dominate 
the diverging large order behavior of the full perturbative series.
This way renormalons carry information about the 
inconsistency of perturbation theory and thus on possible
non-perturbative power corrections arising from the strong coupling regime.

Having no straightforward diagrammatic interpretation
in the non-Abelian case the ``skeleton expansion'' is still just a conjecture. 
There is an obvious difficulty due to gluon self-interaction
vertices. This difficulty can be hopefully resolved by separating 
contributions from such diagrams into several different skeleton terms
in a gauge invariant manner. 
Ref.~\cite{Watson} gives a concrete suggestion for a diagrammatic
construction of the skeleton expansion in QCD, based on the pinch
technique. However, more work is required to establish it beyond the
one-loop level. 
On the other hand, the structure of the single gluon exchange 
term (``leading skeleton'') is strongly motivated by the large 
$N_f$ limit where the Abelian correspondence is transparent.
The non-Abelian analogue of the renormalon integral,
obtained through the ``Naive Non-Abelianization'' \cite{Broadhurst} procedure, 
was used extensively in the last few years in  
all order perturbative resummation (e.g. \cite{BB}--\cite{LTM}) and 
parametrization of power corrections 
(e.g. \cite{W}--\cite{EEC_Broad} and \cite{Grunberg-pow}).

In the present work, we perform an analysis of infrared power corrections 
together with all order perturbative resummation of renormalon-type
diagrams in the ``massive gluon'' approach \cite{BB,BBB,DMW}.
We show that these two aspects of improving the standard perturbative
calculation cannot be dissociated and must be performed together. 

We study one specific example: the average thrust in $e^+e^-$
annihilation, where there is a priori evidence for both large 
perturbative corrections and strong $1/Q$ power corrections.
We perform renormalon resummation at the level of a single gluon
emission, emphasizing the renormalization group invariance of this procedure.
We discuss in detail the ambiguity between the perturbative sum and
the non-perturbative infrared power corrections.  
We also address the complications that arise when applying the inclusive
``massive gluon'' resummation to not-completely-inclusive Minkowskian 
observables, such as the example at hand.

The paper is organized as follows: in sec.~2 we describe the specific
assumptions we make concerning the ``dressed skeleton expansion'' in
QCD and the immediate consequences that follow. We also compare the
``skeleton expansion'' with the BLM scale fixing procedure. 
In sec.~3 we discuss various natural regularizations of 
the perturbative sum, concentrating on Minkowskian quantities.
In sec.~4 we show that the regularization independence of the full QCD
result is achieved only by adding to the regularized perturbative sum
explicit power terms. We also discuss the connection with the
infrared finite coupling approach \cite{DW,DMW}. In sec.~5 we present
the application of the method to the average thrust.
Sec.~6 contains our conclusions.

\section{The ``dressed skeleton expansion'' and BLM}

Consider first a generic Euclidean quantity $D(Q^2)$, which has the 
perturbative expansion
\begin{equation}
D_{\PT}(Q^2)= r_0 a + r_1 a^2+\cdots
\label{Dpt}
\end{equation}
where $a\equiv{\alpha_s\over \pi}$ is the coupling at scale
$\mu_R^2$ in (say) the ${\rm \overline {MS}}$  scheme. Since the leading
coefficient $r_0$ is flavor independent and the next-to-leading
coefficient $r_1$ is linear in $N_f$, one can write
\begin{equation}
r_1= r_0\left[-\beta_0\left(\log{Q^2\over \mu_R^2}+\gamma_1\right)+\delta_1
\right]
\label{r1}
\end{equation}
where 
\beq
\beta_0={1\over 4}\left({11\over 3}N_c-{2\over 3}N_f\right)
\label{beta_0}
\eeq 
is the
one-loop $\beta$ function coefficient, 
\hbox{$\,\beta(a)\,=\,da/d\ln\mu_R^2=-\beta_0a^2+\cdots\,$,}
and $\gamma_1$ and $\delta_1$
are flavor independent. We shall assume, in analogy with QED, 
that $D_{\PT}(Q^2)$ has the ``dressed skeleton expansion''
\begin{equation}
D_{\PT}(Q^2)=D_0^{\PT}(Q^2)+D_1^{\PT}(Q^2)+\cdots
\label{ske}
\end{equation}
where $D_0^{\PT}$ is the contribution of a single dressed gluon exchange, 
$D_1^{\PT}$  comes from a double exchange, etc. This means in
particular that we assume for $D_0^{\PT}$ the representation  
\begin{equation}
D_0^{\PT}(Q^2)\equiv 
\int_{0}^\infty{dk^2
\over k^2}\
\bar{a}_{\PT}(k^2)\, \Phi_D(k^2/Q^2) 
\label{D0}
\end{equation}
where $\Phi_D$ is the ``momentum distribution function'' \cite{Neu}
which depends on the observable under consideration ($D$),
whereas $D_1^{\PT}$ would be given by
\begin{equation}
D_1^{\PT}(Q^2)\equiv 
\int_{0}^\infty{dk_1^2\over k_1^2}\,{dk_2^2\over k_2^2}\,
\bar{a}_{\PT}(k_1^2)\,\bar{a}_{\PT}(k_2^2)\,
\Phi_D^1(k_1^2/Q^2,k_2^2/Q^2)
\label{D1}
\end{equation}
and so on.
The physical ``skeleton coupling'' $\bar{a}_{\PT}$ appearing in both 
(\ref{D0}) and (\ref{D1}) does not depend on the observable under
consideration. It is supposed to be
{\em uniquely} determined\footnote{In QED, it is 
the Gell-Mann Low effective charge.} and is a priori 
{\em different} from the renormalization scheme coupling 
$a\equiv a(\mu_R^2)$. 
We can therefore consider its (a priori non-trivial) expansion in the 
renormalization scheme coupling
\begin{equation}
\bar{a}_{\PT}(k^2)= a \,+\, \left[-\beta_0 \left(\log{k^2\over
\mu_R^2}+c_1\right)+d_1\right]\,a^2\,+\,\cdots
\label{bara}
\end{equation}

Using eq.~(\ref{bara}) in eq.~(\ref{D0}) we obtain
\begin{equation}
D_0^{\PT}(Q^2)=r_0 a+r_1^0 a^2+\cdots
\label{D0pt}
\end{equation}
with $r_0=\phi_0$ and 
\begin{equation}
r_1^0= \phi_0\left[-\beta_0 \left(\log{Q^2\over \mu_R^2}+{\phi_1\over \phi_0}
+c_1\right)+d_1\right]
\label{r11}
\end{equation}
where $\phi_i$ are the log-moments of the momentum distribution function,
\begin{equation}
\phi_i\equiv \int_{0}^\infty{dk^2
\over k^2}\ \left(\log{k^2\over Q^2}\right)^i\
\Phi_D(k^2/Q^2). 
\label{phi_i}
\end{equation}

The only source of $N_f$ dependence in the next-to-leading order 
coefficient $r_1$ are vacuum polarization corrections which dress the 
exchanged gluon, namely corrections that are fully included in
$D_0^{\PT}(Q^2)$. 
It therefore follows that the ``Abelian'' parts proportional to 
$\beta_0$ in $r_1$ and $r_1^0$ are the same, i.e.
\begin{equation}
r_0\gamma_1 =\phi_1+\phi_0 c_1. 
\label{g1}
\end{equation}

The next observation is that the ``leading skeleton'' term
$D_0^{\PT}(Q^2)$ is a renormalization scheme invariant quantity, 
just as the ``skeleton coupling'' $\bar{a}_{\PT}(k^2)$ itself.
The question whether the approximation of $D_{\PT}(Q^2)$ by 
$D_0^{\PT}(Q^2)$ is a good one thus has a renormalization 
scheme invariant meaning. In particular the leading order correction,
which is ${\cal O}(a^2)$, can be written as
\begin{equation}
D_{\PT}(Q^2)\simeq D_0^{\PT}(Q^2)+(r_1-r_1^0)\,a^2+\cdots
\label{nlo}
\end{equation}
where $r_1-r_1^0$ is renormalization scheme invariant.
Note that
\begin{equation}
r_1-r_1^0=r_0\left(\delta_1-d_1\right)
\label{nlocoef}
\end{equation}
and the r.h.s. can be identified as the difference between 
the next-to-leading coefficients which remain after performing BLM scale
setting \cite{BLM} in the two renormalization group invariant quantities 
$D_{\PT}(Q^2)$ and $D_0^{\PT}(Q^2)$ respectively. Such a difference is 
known \cite{BLM, Grunberg-blm} to be renormalization scheme
invariant, although $\delta_1$ and $d_1$ are separately scheme dependent.

It is interesting to compare the ``skeleton expansion'' approach 
which we take here with the standard BLM scale setting procedure
\cite{BLM}:
\begin{description}
\item{1. } Eq.~(\ref{g1}) is equivalent to the statement that the BLM scale 
for the quantities $D_{\PT}(Q^2)$ and $D_0^{\PT}(Q^2)$ is the same:
\beq
\mu^2_{\BLM}=Q^2\exp\left(\frac{\phi_1}{\phi_0}+c_1 \right)
=Q^2\exp(\gamma_1)
\label{BLM_generic_scheme}
\eeq
with $\phi_i$ defined in (\ref{phi_i}).
Note in particular that if BLM is applied in the ``skeleton scheme'',
where $a=\bar{a}_{\PT}$, $c_1=0$ and then the BLM scale
is the center \cite{Neu,mom} of the momentum distribution\footnote{In order to
interpret $\Phi_D$ as a distribution function it has to be positive
definite. Although no general argument is currently known, it turns out
that in practice $\Phi_D$ is positive definite for almost all
investigated physical quantities \cite{mom}.} $\Phi_D(k^2/Q^2)$, 
\beq
\mu^2_{\BLM}=Q^2\exp\left(\frac{\phi_1}{\phi_0}\right)
\label{BLM_skeleton}
\eeq
i.e. it is the average virtuality of the exchanged dressed gluon.
Using the BLM scale, the leading order term $r_0\,a(\mu_{\BLM}^2)$
can be viewed \cite{Neu,mom} as an approximation to the entire all-order sum 
$D_0^{\PT}(Q^2)$. This approximation is good if $\Phi_D(k^2/Q^2)$ is
narrow, and it is exact for
\beq
\Phi_D(k^2/Q^2)=r_0\,\delta\left(\frac{k^2}{Q^2}-\frac{\mu_{\BLM}^2}{Q^2}\right).
\eeq
\item{2. } The usual justification of BLM (in a generic scheme) 
relies on the assumption that
$\delta_1$ is small and so the large $\beta_0$ piece 
$-\beta_0\left(\log{Q^2\over \mu_R^2}+\gamma_1\right)$ 
dominates the full next-to-leading coefficient $r_1$ in eq.~(\ref{r1}). 
This depends on the renormalization scheme, scale and $N_f$.
On the other hand, in the ``skeleton expansion'' approach 
there is no scale setting to perform, and so the accuracy of approximating 
$D_{\PT}(Q^2)$ by $D_0^{\PT}(Q^2)$ is controlled by the magnitude of
a {\em scheme invariant} coefficient (\ref{nlocoef}); the issue now is
whether $d_1$ is a good approximation to $\delta_1$.
Note also that if the quantities in eq.~(\ref{nlocoef}) 
are computed in the ``skeleton scheme'' 
then\footnote{In QED $d_1$ is always zero.} $d_1=0$ and the scheme invariant
parameter that controls the accuracy of the ``skeleton expansion'' 
can be identified as the standard BLM coefficient $r_0\,\delta_1$.
\end{description}

As mentioned in the introduction, it is not yet clear whether a 
``skeleton expansion'' exists in QCD. Thus we do not know  
the identity of the ``physical skeleton coupling'' $\bar{a}_{\PT}$. 
We do know, however, that in the Abelian limit the ``skeleton coupling'' should
coincide with the V-scheme and so the Abelian coefficient $c_1$ is determined.
For instance, in the {\hbox{$\overline{\hbox{MS}}\,$}} scheme, 
$c_1=-{5\over 3}$.
In order to perform the ``leading skeleton'' resummation in practice
we need to specify also the non-Abelian coefficient $d_1$. We shall therefore
consider in this paper three schemes which share the same $c_1$
mentioned above and differ by $d_1$ (all quoted values are in 
{\hbox{$\overline{\hbox{MS}}\,$}}): 
\begin{description}
\item{a) } The ``gluon bremsstrahlung'' coupling \cite{CMW}, where
$d_1^{\brem}=1-{\pi^2\over 4}$. This coupling was used in \cite{DW,DMW}
for parametrization of power corrections to event-shape variables.
\item{b) } The skeleton effective charge found in \cite{Watson} 
using the pinch technique, where \hbox{$d_1^{\pinch}=1$}.
\item{c) } The V-scheme coupling \cite{BLM}, defined by the static heavy quark
potential, where \hbox{$d_1^{\V}=-2$}. 
\end{description}
One might worry that without a precise identification of the
``skeleton coupling'' we introduce back some kind of renormalization scheme
dependence. We shall see that in practice (sec.~5), the inclusion of a
next-to-leading order correction as in (\ref{nlo}) effectively compensates
to a large extent for the ambiguity in $d_1$.
 
\section{Regularization}

We begin with the observation that the integrals representing the
terms $D_i^{\PT}$
in the ``skeleton expansion'', e.g. $D_0^{\PT}$ of eq.~(\ref{D0}), are
ill-defined since they run over the Landau singularity. 
This is the way the infrared renormalon ambiguity of the perturbative
sum appears in the framework of the ``skeleton expansion''.
It is possible to make these renormalon integrals mathematically
well-defined by specifying a formal procedure to avoid the
Landau singularity. However, this would not cure the {\em physical} problem of
infrared renormalons: some additional information associated
with large distances is required in order to obtain the
full QCD result from the perturbative one. 

Still, in order to use the ``skeleton expansion'' in practice one is
bound to {\em define} $D_i^{\PT}$ somehow. We shall concentrate in this
paper on the single dressed gluon term $D_0^{\PT}$ and refer to the
regularized integral of eq.~(\ref{D0}) as the ``sum of perturbation
theory''. We shall consider various possible regularization prescriptions:  
Principal Value (PV) Borel summation,
Analytic Perturbation Theory (APT) approach (``gluon mass integral''),
infrared cutoff method, and truncation of the perturbative series at 
the minimal term. 
Any two such prescriptions differ just by power terms. Thus, the ``dressed 
skeleton'' approach leads us implicitly to the subject of power
corrections. 

It is clear that physics does not depend upon the definition used. 
In practice, having no way to calculate the non-perturbative
contribution we shall just parametrize it properly and fit
the data by the ``sum of perturbation theory'' plus a power term.
The prescription dependence of the two separate contributions 
cancels by construction. While in principle any
regularization can be used, some may be more illuminating then others. 
As we shall see in sec.~3.3 and later in sec.~4 the infrared cutoff 
regularization is of special interest, allowing to separate at once 
short vs. long distance physics and perturbative vs. non-perturbative physics.

The purpose of this section is to derive the general relations between
different regularization prescriptions which are useful in the
analysis of Minkowskian observables.
We begin by discussing the application of the ``skeleton expansion'' to
Minkowskian observables concentrating on the ``leading skeleton''
term. We review the Analytic Perturbation Theory (APT)
or ``gluon mass'' \cite{BB,BBB,DMW,Grunberg-pow} integral 
which seems to be the most convenient regularization  
for a practical calculation of the perturbative sum (sec.~3.1). 
We then (sec.~3.2) discuss the power
corrections that distinguish between the
APT integral and the principal value Borel sum \cite{BB,BBB,Grunberg-pow}. 
In sec.~3.3 we explain
how an Euclidean infrared cutoff can be introduced for Minkowskian observables
and derive the relation between the cutoff regularized perturbative
sum and the APT integral. In sec.~3.4 we generalize the explicit
formulae of sec.~3.2 and 3.3 to the case of a two-loop ``skeleton
coupling'', and finally, in sec.~3.5 we comment of the comparison
between different regularizations. Appendix A describes an alternative
computation method for the cutoff regularization in terms of the APT integral.

\subsection{Minkowskian quantities and the APT integral}

We assume that a ``skeleton expansion'' such as (\ref{ske}) can be 
constructed also for Minkowskian observables. 
For a Minkowskian observable which is related by a dispersion
relation to an Euclidean quantity, linearity of the dispersion
relation implies that the expansion will take the form
\beq
R_{\PT}(Q^2)\,
=\,R_0^{\PT}(Q^2)+ R_1^{\PT}(Q^2)+\cdots
\label{ske_min}
\eeq
where $R_0^{\PT}$ is the leading ``dressed skeleton'' which is related
to $D_0^{\PT}$
by a dispersion relation \cite{Grunberg-pow}. 
Similarly, $R_1^{\PT}$ should be related by a dispersion relation to 
$D_1^{\PT}$ of eq. (\ref{D1}), and so on.
If there is no dispersion relation with an Euclidean quantity, 
the existence of a ``skeleton expansion'' is more doubtful.
In particular, as we discuss in sec.~5, there is presumably no way to
replace the entire perturbative series of not-completely-inclusive
observables such as weighted cross sections by a ``skeleton expansion''.

Let us concentrate now on the ``single dressed gluon approximation''
$R_0^{\PT}$. It turns out that $R_0^{\PT}$ cannot be expressed in the  
``Euclidean'' representation of eq.~(\ref{D0}) as an integral over the
space-like coupling $\bar{a}_{\PT}$. 
Instead, $R_0^{\PT}$ has a ``Minkowskian'' representation\footnote{This 
representation applies only to inclusive enough quantities which do
not resolve the decay products of an emitted gluon \cite{BB,BBB,DMW}.
Then the time-like ``skeleton coupling'' $\bar{a}_{\eff}^{\PT}$ 
can be reconstructed from the higher order terms related to the gluon decay.}
in terms of the time-like discontinuity of the coupling \cite{BB,BBB,Neu,DMW}, 
$R_0^{\PT}=R_{\APT}$ with
\begin{eqnarray}
\label{Rapt} 
R_{\APT}(Q^2)&\equiv& \int_0^\infty{d\mu^2 \over\mu^2}\
\bar{\rho}_{\PT}(\mu^2)\ \left[{\cal F}_R(\mu^2/Q^2)-
{\cal F}_R(0)\right] \nonumber\\
 &\equiv&\int_0^\infty{d\mu^2\over\mu^2}\ \bar{a}_{\eff}^{\PT}(\mu^2)\ 
\dot{{\cal F}}_R(\mu^2/Q^2) 
\end{eqnarray}
where the reason for the name APT shall be explained below and
\begin{equation}
\bar{\rho}_{\PT}(\mu^2) = \frac{1} {2\pi i}
 {\rm Disc}\left\{\bar{a}_{\PT}(-\mu^2)\right\}
\equiv 
\frac{1}{2\pi i}\left[\bar{a}_{\PT}\left(-\mu^2+i\epsilon\right)
-\bar{a}_{\PT}\left(-\mu^2-i\epsilon\right)\right] 
\label{discpt}
\end{equation}
is the time-like ``spectral density''. The corresponding 
Minkowskian ``effective coupling'' $\bar{a}_{\eff}^{\PT}(\mu^2)$ is 
defined by
\begin{equation} 
{d\bar{a}_{\eff}^{\PT}\over d\ln \mu^2}=\bar{\rho}_{\PT}(\mu^2).
\label{aeffpt}
\end{equation}
Note that this time-like coupling is well defined and finite at any
$\mu$ already at the perturbative level, as opposed to the space-like 
coupling that in general has Landau singularities. For instance, in
the one-loop case 
\beq
\bar{a}_{\PT}(k^2)=\frac{1}{\beta_0}\,
\frac{1}{\log \frac{k^2}{\Lambda^2} }
\label{a_oneloop}
\eeq
has a Landau singularity while the corresponding time-like coupling,
\beq
\bar{a}^{\PT}_{\eff}(\mu^2)=\frac{1}{\beta_0}\,
\left[\frac{1}{2}-\frac{1}{\pi}\arctan\left(\frac{1}{\pi}
\log \frac{\mu^2}{\Lambda^2}\right)\right] 
\label{a_eff_oneloop}
\eeq
is finite for $0<\mu^2<\infty$ and has an infrared 
fixed-point: $\bar{a}_{\eff}^{\PT}(0)=1/\beta_0$.

The ``characteristic function''  ${\cal F}_R$ in eq.~(\ref{Rapt})  is computed 
from the  one-loop Feynman diagrams with a finite gluon mass $\mu$,
and $\dot{{\cal F}}_R\equiv - d{\cal F}_R/d\ln \mu^2$. 
It is usually composed of two distinct pieces
\begin{equation}
{\cal F}_R(\mu^2/Q^2)= \left\{\begin{array}{llr}
{\cal F}_{R}^{(-)}({\mu^2/Q^2})&\,\,\,\,\,\,\,\,\,\,& 0<\mu^2<Q^2\\
 {\cal F}_{R}^{(+)} ({\mu^2/Q^2})&                  & \mu^2> Q^2
\end{array}\right.
\label{char}
\end{equation}  
where ${\cal F}_{R}^{(-)}$ is the sum of a real and a virtual contribution,
while ${\cal F}_{R}^{(+)}$ contains only the virtual contribution, 
and may vanish identically as in the case of thrust.
This property prevents \cite{Neu,Grunberg-pow} a representation
of $R_0^{\PT}$ similar to eq.~(\ref{D0}) to be reconstructed from eq. 
(\ref{Rapt}) using analyticity. 

In the Euclidean case, the characteristic function is made instead 
of a {\em single} piece and satisfies a dispersion relation. 
Its discontinuity is \cite{Neu,BBB1} nothing but the Euclidean 
``momentum distribution function'' $\Phi_D$ of eq.~(\ref{D0})
\begin{equation}
\Phi_D(k^2/Q^2)
\equiv-{1\over 2\pi i}\left[{\cal F}_{D}\left({k^2\over Q^2} e^{i\pi}
\right)-{\cal F}_{D}\left({k^2\over Q^2} e^{-i\pi}\right)\right].
\label{phi_D}
\end{equation}
However,  
\begin{equation}
D_{\APT}(Q^2)
\equiv \int_0^\infty {d\mu^2\over\mu^2}\ \bar{a}_{\eff}^{\PT}(\mu^2)\ 
\dot{{\cal F}}_D(\mu^2/Q^2) 
\label{Dapt} 
\end{equation}
differs from $D^{\PT}_0$ of (\ref{D0}) by power terms: while $D^{\PT}_0$ is
ambiguous involving integration over the Landau singularity, $D_{\APT}$ is well
defined thanks to the non-trivial infrared fixed point in
$\bar{a}_{\eff}^{\PT}(\mu^2)$, e.g. (\ref{a_eff_oneloop}) in the
one-loop case.

One can use analytic continuation to express $D_{\APT}$ 
in the Euclidean representation. This yields
\begin{equation}
D_{\APT}(Q^2)=\int_{0}^\infty{dk^2 \over k^2}\
\bar{a}_{\APT}(k^2)\
\Phi_D(k^2/Q^2) 
\label{D_APT}
\end{equation}
where $\bar{a}_{\APT}$ is defined through the following dispersion relation
\begin{eqnarray}
\label{dispapt}
\bar{a}_{\APT}(k^2) &\equiv& -\int_0^\infty{d\mu^2\over
\mu^2+k^2}\
\bar{\rho}_{\PT}(\mu^2) \nonumber\\
&\equiv&k^2\int_0^\infty{d\mu^2\over(\mu^2+k^2)^2}\ \bar{a}_{\eff}^{\PT}
(\mu^2)
\end{eqnarray}
which implies in particular the absence of Landau singularity. 
The name Analytic Perturbation Theory \cite{SS} is attached to this coupling
since it has, by definition, a physical analyticity structure.   
The APT coupling differs from the usual perturbative coupling by power
terms
\begin{equation}
\bar{a}_{\APT}(k^2)=\bar{a}_{\PT}(k^2) + \delta\bar{a}_{\APT}(k^2)
\label{d-apt}
\end{equation}
with 
\begin{equation}
\delta\bar{a}_{\APT}(k^2)=\sum_{n=1}^\infty (-1)^{n+1} b_n 
\left({\Lambda^2\over k^2}\right)^n
\label{dapt-as}
\end{equation} 
and in the one-loop case
\begin{equation}
\delta\bar{a}_{\APT}(k^2)=-{1\over\beta_0}{1\over{k^2\over \Lambda^2}-1}.
\label{d-apt1}
\end{equation}

Returning now to Minkowskian quantities, we keep the subscript  
``APT'' in eq.~(\ref{Rapt}) in order to stress the fact that this
``gluon mass integral'' is well defined and is {\em different} 
from the corresponding Borel sum $R_0^{\PT}$.  
$R_{\APT}$ and $R_0^{\PT}$ share by definition the same perturbative
expansion. The two differ by power corrections which are in general
ambiguous, and thus
$R_{\APT}$ can also be looked at as a particular regularization of 
$R_0^{\PT}$.  

\subsection{APT vs. Borel sum}

It is useful to derive an expression of the Borel sum\footnote{The 
``leading skeleton'' subscript $0$ is dropped in
the following for simplicity: $R_{\PT}$ refers from now on to the
Borel sum of the ``leading skeleton'' ($R_0^{\PT}$ in the previous
section).} $R_{\PT}$ in terms of 
$R_{\APT}$, following \cite{BB,BBB}. The ``renormalization 
scheme invariant'' Borel representation \cite{BY,B2,G3} of
eq.~(\ref{Rapt}) can be written as \cite{Grunberg-pow} 
\begin{equation}
R_{\PT}(Q^2)=\int_0^\infty dz\ \tilde{a}_{\eff}(z)\ 
\left[\int_0^{\infty}{d\mu^2\over \mu^2}\
\dot{{\cal F}}_R(\mu^2/Q^2)\ 
\exp\left(-z\beta_0\ln{\mu^2\over \Lambda^2}\right)\right]
\label{R-borel}
\end{equation}
where $\tilde{a}_{\eff}(z)$ is the (renormalization scheme invariant) 
Borel image of the Minkowskian coupling
\begin{equation}
\bar{a}_{\eff}^{\PT}(\mu^2)=\int_0^\infty dz
\exp\left(-z\beta_0\ln{\mu^2 \over\Lambda^2}\right)\
 \tilde{a}_{\eff}(z).
\label{aeff-borel}
\end{equation}
$\tilde{a}_{\eff}(z)$ is related to the Borel image $\tilde{a}(z)$ 
of the space-like coupling 
\begin{equation}
\bar{a}_{\PT}(k^2)=\int_0^\infty dz 
\exp\left(-z\beta_0\ln{k^2 \over\Lambda^2}\right)\
 \tilde{a}(z) 
\label{a-borel}
\end{equation}
by the relation \cite{BY}
\begin{equation}
\tilde{a}_{\eff}(z)={\sin(\pi\beta_0 z)
\over\pi\beta_0 z}\ \tilde{a}(z).
\label{b-b}
\end{equation}
We define $\Lambda$ to be the Landau singularity of $\bar{a}_{\PT}$ and
assume that the Borel integrals converge for $k^2$ and $\mu^2$ larger
then $\Lambda^2$ and that the Borel representation of $R_{\PT}$ 
converges for large enough $Q^2$.
Eq.~(\ref{R-borel}) is obtained by substituting eq.~(\ref{aeff-borel}) 
into eq.~(\ref{Rapt}), and interchanging the order of integrations.

Next we choose a separation scale $\mu_I>\Lambda$ and write
\begin{equation}
R_{\PT}(Q^2)=R_<^{\PT}(Q^2)+R_>^{\PT}(Q^2)
\label{Rpt-split}
\end{equation} 
where
\begin{equation}R_<^{\PT}(Q^2)\equiv\int_0^\infty 
dz\ \tilde{a}_{\eff}(z)\ 
\left[\int_0^{\mu_I^2}{d\mu^2\over \mu^2}\
\dot{{\cal F}}_R(\mu^2/Q^2)\ 
\exp\left(-z\beta_0\ln{\mu^2\over
\Lambda^2}\right)\right]\label{Rpt<}\end{equation}
contains the infrared renormalon ambiguities, and
\begin{equation}
R_>^{\PT}(Q^2)\equiv\int_0^\infty dz\ \tilde{a}_{\eff}(z)\ 
\left[\int_{\mu_I^2}^{\infty}{d\mu^2\over \mu^2}\
\dot{{\cal F}}_R(\mu^2/Q^2)\ 
\exp\left(-z\beta_0\ln{\mu^2\over
\Lambda^2}\right)\right]
\label{Rpt>}
\end{equation}
is unambiguous.
Since $\mu_I>\Lambda$, we can replace 
$\bar{a}_{\eff}^{\PT}(\mu^2)$ by its Borel representation
for $\mu>\mu_I$, and therefore
\beq
R_>^{\PT}(Q^2)=\int_{\mu_I^2}^\infty{d\mu^2\over\mu^2}
\ \bar{a}_{\eff}^{\PT}(\mu^2)\ \dot{{\cal F}}_R(\mu^2/Q^2)
\equiv R_{\APT}(Q^2)-R_<^{\APT}(Q^2)
\label{Rapt>}
\eeq
where 
\begin{equation}
R_<^{\APT}(Q^2)\equiv\int_0^{\mu_I^2}{d\mu^2\over\mu^2}
 \bar{a}_{\eff}^{\PT}(\mu^2)\ \dot{{\cal F}}_R(\mu^2/Q^2)
\label{Rapt<}.
\end{equation} 
We thus end up with
\begin{equation}R_{\PT}(Q^2)=R_{\APT}(Q^2)-\delta R_{\APT}(Q^2)
\label{Rapt-pt}
\end{equation}
where \cite{Grunberg-pow}
\begin{equation}
\delta R_{\APT}(Q^2)\equiv R_<^{\APT}(Q^2)-R_<^{\PT}(Q^2).
\label{dRapt}
\end{equation}

At the one-loop level $\delta R_{\APT}$ can be expressed quite
elegantly as \cite{BB,BBB}
\begin{equation}
\left.\delta R_{\APT}\right\vert_{\1loop}=-{1\over\beta_0} \left[{\cal
F}_R(-\Lambda^2/ Q^2)-{\cal F}_R(0)\right].
\label{BB_for}
\end{equation}
The principal value Borel sum $R_{\PT\vert\PV}$ is then obtained by
replacing $\delta R_{\APT}$ by its real part in eq.~(\ref{Rapt-pt}).
The imaginary part of $\delta R_{\APT}$ provides a measure of
the summation ambiguity.    

Note that although $R_<^{\APT}$ and $R_<^{\PT}$ in (\ref{dRapt}) are 
separately $\mu_I$ dependent, their difference $\delta R_{\APT}$ does
 not depend on $\mu_I$. This formula is of practical utility
when only the small $\mu^2/Q^2$ expansion of ${\cal F}_R$ is known
analytically (e.g. the average thrust; see sec.~5): $R_{\APT}$ can
then be obtained from~(\ref{Rapt}) by numerical integration over 
${\cal F}_R$, whereas $\delta R_{\APT}$, which involves only scales
below $\mu_I$, can be evaluated at large $Q^2$ using 
the small $\mu^2/Q^2$ expansion of ${\cal F}_R$.

Consider indeed the contribution of a generic term in the small $\mu^2/Q^2$ 
expansion of ${\cal F}_R$, namely assume 
\begin{equation}
{\cal F}_R(\mu^2/Q^2)-{\cal F}_R(0)=\left({\mu^2\over
Q^2}\right)^n\ \left(B^{(n)}_R 
\log{Q^2\over \mu^2}+C^{(n)}_R\right)
\label{F-low}
\end{equation}
where $n$ is not necessarily integer. Eq.~(\ref{F-low}) implies 
\begin{equation}
\dot{{\cal F}}_R(\mu^2/Q^2)= -n\left({\mu^2\over
Q^2}\right)^n\ \left(B^{(n)}_R 
\log{Q^2\over \mu^2}+C^{(n)}_R-{B^{(n)}_R\over n}\right).
\label{Fdot-low}
\end{equation}
We then obtain from eq.~(\ref{Rpt<})
\begin{eqnarray}
\label{Rpt<n}
&R&_{<,n}^{\PT}(Q^2)= \nonumber \\
&-&\left({\mu_I^2\over Q^2}\right)^n
\left( B^{(n)}_R 
\log{Q^2\over \mu_I^2}+ C^{(n)}_R-{B^{(n)}_R\over n}\right) 
\int_0^\infty{ dz\ \tilde{a}_{\eff}(z) \,
 {1\over 1-{z\over z_n}}\,
\exp\left(-z\beta_0\ln{\mu_I^2\over\Lambda^2}\right)}\nonumber\\
&- &\left({\mu_I^2\over Q^2}\right)^n{B^{(n)}_R\over n} 
\int_0^\infty 
dz\ \tilde{a}_{\eff}(z)\, 
 {1\over \left(1-{z\over z_n}\right)^2}\, \exp\left(-z\beta_0\ln{\mu_I^2\over
\Lambda^2}\right)
\end{eqnarray}
where $z_n\equiv {n\over \beta_0}$, and from eq.~(\ref{Rapt<})
\begin{eqnarray}
\label{Rapt<n}
&R&_{<,n}^{\APT}(Q^2)=\nonumber \\
&-&\left({\mu_I^2\over Q^2}\right)^n \left(B^{(n)}_R 
\log{Q^2\over \mu_I^2}+C^{(n)}_R-{B^{(n)}_R\over
n}\right)\int_0^{\mu_I^2}\,n\, {d\mu^2\over\mu^2}\left({\mu^2\over \mu_I^2}
\right)^n\ \bar{a}_{\eff}^{\PT}(\mu^2)\nonumber\\
 &-&\left({\mu_I^2\over Q^2}\right)^n\  {B^{(n)}_R\over
n}\int_0^{\mu_I^2}\,n^2\,{d\mu^2\over\mu^2}
\left({\mu^2\over \mu_I^2}
\right)^n\ \log{\mu_I^2\over \mu^2}\ \bar{a}_{\eff}^{\PT}(\mu^2).
\end{eqnarray}
In both (\ref{Rpt<n}) and (\ref{Rapt<n}) the combination
in front of the first integral (the single Borel pole
integral in (\ref{Rpt<n})) is the same as
in $\dot{{\cal F}}_R$. The double Borel pole integral in
(\ref{Rpt<n}) originates from the logarithmic term in 
(\ref{F-low}) and it can be obtained by differentiating the single 
pole integral with respect to $n$.

One deduces the generic contribution to $\delta R_{\APT}$ at large $Q^2$
\begin{equation}
\delta R_n^{\APT}=-\left({\Lambda^2\over
Q^2}\right)^n\left[\left(B^{(n)}_R 
\log{Q^2\over \Lambda^2}+ C^{(n)}_R-{B^{(n)}_R\over n}\right) b_n
+{B^{(n)}_R\over n}\ b_n^{1}\right]
\label{dRaptn}
\end{equation}
with
\begin{equation}
b_n\equiv
\int_0^{\Lambda^2}\,n\,{d\mu^2\over\mu^2}\left({\mu^2\over
\Lambda^2}\right)^n
\bar{a}_{\eff}^{\PT}(\mu^2)-\int_0^\infty dz\  
\tilde{a}_{\eff}(z)\
\frac{1}{1-\frac{z}{z_n}}
\label{bn-apt}
\end{equation}
and
\begin{equation}
b_n^{1}\equiv \int_0^{\Lambda^2}\,n^2\,{d\mu^2\over\mu^2}
\left({\mu^2\over \Lambda^2}\right)^n\
\ln{\Lambda^2\over\mu^2}\ \bar{a}_{\eff}^{\PT}(\mu^2)
-\int_0^\infty dz\  
 \tilde{a}_{\eff}(z)\
\frac{1}{\left(1-\frac{z}{z_n}\right)^2}
\label{barbn-apt}
\end{equation}
where we have used the fact that $\delta R_n^{\APT}$ does not depend
on $\mu_I$ and thus have set $\mu_I=\Lambda$.
Note that $b_n^{1}$ is related to $b_n$ by
\beq
b_n^{1}=n^2\frac{d}{dn}\left(\frac{1}{n}\,b_n\right).
\eeq
 
The integrals (\ref{bn-apt}) and (\ref{barbn-apt}) are in general
difficult to calculate. 
However, in the one-loop case ($\tilde{a}(z)\equiv 1$) we can obtain
the final results for $b_n$ and $b_n^{1}$ by comparing 
(\ref{dRaptn}) with (\ref{BB_for}):
\begin{equation}
\left. b_n\right\vert_{\1loop}={1\over\beta_0}\,e^{\pm i \pi n}
\label{bn-apt1}
\end{equation}
and (deriving with respect to $n$)
\begin{equation}b_n^{1}\vert_{\1loop}={1\over\beta_0}\,
e^{\pm i\pi n} \left(1\pm i\pi n\right).
\label{barbn-apt1}
\end{equation}

Eq.~(\ref{bn-apt1}) can also be checked for $n$ integer with the
observation \cite{Grunberg-pow} that the $b_n$'s of eq.~(\ref{bn-apt}) then
coincide with the coefficients in the large $k^2$
expansion of $\delta\bar{a}_{\APT}(k^2)$  defined in
eq.~(\ref{dapt-as}).

\subsection{Infrared cutoff regularization}

Let us now turn to the infrared cutoff regularization. 
Consider first an Euclidean quantity $D(Q^2)$. A 
natural regularization of eq.~(\ref{D0}) is obtained by introducing an infrared
cutoff $\mu_I$ (above the Landau singularity)
\begin{equation}
D_{\UV}^{\PT}(Q^2)\equiv \int_{\mu_I^2}^\infty{dk^2\over
k^2}\ \bar{a}_{\PT}(k^2)\
\Phi_D(k^2/Q^2)
\label{Dptuv1}
\end{equation}
which removes the long distance part of the perturbative contribution
\begin{eqnarray}
\label{Dptuv2}
D_{\UV}^{\PT}(Q^2)&=&\int_{0}^\infty
{dk^2\over k^2}\
\bar{a}_{\PT}(k^2)\ \Phi_D(k^2/Q^2)-
\int_{0}^{\mu_I^2}{dk^2\over k^2}\
\bar{a}_{\PT}(k^2)\ \Phi_D(k^2/Q^2)\nonumber\\
&\equiv&D_{\PT}(Q^2)-D_{\IR}^{\PT}(Q^2).
\end{eqnarray}
The regularized perturbative sum (\ref{Dptuv1}) is fully under control
in perturbation theory, provided $\mu_I/\Lambda$ is large enough. One
can then safely evaluate the ``leading skeleton'' term using a one 
or two loop approximation to the running
``skeleton coupling''. The accuracy of
this approximation can be estimated comparing the results 
obtained at the one and two loop levels.
In this respect the cutoff regularization is of special interest: 
other regularizations, e.g. the principal value Borel sum,
involve properties of the coupling in the infrared \hbox{$0<k^2<\mu_I^2$}. 
It is then hard to explain why the one-loop or two-loop couplings 
are of any relevance.

For Minkowskian quantities, we have seen that a representation such as
eq.~(\ref{D0}) does not exist. One could imagine a regularization 
in the form (\ref{Rpt>}). However, as opposed to a separation 
between large and small space-like momentum, 
the separation between large and small time-like momentum 
(\ref{Rpt-split}) does not correspond to a separation between 
short and long distances. 

It is natural to define instead \cite{Mil,DSW,Grunberg-pow}, 
just as for Euclidean quantities, 
\begin{equation}
R_{\UV}^{\PT}(Q^2)\equiv
R_{\PT}(Q^2)-R_{\IR}^{\PT}(Q^2)
\label{Rptuv}
\end{equation}
where $R_{\PT}$ is the Borel sum (eq.~(\ref{R-borel})), and
\begin{equation}
R_{\IR}^{\PT}(Q^2)\equiv \int_0^{\mu_I^2}{dk^2\over k^2}
 \bar{a}_{\PT}(k^2)\ \Phi_R(k^2/Q^2).
\label{Rptir}
\end{equation}
Here $\Phi_R(k^2/Q^2)$ is  (minus) the discontinuity  at $\mu^2=-k^2<0$
of the ``small gluon mass'' piece ${\cal F}_{R}^{(-)}$ of the
characteristic function (cf. eq.~(\ref{phi_D}))
\begin{equation}
\Phi_R(k^2/Q^2)\equiv-{1\over 2\pi i}
\left[{\cal F}_{R}^{(-)}\left({k^2\over Q^2} e^{i\pi}
\right)-{\cal F}_{R}^{(-)}\left({k^2\over Q^2}
e^{-i\pi}\right)\right].
\label{phi}
\end{equation}

The regularized perturbative sum $R_{\UV}^{\PT}$ defined
in eq.~(\ref{Rptuv}) has a similar structure to its Euclidean analogue 
$D_{\UV}^{\PT}$ of eq.~(\ref{Dptuv2}). The only difference is
that $R_{\UV}^{\PT}$ (like $R_{\PT}$) cannot be
written as an integral over the space-like ``skeleton coupling'' 
$\bar{a}_{\PT}$. For Euclidean quantities, the prescription of
eq.~(\ref{Rptuv}, \ref{Rptir}) reproduces eq.~(\ref{Dptuv1}). 
In \cite{Grunberg-pow}, it was shown explicitly using a Borel representation
that $R_{\UV}^{\PT}$ is indeed free of any ambiguity from infrared
renormalons. We have
\begin{equation}
R_{\UV}^{\PT}(Q^2)=\int_0^\infty 
dz\ \tilde{a}_{\eff}(z)\ 
\left[\int_0^{\infty}{d\mu^2\over \mu^2}\
\dot{{\cal F}}_{R}^{\UV}(\mu^2,Q^2)\ \exp
\left(-z\beta_0\ln{\mu^2\over
\Lambda^2}\right)\right]
\label{Rregb}
\end{equation}
where
\begin{equation}
{\cal F}_{R}^{\UV}(\mu^2,Q^2)
\equiv
{\cal F}_R(\mu^2/Q^2)-
\int_0^{\mu_I^2}{dk^2\over k^2+\mu^2}\ 
\Phi_R(k^2/Q^2)
\label{Freg}
\end{equation}
is the ``infrared regularized'' characteristic function, and
\beq
\dot{{\cal F}}_{R}^{\UV}(\mu^2,Q^2)\equiv - \frac{\partial{\cal
F}_{R}^{\UV}(\mu^2,Q^2)}{\partial\ln \mu^2}.
\eeq
The effect of the subtracted term in eq.~(\ref{Freg}) is to remove
any potential non-analytic term in the small $\mu^2$ (``gluon mass'')
expansion of $\dot{{\cal F}}_{R}^{\UV}(\mu^2,Q^2)$ (now distinct from 
its large $Q^2$ behavior, since there is a 
third scale $\mu_I$ involved), by introducing an infrared cutoff $\mu_I$ on the
space-like gluon propagator momentum. This implies the cancellation
of infrared renormalons, which are related only to non-analytic
terms \cite{BBZ,BB,BBB,DMW} in the small $\mu^2$ expansion of
$\dot{\cal F}(\mu^2/Q^2)$.
The Borel representation eq.~(\ref{Rregb}) is not so practical for
concrete calculations, especially in cases (like the one of the
average thrust which will be considered in sec.~5) where the
characteristic function is not known analytically. 
In this section we describe an alternative method to compute
$R_{\UV}^{\PT}$, based on the ``gluon mass'' regularization 
(\ref{Rapt}).

Using eq.~(\ref{Rapt-pt}) one can indeed write
\begin{equation}
R_{\UV}^{\PT}(Q^2)=R_{\APT}(Q^2)-\Delta R(Q^2)
\label{Rptuv1}
\end{equation}
with
\begin{eqnarray}
\label{DR}
\Delta R(Q^2)&\equiv& R_{\IR}^{\PT}(Q^2)+\delta R_{\APT}(Q^2)\nonumber
             \\
             &=     & R_{\IR}^{\PT} - R_<^{\PT} + R_<^{\APT}
\end{eqnarray}
where the overall $\mu_I$-dependence is that of $R_{\IR}^{\PT}$. 
$R_{\UV}^{\PT}$ should be free of any renormalon ambiguity.
Since $R_{\APT}$ in (\ref{Rptuv1}) is unambiguous, so is $\Delta R$.  
We further know that $R_<^{\APT}$ in (\ref{DR}) is unambiguous and
thus it is clear that the ambiguities present in $R_{\IR}^{\PT}$ and
$R_<^{\PT}$ should cancel. 
We can therefore calculate $\Delta R$ as a sum of two well-defined
contributions\footnote{An alternative method to calculate $\Delta R$
based on the APT coupling is described in the appendix.}: 
\begin{description}
\item{a) } $\left(R_{\IR}^{\PT} - R_<^{\PT}\right)$ which is  
a Borel integral free of renormalon ambiguities.
\item{b) } $R_<^{\APT}$ which is the ``small gluon mass'' integral.
\end{description} 

Let us now calculate $\left(R_{\IR}^{\PT} - R_<^{\PT}\right)$
explicitly for a generic term such as (\ref{F-low}) in the small $\mu^2$
expansion of ${\cal F}_R$ and check that it is indeed unambiguous.
Using (\ref{phi}) we obtain
\begin{equation}
\Phi_R(k^2/Q^2)= -\left({k^2\over Q^2}\right)^n\ 
\left[{\sin \pi n\over\pi}\left( B^{(n)}_R \log{Q^2\over k^2}
+ C^{(n)}_R\right)-B^{(n)}_R \cos \,\pi n\right]
\label{phi-low}
\end{equation}
which implies that
\begin{eqnarray}
\label{Rirptn}
&R&_{\IR,n}^{\PT}(Q^2)= \nonumber \\
&-&\left({\mu_I^2\over Q^2}\right)^n
\left[\left( B^{(n)}_R \log{Q^2\over \mu_I^2}+ C^{(n)}_R\right)  \,
{\rm sinc}\ \pi n - {B^{(n)}_R\over n} \cos \,\pi n\right]\,
I_n\left(\mu_I^2/\Lambda^2\right)\nonumber\\
&-&\left({\mu_I^2\over Q^2}\right)^n {B^{(n)}_R\over n}\ {\rm sinc}\ \pi n\,\,
I^1_n\left(\mu_I^2/\Lambda^2\right)
\end{eqnarray}
where ${\rm sinc}\ x\equiv {\sin x\over x}$,
\begin{equation}
I_n\left(\mu_I^2/\Lambda^2\right)\equiv\int_0^{\mu_I^2}\,n\,{dk^2\over k^2}
\left({k^2\over \mu_I^2}\right)^n \bar{a}_{\PT}(k^2)
\label{In}
\end{equation}
and
\begin{equation}
I^1_n\left(\mu_I^2/\Lambda^2\right)
\equiv\int_0^{\mu_I^2}\,n^2\,{dk^2\over k^2}
\left({k^2\over \mu_I^2}\right)^n \log{\mu_I^2\over k^2
}\,\, \bar{a}_{\PT}(k^2).
\label{barIn}
\end{equation}
Replacing $\bar{a}_{\PT}(k^2)$ by its Borel representation
(\ref{a-borel}), one gets
\beq
I_n\left(\mu_I^2/\Lambda^2\right)=\int_0^\infty 
dz\ \tilde{a}(z)\, {1\over 1-{z\over z_n}}\,
\exp\left(-z\beta_0\ln{\mu_I^2\over\Lambda^2}\right)
\label{In_borel}
\eeq
and
\beq
I^1_n\left(\mu_I^2/\Lambda^2\right)=\int_0^\infty 
dz\ \tilde{a}(z)\,{1\over \left(1-{z\over z_n}\right)^2}\,
\exp\left(-z\beta_0\ln{\mu_I^2\over\Lambda^2}\right).
\label{I1n_borel}
\eeq
Note that both $R_{<,n}^{\PT}$ (eq.~(\ref{Rpt<n})) and $R_{\IR,n}^{\PT}$ 
have simple and double renormalons poles
at $z=z_n\equiv {n\over \beta_0}$. 
As mentioned before, the double poles reflect the log-enhanced small 
$\mu^2$ behavior of ${\cal F}_R$.
 
As an example, in the one-loop case 
where $\tilde{a}(z)\equiv 1$ 
the integrals in eq.~(\ref{In_borel}) and
(\ref{I1n_borel}) yield 
\begin{equation}
I_n\vert_{\1loop}=\,-\,{n\over \beta_0} 
e^{-nt_I}\, {\rm Ei}_1\left(-n t_I\right)
\label{In_Ei}
\end{equation}
and
\begin{equation}
\left. I^1_n\right\vert_{\1loop}=\,-\,{n\over \beta_0} 
\left[n t_I\,e^{-nt_I}\, {\rm Ei}_1\left(-nt_I\right) + 1 \right] 
\label{barIn_Eilog}
\end{equation}
where $t_I \equiv \ln\left(\mu_I^2/\Lambda^2\right)$ and
the exponential integral function is defined in the complex plane by
\beq
{\rm Ei}_k (x)\equiv \int_1^\infty \, e^{-xz}\,\frac{dz}{z^k}.
\label{Ei_def} 
\eeq
This function has a cut on the positive real axis, and thus $I_n$ and
$I_n^1$ are {\em ambiguous} as expected from eq.~(\ref{In_borel}) and
(\ref{I1n_borel}).

On the other hand, using eq.~(\ref{b-b}) and (\ref{Rpt<n}) one deduces
\begin{eqnarray}
\label{Rptir-<-as1}
 &R&_{\IR,n}^{\PT}(Q^2)-R_{<,n}^{\PT}(Q^2)=  \\ \nonumber
&-&\left({\mu_I^2\over Q^2}\right)^n \left( B^{(n)}_R 
\log{Q^2\over \mu_I^2}+ C^{(n)}_R\right) \, p_n
\left({\mu_I^2}/{\Lambda^2} \right)
-\left({\mu_I^2\over Q^2}\right)^n\frac{B^{(n)}_R}{n} \,
p^1_n\left({\mu_I^2}/{\Lambda^2}\right) 
\end{eqnarray}
where
\beq
p_n\left({\mu_I^2}/{\Lambda^2}\right)=\int_0^\infty dz\ \tilde{a}(z) \, 
{1\over 1-{z\over z_n}}\,\left[{{\rm sinc}\,\pi n-{\rm sinc}\,
\pi \beta_0 z }\right] \, \exp\left(-z\beta_0\ln{\mu_I^2\over\Lambda^2}\right)
\label{p}
\eeq
and
\begin{eqnarray}
\label{p1}
&p^1_n&\left({\mu_I^2}/{\Lambda^2}\right)=  \\
&&\int_0^\infty dz\ \tilde{a}(z)  \,{1\over 1-{z\over z_n}}\,
\left[ {\rm sinc}\, \pi\beta_0 z-\cos \,\pi n 
- {{\rm sinc}\, \pi\beta_0 z-{\rm sinc}\, \pi n\over 1-{z\over z_n}}
\right]\, \exp\left(-z\beta_0\ln{\mu_I^2\over\Lambda^2}\right) \nonumber
\end{eqnarray}
Explicit calculation of the limit of the integrands in $p_n$ and $p_n^1$
for $z \longrightarrow z_n$ shows that the {\em renormalon pole cancels}.

Note that the combination appearing in front 
of $p_n$ in (\ref{Rptir-<-as1}) is the same as in ${\cal
F}_R$. Alternatively, we can
rewrite eq.~(\ref{Rptir-<-as1}) such that the combination in front of
$p_n$ will be that of $\dot{{\cal F}}_R$, namely
\begin{eqnarray}
\label{Rptir-<-as2}
&R&_{\IR,n}^{\PT}(Q^2)-R_{<,n}^{\PT}(Q^2) \\
&& =-\left({\mu_I^2\over Q^2}\right)^n \left( B^{(n)}_R 
\log{Q^2\over \mu_I^2}+ C^{(n)}_R-{B^{(n)}_R\over n}\right) \, p_n
\left({\mu_I^2}/{\Lambda^2}\right) -
\left({\mu_I^2\over Q^2}\right)^n{B^{(n)}_R\over n} \, \tilde{p}_n^1
\left({\mu_I^2}/{\Lambda^2}\right) \nonumber
\end{eqnarray}
where
\begin{eqnarray}
&\tilde{p}_n^1&\left({\mu_I^2}/{\Lambda^2}\right)=\\
&&\int_0^\infty dz\ \tilde{a}(z) \,{1\over 1-{z\over z_n}}\,
\left[ {\rm sinc}\ \pi n -\cos\, \pi n - {{\rm sinc}\ \pi\beta_0 z
-{\rm sinc}\ \pi n\over 1-{z\over z_n}} \right]\,
\exp\left(-z\beta_0\ln{\mu_I^2\over \Lambda^2}\right) \nonumber
\end{eqnarray}
It is now straightforward to see that the expression in the square
brackets vanishes in the limit $z \longrightarrow z_n$. 

In the one-loop case, where $\tilde{a}(z)\equiv 1$, all the Borel
integrals can be evaluated analytically.
We obtain from (\ref{p}) 
\beq
\left. p_n\right\vert_{\1loop}=\frac{1}{\beta_0 \,\pi}{\rm Re}\left\{i\,
{\rm Ei}_1\left(n\,(i\pi-t_I)\right)\,e^{n\,(i\pi-t_I)}\right\} 
+\frac{1}{\beta_0}
\left[\frac{1}{2}-\frac{1}{\pi}\arctan\left(\frac{t_I}{\pi}\right)\right].
\label{p_1loop}
\eeq
The exponential integral function in the first term in (\ref{p_1loop}) 
is defined in (\ref{Ei_def}) and it is unambiguous for complex arguments.
We identify (see (\ref{a_eff_oneloop})) the second term 
in (\ref{p_1loop}) as the time-like
coupling at the cutoff scale, namely $\bar{a}^{\PT}_{\eff}(\mu_I^2)$.
Note that this term is independent of $n$.
Similarly, we obtain from (\ref{p1})
\beq
\left. p^1_n\right\vert_{\1loop}
=\,\frac{n}{\beta_0 \,\pi}{\rm Re}\left\{ (-i)\,
{\rm Ei_1}\left(n\,(i\pi-t_I)\right)\,e^{n\,(i\pi-t_I)}\,(i\pi-t_I)\right\}. 
\label{p1_1loop}
\eeq 

The second ingredient required for the calculation of $\Delta R$ in
(\ref{DR}) is the ``small gluon mass'' integral
$R_{<,n}^{\APT}$ which is defined in (\ref{Rapt<}) and written explicitly in 
(\ref{Rapt<n}). In order to calculate $R_{<,n}^{\APT}$
it is convenient to first integrate  eq.~(\ref{Rapt<}) by parts to get an
expression in terms of the discontinuity of the coupling
\begin{equation}
R_<^{\APT}(Q^2)=-\left[{\cal F}_R(\mu_I^2/ Q^2)-{\cal F}_R(0)\right]
\bar{a}_{\eff}^{\PT}(\mu_I^2)+\int_0^{\mu_I^2}{d\mu^2\over\mu^2}
\left[{\cal F}_R(\mu^2/ Q^2)-{\cal F}_R(0)\right]
\bar{\rho}_{\PT}(\mu^2)
\label{Rapt<1}
\end{equation}
We take again a generic term (\ref{F-low}) in ${\cal F}_R(\mu^2/Q^2)$ 
and perform the integral analytically in the one-loop case. 
The result is
\begin{eqnarray}
\label{Rapt<_1loop}
 &R&_{<,n}^{\APT}(Q^2)=  \\ \nonumber
&-&\left({\mu_I^2\over Q^2}\right)^n \left( B^{(n)}_R 
\log{Q^2\over \mu_I^2}+ C^{(n)}_R\right) \, h_n
\left({\mu_I^2}/{\Lambda^2} \right)
-\left({\mu_I^2\over Q^2}\right)^n\frac{B^{(n)}_R}{n} \,
h^1_n\left({\mu_I^2}/{\Lambda^2}\right) 
\end{eqnarray}
where $h_n=-p_n\,+\,q_n\,+\,g_n$ and
$h^1_n=-p^1_n\,+\,q^1_n\,+\,g^1_n$. The functions $q_n$ and $q_n^1$,
as well as $g_n$ and $g_n^1$, will be given a simple interpretation
below. In the one-loop case
\begin{eqnarray}
\label{g}
g_n\vert_{\1loop}&=&\,\frac{1}{\beta_0}\,\cos \,\pi n\, e^{-nt_I}\\
\nonumber
\left.\,g^1_n\right\vert_{\1loop}&=&\,\frac{n}{\beta_0}\,(\pi\sin n
\pi+t_I \cos\, \pi n)e^{-nt_I}
\end{eqnarray}
and
\begin{eqnarray}
\label{q}
q_n\vert_{\1loop}&=&\,\frac{1}{\beta_0} \,\frac{\sin \pi n}{\pi}
\,e^{-nt_I}\,{\rm Ei}(nt_I)  \\  \nonumber 
\left.\,q^1_n\right\vert_{\1loop}&=&\frac{1}{\beta_0}
\left[ -\,\frac{\sin \pi n}{\pi}\,+\,
\,e^{-nt_I}\,{\rm Ei}(nt_I)\,n\,\left(t_I\,\frac{\sin \pi n}{\pi}\,-\cos \,\pi n
\right)\right]
\end{eqnarray}
where ${\rm Ei}(x)$ in (\ref{q}) 
is the Cauchy principal value exponential integral function, 
defined for real arguments such that
for negative $x$, ${\rm Ei}(x)=-{\rm Ei}_1 (-x)$ and for positive $x$,
${\rm Ei}(x)={\rm Re}\left\{ -{\rm Ei}_1 (-x)\right\}$, where ${\rm Ei}_k$
is defined in (\ref{Ei_def}).
 
Combining eqs. (\ref{Rptir-<-as1}) and (\ref{Rapt<_1loop}) according to
\beq
\Delta R= \left(R_{\IR}^{\PT}-R_<^{\PT}\right)\,+\,R_<^{\APT}
\label{delta_R_sep}
\eeq
we find that the generic large $Q^2$ contribution $\Delta R_n$   
is given by
\begin{eqnarray}
\label{DR_1loop_ph}
\Delta R_n(Q^2)&=& - \left({\mu_I^2\over Q^2}\right)^n \left( B^{(n)}_R 
\log{Q^2\over \mu_I^2}+ C^{(n)}_R\right) \, 
\left[p_n\left({\mu_I^2}/{\Lambda^2} \right)+
h_n\left({\mu_I^2}/{\Lambda^2} \right)\right]\nonumber \\
&-&\left({\mu_I^2\over Q^2}\right)^n\frac{B^{(n)}_R}{n} \,
\left[p^1_n\left({\mu_I^2}/{\Lambda^2}\right)+ 
h^1_n\left({\mu_I^2}/{\Lambda^2}\right)\right]. 
\end{eqnarray}

Finally, we can rewrite the result as indicated in the first line of
(\ref{DR}) separating $\Delta R$ into $R_{\IR}^{\PT}$ and 
$\delta R_{\APT}=R^{\APT}_<-R^{\PT}_{<}$.
As explained above, $R_{\IR}^{\PT}$ and $\delta R_{\APT}$ are both
imaginary and ambiguous. Nevertheless, knowing that the imaginary parts
cancel in $\Delta R$, we can write 
\begin{equation}
\Delta R_n={\rm Re}\left\{R_{\IR,n}^{\PT}\right\}+
{\rm Re} \left\{\delta R^{\APT}_{n}\right\}.
\label{dR-1}
\end{equation}
This separation of $\Delta R$ makes a clear connection between
the cutoff regularization and the Borel sum principal value
regularization $R_{\PT\vert\PV}$, 
\begin{equation}
R_{\UV}^{\PT}=R_{\APT}-{\rm Re} \left\{\delta
R_{\APT}\right\}-{\rm Re}\left\{R_{\IR}^{\PT}\right\}
=R_{\PT\vert\PV}-{\rm Re}\left\{R_{\IR}^{\PT}\right\}.
\label{Rptuv-1}
\end{equation}
We now make the following crucial identification:
\begin{eqnarray}
\label{qn_form}
 &{\rm Re}&\left\{R_{\IR,n}^{\PT}\right\}=  \\ \nonumber
&-&\left({\mu_I^2\over Q^2}\right)^n \left( B^{(n)}_R 
\log{Q^2\over \mu_I^2}+ C^{(n)}_R\right) \, q_n
\left({\mu_I^2}/{\Lambda^2} \right)
-\left({\mu_I^2\over Q^2}\right)^n\frac{B^{(n)}_R}{n} \,
q^1_n\left({\mu_I^2}/{\Lambda^2}\right) 
\end{eqnarray}
where the one-loop functions $q_n$ and $q_n^1$ of eq.~(\ref{q}) can be 
obtained directly from (\ref{Rirptn}) with (\ref{In_Ei}) and
(\ref{barIn_Eilog}), while
\begin{eqnarray}
\label{gn_form}
 &{\rm Re}& \left\{\delta R^{\APT}_{n}\right\}=  \\ \nonumber
&-&\left({\mu_I^2\over Q^2}\right)^n \left( B^{(n)}_R 
\log{Q^2\over \mu_I^2}+ C^{(n)}_R\right) \, g_n
\left({\mu_I^2}/{\Lambda^2} \right)
-\left({\mu_I^2\over Q^2}\right)^n\frac{B^{(n)}_R}{n} \,
g^1_n\left({\mu_I^2}/{\Lambda^2}\right) 
\end{eqnarray}
where $g_n$ and $g^1_n$ of eq.~(\ref{g}) can be obtained directly 
from eq.~(\ref{dRaptn}) with (\ref{bn-apt1}) and (\ref{barbn-apt1}). 
Adding (\ref{qn_form}) and (\ref{gn_form}) and using the relations
$p_n+h_n=q_n+g_n$ and $p_n^1+h_n^1=q_n^1+g_n^1$ we recover (\ref{DR_1loop_ph}).

We emphasize that $q_n$ and $q^1_n$ have formal power series expansions
in terms of $1/t_I$ or $\bar{a}_{\PT}(\mu_I^2)$.
The leading term in this expansion, which is a valid approximation at 
large enough $\mu_I$ can be obtained by replacing $\bar{a}_{\PT}(k^2)$ with
$\bar{a}_{\PT}(\mu_I^2)$ inside the integrals in eqs. (\ref{In}) and
(\ref{barIn}), 
\begin{eqnarray}
&&\Delta R_n(Q^2)\simeq
R_{\IR,n}^{\PT}(Q^2)\label{Rirpt-lead}\simeq\\
&&-\left({\mu_I^2\over Q^2}\right)^n
\bar{a}_{\PT}(\mu_I^2) \left[ \left(B^{(n)}_R 
\log{Q^2\over \mu_I^2}+C^{(n)}_R\right) \,
{\rm sinc}\ \pi n + {B^{(n)}_R\over n} 
({\rm sinc}\ \pi n-\cos \pi n)\right]\nonumber.
\end{eqnarray}
On the other hand $g_n$ and $g_n^1$ do not have similar 
expansions in $1/t_I$, since they are proportional to a power of
$\Lambda^2/\mu_I^2$. Substituting $g_n$ and $g_n^1$ in
(\ref{gn_form}) one obtains a cutoff independent result for 
${\rm Re} \left\{\delta R^{\APT}_{n}\right\}$, 
\begin{eqnarray}
 &{\rm Re}& \left.\left\{\delta
R^{\APT}_{n}\right\}\right\vert_{\1loop}
= \nonumber \\
&-&\left({\Lambda^2\over Q^2}\right)^n \,\frac{1}{\beta_0}\,
\left[\left( B^{(n)}_R 
\log{Q^2\over \Lambda^2}+ C^{(n)}_R\right) \, \cos \pi n\,+\,
{B^{(n)}_R}\,\pi\,\sin \pi n\right].
\label{delta_R_n_1loop}
\end{eqnarray}
In general, if $\mu_I\gg\Lambda$, 
\beq
\left\vert {\rm Re}\left\{\delta R^{\APT^{\,}}_{n_{\,}}\right\}\right\vert
\ll
\left\vert
{\rm Re}\left\{R_{\IR,n}^{\PT}\right\}\right\vert,
\label{delta_R_APT_small}
\eeq
since ${\rm Re}\left\{\delta R^{\APT}_{n}\right\}$
behaves as
$\left(\Lambda^2/Q^2\right)^n\,=\,
\left(\mu_I^2/Q^2\right)^n\,\left(\Lambda^2/\mu_I^2\right)^n$ while
${\rm Re}\left\{R_{\IR,n}^{\PT}\right\}$ behaves 
as $\left(\mu_I^2/Q^2\right)^n\,\bar{a}_{\PT}(\mu_I^2)$.
The only exception to (\ref{delta_R_APT_small}) is an {\em analytic} term 
(integer $n$ with no logarithmic terms: $B_R^{(n)}=0$) where ${\rm
Re}\left\{R_{\IR,n}^{\PT}\right\}$ vanishes identically.
Note, on the other hand, 
that if the leading $n$ is half-integer and there are no logarithmic 
terms ($B^{(n)}_R=0$) then ${\rm Re}\left\{\delta R^{\APT}_{n}\right\}$
vanishes identically. In this case $R_{\APT}$ coincides with the 
principal value Borel sum, up to some sub-leading power corrections which 
can be ignored if $Q^2$ is large enough. 

\subsection{Generalization to two-loop running coupling}

\setcounter{footnote}{0}
In the ``skeleton expansion'' approach (\ref{ske}) all the diagrams that 
correspond to dressing a single gluon are formally contained in the first 
term $D_0^{\PT}$. Identifying the coupling in (\ref{D0}) or
(\ref{Rapt}) with the one-loop coupling, thus amounts to some approximation
already at the level of the leading term in this expansion.
In order to improve this approximation, or at least to have some
information about its accuracy, it is important to calculate the integral
also with the two-loop running coupling.

Since the ``skeleton expansion'' is not systematically defined, the
justification of (\ref{D0}) or (\ref{Rapt}) with the
particular function $\Phi_D(k^2/Q^2)$ or ${\cal F}_R(\mu^2/Q^2)$ 
is based on the 
large $N_f$ limit and the ``Naive Non-Abelianization'' procedure. 
In this approximation the coupling is strictly one-loop, and so 
using eq.~(\ref{D0}) or (\ref{Rapt}) beyond one-loop is
not really justified. Still, having in mind the picture of the
``skeleton expansion'', we regard 
the running coupling $\bar{a}_{\PT}$ inside the integral as
an all-order coupling and eventually as a (non-perturbative) infrared
regular coupling. The first stage is to promote it from the
one-loop to the two-loop level. 

Technically, performing the infrared cutoff regularization with a 
two-loop running
coupling is more involved, but as we shall see in this section, 
it is achievable.  
One possibility is to use the exact
expression \cite{G3} for the renormalization scheme invariant Borel 
transform $\tilde{a}(z)$ of eq.~(\ref{a-borel}), but
the resulting integrals are not obviously tractable. 
Alternatively, we use the observation that for any coupling which has 
only\footnote{Specifically we require absence of complex 
Landau singularities.} a space-like Landau cut 
ending at $k^2=-\Lambda^2$ the following dispersion relation is satisfied
\begin{equation}
\bar{a}_{\PT}(k^2)=
-\int_{-\Lambda^2}^\infty{d\mu^2\over\mu^2+k^2}\ 
\bar{\rho}_{\PT}(\mu^2).
\label{disp-pt}
\end{equation}
This holds in particular for the two-loop coupling with $\beta_1>0$.
It then follows \cite{G5} from eqs. (\ref{d-apt}) and (\ref{dispapt}) that
\begin{equation}
\delta\bar{a}_{\APT}(k^2)=\int_{-\Lambda^2}^0{d\mu^2\over\mu^2+k^2}\ 
\bar{\rho}_{\PT}(\mu^2).
\label{disp-dapt}
\end{equation}
We therefore find that the $b_n$'s in eq.~(\ref{dapt-as}) are given by
\begin{equation}
b_n= e^{\pm i \pi n}\, \int_{-\Lambda^2}^0{d\mu^2\over\mu^2}\ 
\left(-{\mu^2\over \Lambda^2}\right)^n \bar{\rho}_{\PT}(\mu^2)
\label{bn-apt2}
\end{equation}
where we exhibited the phase arising from the negative integration range.
As mentioned above, for integer $n$, the $b_n$'s in
eq.~(\ref{dapt-as}) and thus also in eq.~(\ref{bn-apt2}) coincide with
those of eq.~(\ref{bn-apt}). We assume that the identity between 
(\ref{bn-apt2}) and (\ref{bn-apt}) holds also for non-integer $n$.

Consider now the case of a pure power contribution (i.e. no logs:
$B_R^{(n)}=0$) to ${\cal F}_R$. Then eq.~(\ref{dRaptn}) gives
\begin{equation}\delta R_n^{\APT}=-C^{(n)}_R\ b_n\left({\Lambda^2\over Q^2
}\right)^n
\label{dRapt-low1}\end{equation}
On the other hand, eq.~(\ref{Rirptn}) gives
\begin{equation}
R_{\IR,n}^{\PT}=-C^{(n)}_R \left({\mu_I^2\over Q^2 }\right)^n {\rm sinc}\
\pi n\ I_n\left({\mu_I^2}/{\Lambda^2}\right) 
\label{Rirpt-low1}
\end{equation}

It was shown in \cite{G4,DU} that if $\bar{a}_{\PT}$ 
satisfies the two-loop renormalization group equation
\begin{equation}
{d\bar{a}_{\PT}\over d\ln k^2} 
= -\beta_0 \,\bar{a}_{\PT}^2 -\beta_1 \,\bar{a}_{\PT}^3,
\label{a-2loop}
\end{equation}
the {\em standard}\, Borel representation of $I_n$ is
\begin{equation}
I_n=\int_0^{\infty} dz\ \exp\ \left(- \frac{z}{a_I}\right) 
\frac{\exp\left(-\frac{\beta_1}{\beta_0}\ z\right)}{\left(1 -
\frac{z}{z_n}\right)^{1+\delta_n}}
=\int_0^{\infty} dz\ \exp\ \left(- \frac{z}{\tilde{a}_I}\right) 
\frac{1}{\left(1 - \frac{z}{z_n}\right)^{1+\delta_n}}
\label{In-2loop}
\end{equation}
where $a_I\equiv \bar{a}_{\PT}(\mu_I^2)$, 
\beq
\delta_n\equiv {\beta_1\over \beta_0}z_n=n\frac{\beta_1}{\beta_0^2}
\label{delta_n}
\eeq
and 
\begin{equation}
{1\over \tilde{a}_I}\equiv {1\over a_I}+{\beta_1\over \beta_0}.
\label{atilde-I}
\end{equation}
The integral in eq.~(\ref{In-2loop}) can be expressed
in terms of the incomplete gamma function, 
\begin{equation}
I_n=-z_n \left(-{z_n\over \tilde{a}_I}\right)^{\delta_n}
\exp\left(-{z_n\over \tilde{a}_I}\right)\,\Gamma\left(-\delta_n,-{z_n\over
\tilde{a}_I}\right)
\label{In-2loop1}
\end{equation}
where 
\beq
\Gamma(z,x)\equiv \int_x^{\infty}\,{dt\over t}\, t^z \,e^{-t}. 
\label{Gamma_def}
\eeq

Let us first compute the (ambiguous) imaginary part of $I_n$. It is
convenient to write
\begin{eqnarray}
\label{In-split}
I_n&=&\int_0^{z_n} dz\ \exp\ \left(-
\frac{z}{\tilde{a}_I}\right) 
\frac{1}{\left(1 - \frac{z}{z_n}\right)^{1+\delta_n}}
+\int_{z_n}^{\infty} dz\ \exp\ \left(- \frac{z}{\tilde{a}_I}\right) 
\frac{1}{\left(1 - \frac{z}{z_n}\right)^{1+\delta_n}}\nonumber\\
&\equiv&I_{n}^{(-)}+I_{n}^{(+)}
\end{eqnarray}
where only $I_{n}^{(+)}$ contributes to the imaginary part. We 
have\footnote{This simple form can be obtained 
from eq.~(\ref{In-2loop1}) by putting to zero the second argument of
the incomplete gamma function.} \cite{G6}
\begin{equation}
I_{n}^{(+)}=-z_n \left(-{z_n\over \tilde{a}_I}\right)^{\delta_n}
\exp\left(-{z_n\over \tilde{a}_I}\right)\, \Gamma(-\delta_n),
\label{In+}
\end{equation}
and therefore
\begin{eqnarray}
\label{Im-In}
{\rm Im}\left\{I_n\right\}={\rm Im}\left\{I_{n}^{(+)}\right\}
&=&\mp \sin \pi \delta_n\ z_n 
\left({z_n\over \tilde{a}_I}\right)^{\delta_n}
\exp\left(-{z_n\over \tilde{a}_I}\right)\,\Gamma(-\delta_n)\nonumber\\
&=&\pm \pi\ z_n\ {1\over \Gamma(1+\delta_n)}\left({z_n\over
\tilde{a}_I}\right)^{\delta_n}
\exp\left(-{z_n\over \tilde{a}_I}\right)
\end{eqnarray}
where the identity 
\beq
\Gamma(-\delta_n)\equiv{-\pi\over\sin 
\pi\delta_n}\,{1\over \Gamma(1+\delta_n)}
\label{Gamma_iden}
\eeq
has been used in the second line.

Integrating the two-loop renormalization group equation
eq.~(\ref{a-2loop}) gives
\begin{equation}
\left({z_n\over \tilde{a}_I}\right)^{\delta_n}
\exp\left(-{z_n\over \tilde{a}_I}\right)=\delta_n^{\delta_n}\
e^{-\delta_n}\left({\Lambda^2\over\mu_I^2}
\right)^n
\label{atilde-2loop}
\end{equation}
where $\Lambda$ is the tip of the Landau cut. From (\ref{Rirpt-low1}),
(\ref{Im-In}) and (\ref{atilde-2loop}) it follows that
\begin{equation}
{\rm Im}\left\{R_{\IR,n}^{\PT}\right\}
=\pm C^{(n)}_R\ {\rm sinc}\ \pi n\
\pi\ z_n\ {1\over \Gamma(1+\delta_n)}\ \delta_n^{\delta_n}\
e^{-\delta_n}\left({\Lambda^2\over Q^2}\right)^n
\label{im-rptir}
\end{equation}

\setcounter{footnote}{0}
We can now compute $b_n$ of eq.~(\ref{bn-apt2}) observing (see
sec.~3.3) that this imaginary part must 
cancel with the one coming from $\delta R_{\APT}$, namely 
(eq.~(\ref{dRapt-low1}))
\begin{equation}
{\rm Im}\left\{\delta R_n^{\APT}\right\}=-C^{(n)}_R\
{\rm Im}\left\{ b_n\right\}\left({\Lambda^2\over Q^2 }\right)^n.
\label{Im-dRapt}
\end{equation}
From (\ref{bn-apt2}), which is assumed\footnote{What we actually use
here is only the assumption that the phase of $b_n$ is $e^{\pm
i\pi n}$.} to be valid also for non-integer
$n$, we obtain
\begin{equation}
{\rm Im} \left\{b_n\right\}=\pm \sin \pi n
\int_{-\Lambda^2}^0{d\mu^2\over\mu^2}\ 
\left(-{\mu^2\over \Lambda^2}\right)^n \bar{\rho}_{\PT}(\mu^2).
\label{Im-bn-apt}
\end{equation}
Comparing eq.~(\ref{im-rptir}) with eq.~(\ref{Im-dRapt}), 
the cancellation condition gives 
\begin{equation}
\int_{-\Lambda^2}^0{d\mu^2\over\mu^2}\ 
\left(-{\mu^2\over \Lambda^2}\right)^n \bar{\rho}_{\PT}(\mu^2)={1\over \beta_0}
{1\over \Gamma(1+\delta_n)}\ \delta_n^{\delta_n}\
e^{-\delta_n}
\label{disc}
\end{equation}
where the sign is determined by the observation that
$\bar{\rho}_{\PT}(\mu^2)$ is negative also when
$\mu^2$ is negative, which implies
\begin{equation}
b_n\vert_{\2loop}=e^{\pm i\pi n}{1\over \beta_0}
{1\over \Gamma(1+\delta_n)}\ \delta_n^{\delta_n}\
e^{-\delta_n}.
\label{bn-apt3}
\end{equation}  
Taking the real part of $\delta R_{\APT}$ 
(eq.~(\ref{dRapt-low1})) and using eq.~(\ref{bn-apt3}) we get
\begin{equation}
\left.{\rm Re}\left\{\delta R_n^{\APT}\right\}\right\vert_{\2loop}
=-{C^{(n)}_R\over \beta_0}
\left({\Lambda^2\over Q^2 }\right)^n \cos \pi n\ 
{1\over \Gamma(1+\delta_n)}\ \delta_n^{\delta_n}\
e^{-\delta_n},
\label{rebn-apt2}
\end{equation}
which generalizes eq.~(\ref{delta_R_n_1loop}) with $B_R^{(n)}=0$ to two-loops. 
Just like in the one-loop case, when $n$ is half-integer (\ref{rebn-apt2}) 
vanishes and then $R_{\APT}$ coincides with the principal value Borel sum.

To compute $\Delta R$ (eq.~(\ref{dR-1})), we next need 
${\rm Re}\left\{R_{\IR}^{\PT}\right\}$.
We have
\begin{equation}
{\rm Re}\left\{R_{\IR,n}^{\PT}\right\}
=-C^{(n)}_R \left({\mu_I^2\over Q^2 }\right)^n 
{\rm sinc}\ \pi n\ {\rm Re}\left\{I_n\right\} 
\label{reRirpt-low1}
\end{equation}
with (eq.~(\ref{In-split})) ${\rm Re}\left\{I_n\right\}
=I_{n}^{(-)}+{\rm Re}\left\{I_{n}^{(+)}\right\}$. But
\begin{equation}
I_{n}^{(-)}=z_n \left(-{z_n\over \tilde{a}_I}\right)^{\delta_n}
\gamma\left(-\delta_n,-{z_n\over \tilde{a}_I}\right)\exp\left(-{z_n\over
\tilde{a}_I}\right)
\label{In(-)-2loop1}
\end{equation}
where 
\beq
\gamma(z,x)\equiv\int_0^x{dt\over t} \,t^z\, e^{-t}=\Gamma(z)-\Gamma(z,x),
\eeq
hence
\begin{eqnarray}
\label{reIn(-)-2loop1}
{\rm Re}\left\{I_n\right\}
&=&z_n \left(-{z_n\over \tilde{a}_I}\right)^{\delta_n}
\gamma\left(-\delta_n,-{z_n\over \tilde{a}_I}\right)\exp\left(-{z_n\over
\tilde{a}_I}\right)\nonumber \\
&-& z_n\cos \pi\delta_n\, \Gamma(-\delta_n)\left({z_n\over
\tilde{a}_I}\right)^{\delta_n}
\exp\left(-{z_n\over \tilde{a}_I}\right)
\end{eqnarray}
where we also used eq.~(\ref{In+}). It follows that
\begin{eqnarray}
\label{reRirpt-low2}
\left.{\rm Re}\left\{R_{\IR,n}^{\PT}\right\}\right\vert_{\2loop}&
=&-{C^{(n)}_R\over \beta_0} {\sin \pi n\over \pi}
 \left({\mu_I^2\over Q^2 }\right)^n\left(-{z_n\over
\tilde{a}_I}\right)^{\delta_n}
\gamma\left(-\delta_n,-{z_n\over \tilde{a}_I}\right)\exp\left(-{z_n\over
\tilde{a}_I}\right)\nonumber\\
& &+{C^{(n)}_R\over \beta_0} {\sin \pi n\over \pi}\left({\Lambda^2\over Q^2
}\right)^n \cos \pi\delta_n\, 
\Gamma(-\delta_n)\ \delta_n^{\delta_n}\, e^{-\delta_n}
\end{eqnarray}
where we used eq.~(\ref{atilde-2loop}) to simplify the $\mu_I$
independent part.
We checked that taking the limit $\beta_1\longrightarrow 0$ 
in eq.~(\ref{reRirpt-low2}), the singularities in the two terms cancel
and the one-loop result of eq.~(\ref{qn_form}) with~(\ref{q}) is recovered.
Combining (\ref{reRirpt-low2}) and (\ref{rebn-apt2}) we end up with
\begin{eqnarray}
\label{dR-low2}
\left.\Delta R_n\right\vert_{\2loop}&
=&-{C^{(n)}_R\over \beta_0} {\sin \pi n\over \pi}
 \left({\mu_I^2\over Q^2 }\right)^n\left(-{z_n\over
\tilde{a}_I}\right)^{\delta_n}
\gamma\left(-\delta_n,-{z_n\over
\tilde{a}_I}\right)\exp\left(-{z_n\over \tilde{a}_I}\right) \\
& &-{C^{(n)}_R\over \beta_0} \left({\Lambda^2\over Q^2 }\right)^n {1\over
\Gamma(1+\delta_n)}\ 
\delta_n^{\delta_n}\ e^{-\delta_n} \left(\cos \pi n\, +\, \sin \pi n \, {\cos
\pi\delta_n\over
\sin \pi\delta_n}\right) \nonumber
\end{eqnarray}
where the identity (\ref{Gamma_iden}) has been used in the second term. 
The extension of these results to the more general case where a log term 
is present in the small gluon mass expansion of ${\cal F}_R$ can be dealt
with using the $d/dn$ trick.

\subsection{Comparing different regularizations}

It follows from eq.~(\ref{Rptuv-1}) that $R_{\UV}^{\PT}$
and $R_{\PT\vert\PV}$ differ by {\em infrared} renormalon power 
terms contained in $\rm{Re}\left\{R_{\IR}^{\PT}\right\}$, which are related to 
the {\em non-analytic} terms \cite{BB,BBB,BBZ,DMW} in the small $\mu^2$ 
expansion of ${\cal F}_R$.
This implies that these two regularizations are in fact equivalent for
the analysis of power corrections of infrared origin, as we shall see 
in the next section. The same statement can be made for the effective
regularization obtained by truncating the perturbative series in a given
renormalization scheme at the minimum term, which turns out to be 
numerically close to the principal value regularization\footnote{
Truncation at the minimal term is a priori a scheme dependent procedure,
but we find that this scheme dependence is small.}
(see sec.~5.4 and fig.~\ref{BLM} and~\ref{K_110_reg}).  

On the other hand, an arbitrary regularization may differ from the
principal value Borel sum by power terms of {\em ultraviolet} origin
which arise from {\em analytic} terms in the small $\mu^2$ expansion
of ${\cal F}_R$. Examples of the latter kind are the APT
regularization, where the contribution of analytic terms to $\delta
R_{\APT}$ is apparent in (\ref{delta_R_n_1loop}) and
(\ref{rebn-apt2}), and the cutoff regularization in the Minkowskian
representation $R^{\PT}_>$, eq.~(\ref{Rpt>}).  
This property makes these regularizations inconvenient for the analysis 
of infrared power corrections.
We stress, however, that when the leading power corrections are of the
$1/Q$ type (assuming $n=\frac{1}{2}$ and $B^{\left(\frac{1}{2}\right)}_R=0$) 
the APT regularization becomes quite convenient since it
coincides (see the end of sec.~3.3) with the principal value Borel sum
up to some sub-leading power corrections which can be usually ignored. 

\section{Power corrections}

Let us assume that the perturbative analysis of the ``leading
skeleton'' term $R_0^{\PT}$ in eq.~(\ref{Rapt}) indicates a leading
renormalon at $z=z_n=n/\beta_0$ in the Borel plane, which implies that
the Borel sum ambiguity is ${\cal O}(1/Q^{2n})$. Since the full
(non-perturbative) QCD result should be unambiguous, it differs at
large $Q^2$ from a generic regularization of the perturbative 
sum ($R^{\PT}_r$) by a ${\cal O}(1/Q^{2n})$ power correction (see  
refs.~\cite{david,BBK}).
In principle, the non-perturbative corrections should be
calculable in QCD. In the absence of such a calculation 
one will naturally attempt to fit experimental data by
\begin{equation}
R =  R^{\PT}_r + {\lambda_r\over Q^{2n}},
\label{pow-r}
\end{equation} 
neglecting, for simplicity, possible sub-leading power corrections. 
In (\ref{pow-r}) both the regularized sum of perturbation 
theory $R^{\PT}_r$ and the fitted coefficient $\lambda_r$
depend on the regularization method $r$ used. Their sum should of course 
be independent of $r$ if the fit is successful. 
If two regularization methods $r$ and $r'$ differ by a (known) 
infrared power term $\lambda_{r,r'}/Q^{2n}$ they are equivalent, 
in the sense that
\begin{equation}
\lambda_{r'}=\lambda_r+\lambda_{r,r'}
\label{lambda-rr}
\end{equation}
i.e. going from one regularization to the other amounts to a 
straightforward redefinition of the power term coefficient. In
particular we have seen that
this holds for the principal value Borel sum 
$R_{\PT\vert\PV}$ and the momentum cutoff regularization $R_{\UV}^{\PT}$.
Nevertheless, the latter allows for a possible physical interpretation 
of the power corrections in terms 
of an infrared finite coupling \cite{DW,DMW}.

In the approach of \cite{DW, DMW} (see also \cite{Neu}) 
one assumes the existence of a non-perturbative coupling 
\begin{equation}
\bar{a}(k^2)=\bar{a}_{\PT}(k^2)+\delta\bar{a}(k^2)
\label{a-full}
\end{equation} 
regular in the infrared region. 
As opposed to $\bar{a}_{\PT}$, the non-perturbative coupling 
$\bar{a}$ is assumed to satisfy the dispersion relation\footnote{While
the APT coupling (\ref{dispapt}) has all the assumed properties for
the non-perturbative coupling $\bar{a}$, we do {\em not} imply that it
is the correct model.}
\begin{eqnarray}
\label{disp}
\bar{a}(k^2) &=& -\int_0^\infty{d\mu^2\over \mu^2+k^2}\
\bar{\rho}(\mu^2) \nonumber\\
&=&k^2\int_0^\infty{d\mu^2\over(\mu^2+k^2)^2}\ \bar{a}_{\eff}
(\mu^2),
\end{eqnarray}
where $\bar{\rho}(\mu^2)$ is the discontinuity of $\bar{a}$, defined similarly 
to its perturbative part (\ref{discpt}), and $\bar{a}_{\eff}(\mu^2)$ 
is defined  by
\begin{equation} 
{d\bar{a}_{\eff}\over d\ln \mu^2}\equiv \bar{\rho}(\mu^2).
\label{aeff}
\end{equation}
In the framework of the ``skeleton expansion''
such a non-perturbative extension appears quite natural:
if the coupling is regular in the infrared each term $D_i$ in
(\ref{ske}) is well defined. 
We shall therefore assume here that it is the 
``skeleton coupling'' which plays the role
of the infrared finite coupling of \cite{DW,DMW}.
Let us consider, as before, only the first term in this expansion,
the equivalent of (\ref{D0}), 
\begin{equation}
D(Q^2)  =\int_{0}^\infty{dk^2 \over k^2}
\,\bar{a}(k^2)\,\Phi_D(k^2/Q^2). 
\label{D0np}
\end{equation}

One can write \cite{Grunberg-pow} 
\begin{eqnarray}
\label{D-split}
D(Q^2)&=&\int_{0}^{\mu_I^2}{dk^2 \over k^2}\
\bar{a}(k^2)\ \Phi_D(k^2/Q^2) \nonumber \\
&+& \int_{\mu_I^2}^\infty{dk^2\over k^2}\ \bar{a}_{\PT}(k^2)\
\Phi_D(k^2/Q^2)\,+\,
\int_{\mu_I^2}^\infty{dk^2\over k^2}\ \delta\bar{a}(k^2)\
\Phi_D(k^2/Q^2)\nonumber\\
&\equiv&D_{\IR}(Q^2)+D_{\UV}^{\PT}(Q^2)+\delta D_{\UV}(Q^2)
\end{eqnarray}
where $\mu_I$ is for the moment an arbitrary infrared cutoff. Following 
\cite{DW, DMW, Neu}, we shall assume one can choose
$\mu_I$ in such a way that at scales above  $\mu_I$ the full coupling
$\bar{a}$ is well 
approximated by its perturbative piece $\bar{a}_{\PT}$. One can then
neglect the last ultraviolet piece 
in eq.~(\ref{D-split})
\begin{eqnarray}
\label{D-split1}
D(Q^2)&\simeq&\int_{0}^{\mu_I^2}{dk^2
\over k^2}\
\bar{a}(k^2)\ \Phi_D(k^2/Q^2)
+ \int_{\mu_I^2}^\infty{dk^2\over k^2}\ \bar{a}_{\PT}(k^2)\
\Phi_D(k^2/Q^2)\nonumber\\
&\equiv&D_{\IR}(Q^2)+D_{\UV}^{\PT}(Q^2).
\end{eqnarray}

The infrared cutoff regularization thus appears naturally in the
present framework.
The power corrections  arise only from the infrared piece
$D_{\IR}(Q^2)$\footnote{See \cite{Grunberg-pow} for a discussion of
the more general case where the contribution
of $\delta D_{\UV}$ is kept.}. This piece yields,
for large $Q^2$, non-perturbative ``long distance''
power contributions  which correspond to the standard condensates
for observables that admit an operator product expansion.  
If the Feynman diagram kernel 
$\Phi_D(k^2/Q^2)$ is 
${\cal O}\left[(k^2/Q^2)^n\right]$ at
small
$k^2$, this piece contributes an  
${\cal O}\left[(\Lambda^2/Q^2)^n\right]$
term related to a dimension $n$ condensate, with the normalization given by a
small virtuality moment of the infrared regular coupling
$\bar{a}(k^2)$ (see eq.~(\ref{Rir_n}) below).

The generalization of this approach to (inclusive enough) 
Minkowskian quantities has been given in \cite{DMW}.  
One simply extends eq.~(\ref{Rapt}) to the full non-perturbative
coupling, to obtain at the single gluon exchange level 
\begin{eqnarray}
 \label{R} 
R(Q^2)&=& \int_0^\infty{d\mu^2 \over\mu^2}\
\bar{\rho}(\mu^2)\ \left[{\cal F}_R(\mu^2/Q^2)-
{\cal F}_R(0)\right] \nonumber\\
 &=&\int_0^\infty{d\mu^2\over\mu^2}\ \bar{a}_{\eff}(\mu^2)\ 
\dot{{\cal F}}_R(\mu^2/Q^2).
\end{eqnarray}
The analogue of eq.~(\ref{D-split}) is \cite{Mil,DSW,Grunberg-pow} 
\begin{equation} R(Q^2)= R_{\IR}(Q^2)
+R_{\UV}^{\PT}(Q^2)+\delta R_{\UV}(Q^2)
\label{R-split}
\end{equation}
with
\begin{equation}
R_{\IR}(Q^2)\equiv \int_0^{\mu_I^2}{dk^2\over k^2}
\ \bar{a}(k^2)\ \Phi_R(k^2/Q^2)
\label{Rir}
\end{equation}
and $R_{\UV}^{\PT}(Q^2)$ defined in eq.~(\ref{Rptuv},\,\ref{Rptir}).
Assuming again that the ultraviolet piece
$\delta R_{\UV}$ can be neglected, we end up with
\begin{equation} R(Q^2)\simeq R_{\IR}(Q^2)
+R_{\UV}^{\PT}(Q^2).
\label{R-split1}
\end{equation}
In practice \cite{DW,DMW}, 
one expands $\Phi_R(k^2/Q^2)$ at small $k^2$ and obtains 
an approximation to $R_{\IR}(Q^2)$ based on the leading power correction.
Consider for example the case of a leading renormalon at a 
half integer $n$ in (\ref{F-low}) with no logarithms ($B_R^{(n)}=0$).
Then $\Phi_R(k^2/Q^2)$ in eq.~(\ref{phi}) is given by (see eq.~(\ref{phi-low}))
\beq
\Phi_R(k^2/Q^2)=-\left(\frac{k^2}{Q^2}\right)^n\,C_R^{(n)}\,
\frac{\sin{\pi n}}{\pi} 
\eeq
and
\begin{equation}
R_{\IR}(Q^2)= -\left(\frac{\mu_I^2}{Q^2}\right)^n\,C_R^{(n)}\,
\frac{\sin{\pi n}}{\pi n}  \left[
\int_0^{\mu_I^2}n\,{dk^2\over k^2}\,\left(\frac{k^{2}}{\mu_I^{2}}\right)^n
\ \bar{a}(k^2) \right]
\label{Rir_n}
\end{equation}
where the integral in the square brackets is a specific 
moment of the {\em universal}
infrared regular ``skeleton coupling'' which serves as a
non-perturbative parameter and the observable dependent coefficient 
in front comes out of the perturbative calculation of ${\cal F}(\mu^2/Q^2)$.

In this approach, the infrared cutoff $\mu_I$ itself 
acquires some physical meaning. 
It is bound to be small enough such that the approximation of
$\Phi_R(k^2/Q^2)$ by the leading order term in the small $k^2$
expansion will be valid, allowing e.g. the parametrization of $R_{\IR}$ in
the form (\ref{Rir_n}). On the other hand, $\mu_I$ should be large
enough such that $\bar{a}$ will be approximated by the perturbative
coupling $\bar{a}_{\PT}$ above $\mu_I$. This would ensure that
$R^{\PT}_{\UV}$ is under perturbative control. 
Moreover, the universality of the ``skeleton coupling'' $\bar{a}$ 
requires that for different observables a {\em common} $\mu_I$ 
is chosen consistently with the requirements above. 
Such a choice would allow to compare the non-perturbative parameters
obtained by fitting experimental data with (\ref{R-split1}) for 
different observables that share the same leading renormalon behavior. 
Note that it is possible that the data is well fitted by what
appears to be an entirely ``perturbative'', but regularized, 
ansatz such as $R_{\PT\vert\PV}$, or even $R_{\UV}^{\PT}$ with an 
unrealistically low choice of $\mu_I$ (such as $\mu_I=\Lambda$!). However,
this would not deter the alternative interpretation of the same data 
in terms of eq.~(\ref{R-split1}), but this time with
a more ``realistic'' larger value of $\mu_I$: we indeed saw above 
that the fit result cannot depend on the choice
of $\mu_I$. On the other hand, the arbitrariness of the regularization
implies one
cannot fix the correct ``physical'' $\mu_I$ studying of a single observable.

\section{Application: average thrust}

As an example of the method proposed we analyze here a specific
observable, the average thrust $\left< T\right>$ in $e^+e^-$
annihilation.
The thrust characterizes how ``pencil like'' the event is. It is
defined as
\beq
T=\frac{\sum_i \left\vert \vec{p}_i \cdot \vec{n}_T\right\vert}
{\sum _i \left\vert \vec{p}_i \right\vert},
\label{T_def}
\eeq
where $i$ runs over all the particles in the final state,
$\vec{p}_i$ are the 3-momenta of the particles and $\vec{n}_T$ is the
thrust axis which is defined for a given event such that $T$ is
maximized. 
For a ``pencil like'' 2-jet event, $T$ approaches $1$. It is therefore
natural to define $t\equiv 1-T$, such that $t$ vanishes in this limit.

The definition (\ref{T_def}) guarantees that $T$ does not change due
to emission of extremely soft gluons, i.e. it is {\em infrared safe}.
In addition $T$ does not change due to a collinear split of a particle,
i.e. it is {\em collinear safe}. These properties suggest
\cite{Sterman_Weinberg} that the
thrust distribution and the average thrust can be calculated in 
perturbative QCD from {\em parton} momenta and compared with 
experimental measurements where the thrust for a given event is
obtained from {\em hadron} momenta. 
The gap between partons and hadrons may result in a non-perturbative
modification of the perturbative result due to confinement effects.

\begin{figure}[htb]
\begin{center}
\mbox{\kern-0.5cm
\epsfig{file=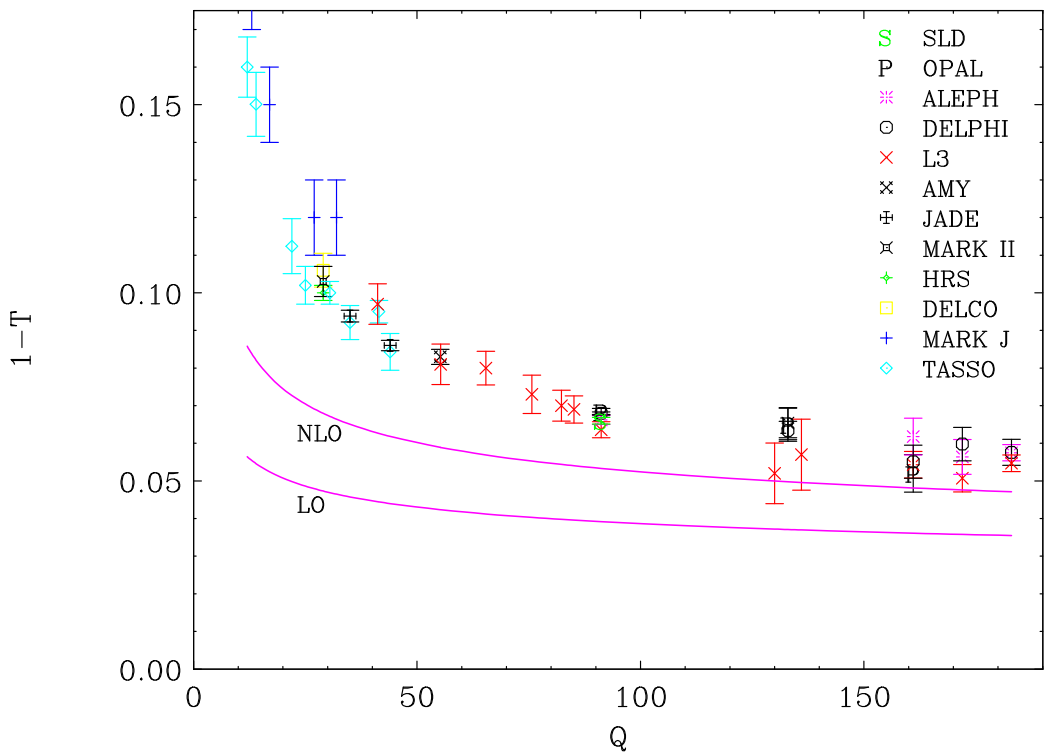,width=10.0truecm,angle=90}
}
\end{center}
\caption{Average of $1-{\rm thrust}$ as a function of the center
 of mass energy Q, according to the available experimental data
 \cite{experiment} and the leading order (LO) and next-to-leading order (NLO)
 perturbative QCD calculation in the $\overline{\rm MS}$ scheme with
 $\mu_R=Q$, given $\alpha_s^{\MSbar}({\rm M_Z})=0.117$.
 }
\label{pert}
\end{figure}

Like other event shape parameters, measurements of the average thrust
do not agree \cite{Moriond_Stenzel} with the next-to-leading order
perturbative calculation \cite{Ellis,EVENT}
\beq 
\left<t\right>_{\NLO}(Q^2) = \frac{C_F}{2}\,
\left[t_0 \,a_{\MSbar}(Q^2) \,+\, t_1 \,a_{\MSbar}^2(Q^2)\right]
\label{t_MSbar}
\eeq
where $C_F=\frac{N_c^2-1}{2N_c}=\frac{4}{3}$ and
\begin{eqnarray}
\label{t_0_and_t1}  
t_0 &=& 1.5776 \\
t_1 &=& 23.7405-1.689\, N_f. \nonumber
\end{eqnarray}
The significant discrepancy is shown in fig.~\ref{pert}. We note that
the experimental data points (there are 44 data points altogether) 
are fairly consistent between different
experiments, and show a stronger variation with $Q$.

There are theoretical indications that {\em perturbative} corrections at the
next-to-next-to-leading order and beyond are large: the
next-to-leading order term is quite a significant correction with
respect to the leading order and there is a large renormalization
scale dependence.   
However, the usual explanation of the discrepancy between the data and the 
next-to-leading order calculation is that hadronization related power 
corrections are important. It is intuitively clear from the definition
of the thrust that emission of soft gluons implies a change in the
thrust which is linear in the gluon momentum, and thus power like
effects that fall as $1/Q$ can appear. 
Fits to the experimental data \cite{Moriond_Stenzel} confirm the
existence of $1/Q$ power corrections.   
Event shape variables are quite special in having such strong power 
corrections: other QCD observables, like the ones for which higher 
twist terms can be analyzed using an operator product expansion, 
usually have power corrections that fall as $1/Q^2$ or faster.

Traditionally, extraction of $\alpha_s$ from experimental data is
based on Monte-Carlo simulations that effectively generate
``hadronization'' power corrections, which are added 
to the next-to-leading order perturbative expression (\ref{t_MSbar}).
In the last 5 years there have been several theoretical attempts to
understand better the subject of power corrections in event shape
variables. These includes analysis of the interplay between
renormalons and power corrections 
\cite{Manohar_Wise,Nason_Seymour,W,Korchemsky_Sterman} 
and renormalon inspired approaches to parametrize power corrections, 
based on either a universal infrared regular physical coupling
\cite{DW,DMW} or a less restrictive observable-dependent
non-perturbative shape function \cite{Shape_function}.
 
Since the average thrust is a priori expected to have both 
large perturbative corrections and significant power-like corrections,
we find this observable quite appropriate to serve as an example of our
approach. 
  
\subsection{The characteristic function for the average thrust}

The observable dependent ingredient in the renormalon resummation
program is the characteristic function ${\cal F}_R(\mu^2/Q^2)$, which
is defined by the leading order perturbative result for a gluon of
mass $\mu^2$. In this section we shall calculate the characteristic 
function for the average thrust, ${\cal F}_T(\mu^2/Q^2)$. 

As mentioned in sec.~3, Minkowskian quantities usually have two
different analytic functions: ${\cal F}^{(-)}(\epsilon)$ 
for $\epsilon\equiv\mu^2/Q^2<1$ and ${\cal F}^{(+)}(\epsilon)$ for
$\epsilon>1$. The reason is that the former includes both real and 
virtual gluon diagrams while the latter includes only virtual gluon diagrams.
In case of event shape variables like the thrust, virtual 
corrections do not contribute at one-loop, and so 
${\cal F}_T^{(+)}(\epsilon)=0$.
What remains to calculate is ${\cal F}_T^{(-)}(\epsilon)$ 
which is entirely due to real gluon emission.
Let us simplify the notation and define 
${\cal F}(\epsilon)\equiv {\cal F}^{(-)}_T(\epsilon)$. The ``gluon mass''
integral (\ref{Rapt}) is now performed up to $\mu^2=Q^2$ 
\beq
\label{Rapt_IR} 
R_{\APT}(Q^2)=\int_0^{1}{d\epsilon \over\epsilon}\
\bar{a}^{\PT}_{\eff}(\epsilon Q^2)\ \dot{{\cal F}}(\epsilon) 
\eeq
which implies that there are strictly no ultraviolet renormalons in
this case:
the large order behavior of $R_{\APT}(Q^2)$ when expanded in some scheme
is determined just by the leading non-analytic terms in the 
small $\epsilon$ expansion of ${\cal F}(\epsilon)$, which are the
leading infrared renormalons. The absence of ultraviolet renormalons
is a direct consequence of the absence of virtual corrections at the
order considered. Higher terms in the ``skeleton expansion'' may give rise
to ultraviolet renormalons in the full perturbative series of 
$\left<t \right>$.   

The characteristic function 
${\cal F}(\epsilon)$ was calculated numerically in \cite{DMW},
and its leading term in the small $\epsilon$ expansion was 
obtained there analytically, ${\dot {\cal
F}}(\epsilon)=4\sqrt{\epsilon}$, which indeed indicates $1/Q$ corrections.
We shall repeat the calculation here.
We find it important to have, if not an analytic expression for 
${\cal F}(\epsilon)$, then at least several leading terms in its asymptotic 
expansion for small $\epsilon$. The sub-leading terms are required 
to verify the convergence of the power correction series (we shall see that in
practice only $1/Q$ terms are important).

In order to explain how ${\cal F}(\epsilon)$ is computed
we first briefly review the kinematics and the calculation of the thrust for
three partons in the final state. 
Let us denote the primary photon 4-momentum by $Q$, 
the 4-momenta of the quark and anti-quark by $p_1$ and $p_2$ ($p_i^2=0$), 
and the 4-momenta of the ``massive gluon'' by $p_3$
($p_3^2=\mu^2=\epsilon Q^2$). 
Following \cite{Ellis} we define $y_{ij}\equiv(p_i+p_j)^2/Q^2$, which 
implies $y_{ij}>0$, and $x_i \equiv {2 Q \cdot p_i}/{Q^2}$.

It follows from energy-momentum conservation that 
\beq
x_1+x_2+x_3=2
\label{sum_x}
\eeq
and from the assumed virtualities of the particles that 
\begin{eqnarray}
\label{y_x_relation}
y_{12}+x_3-\epsilon&=&1\nonumber \\
y_{13}+x_2&=&1\nonumber \\
y_{23}+x_1&=&1.
\end{eqnarray}
In the center of mass frame, where $Q=(Q,0,0,0)$, $x_i$ becomes the
energy fraction of the $i$ parton, $x_i\,=\,\frac{2}{Q} E_i$.
It then follows that
\begin{eqnarray}
\label{vert_pi}
\left\vert \vec{p}_i\right\vert&=&\frac{Q}{2}x_i \,\,\,\,\,\,\,\,\,\,\,
\,\,\,\,\,\,\,\,\,\,\,\,\,\,\,\, i=1,2 \nonumber \\
\left\vert \vec{p}_3\right\vert&=&\frac{Q}{2}\sqrt{x_3^2-4\epsilon}.
\end{eqnarray}

\begin{figure}[htb]
\begin{center}
\mbox{\kern-0.5cm
\epsfig{file=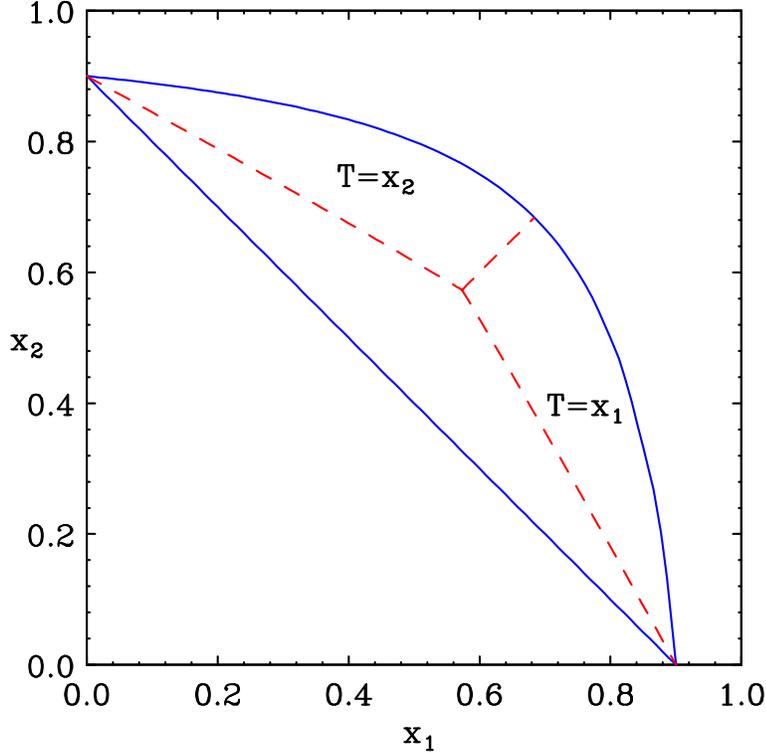,width=10.0truecm,angle=90}
}
\end{center}
\caption{Phase-space for the emission of a virtual gluon with
 $\mu^2\,=\,\epsilon\,Q^2\,=\,0.1\,Q^2$ 
in the plane of the quark and anti-quark 
energy fractions ($x_{1,2}$). Continuous lines represent phase-space
 limits: the upper (curved) line corresponds to the softest gluons
 (\ref{large_epsilon}) while the lower (linear) line corresponds to
 the hardest (\ref{small_epsilon}). Dashed lines represent the
 separation of phase-space according to which particle carries the
 largest momentum and thus determines the thrust axis 
(cf. eq.~(\ref{thrust})): in the upper
 left region $T=x_2$, in the upper right region $T=x_1$ and in the
 lower region $T=\sqrt{x_3^2-4\epsilon}$. 
 }
\label{phase_space}
\end{figure}

The phase space limitations are the following:
\begin{description}
\item{ a)} Hard gluon limit,
\beq 
\label{small_epsilon}
x_1+x_2\geq 1-\epsilon,
\eeq
which follows from the condition $y_{12}>0$ together 
with (\ref{sum_x}) and (\ref{y_x_relation}). 
\item{ b)} Soft gluon limit,
\beq
(1-x_1)(1-x_2)\geq \epsilon,
\label{large_epsilon}
\eeq 
which is obtained in the center of mass frame, 
using (\ref{vert_pi}), from the condition
$\vert \vec{p}_2\vert =\vert \vec{p}_1+\vec{p}_3 \vert\leq \vert \vec{p}_1\vert
+\vert \vec{p}_3 \vert$.
\end{description}
The phase space limitations are shown in fig.~\ref{phase_space} for
the case $\epsilon=0.1$.
The region near the outer (curved) line corresponding to 
(\ref{large_epsilon}) represents soft gluons: for a given
``gluon mass'' $\epsilon Q^2$ and quark energy fraction ($x_1$), the
smallest possible gluon energy is obtained when the inequality
(\ref{large_epsilon}) is saturated. 
The region near the inner (linear) line corresponding to (\ref{small_epsilon}) 
represents hard gluons with maximal energy $x_3=1+\epsilon$.
Note that as $\epsilon$ increases the relevant phase space shrinks
and approaches the region of small $x_1$ and $x_2$ (the lower left 
corner in fig.~\ref{phase_space}).

The characteristic function ${\cal F}(\epsilon)$ is obtained
from the following integral over phase space \cite{DMW},
\beq
{\cal F}(\epsilon)=\int_{\rm \tiny{phase\,\, space}} dx_1 dx_2 \, {\cal
M}(x_1,x_2,\epsilon) \, t(x_1,x_2,\epsilon),
\label{F_def}
\eeq
where $\frac{C_F}{2}\,{\cal M}\,a$ is the squared tree level matrix 
element for the production of a quark, an anti-quark and a gluon of ``mass''
$\mu^2=\epsilon Q^2$ and 
\beq
{\cal M}(x_1,x_2,\epsilon)=
\frac{(x_1+\epsilon)^2+(x_2+\epsilon)^2}{(1-x_1)(1-x_2)}
-\frac{\epsilon}{(1-x_1)^2}-\frac{\epsilon}{(1-x_2)^2}.
\label{M}
\eeq
    
The last ingredient for the calculation of ${\cal F}(\epsilon)$ is the
expression for the thrust.
For two particles in the final state, the thrust axis $\vec{n}_T$ coincides 
with the line along which they move and $T=1$. 
For three particles in the final state, it coincides with
the direction of the particle carrying the largest momentum, 
$\vert\vec{p}_i\vert$.
By momentum conservation in the center of mass frame, the two other 
particles have the sum of momenta $\vec{p}_{j}+\vec{p}_k=-\vec{p}_i$.
The numerator in (\ref{T_def}) then equals $2\vert \vec{p}_i\vert$.
For three {\em massless} particles the denominator in (\ref{T_def})
equals $\vert\vec{p}_1\vert+\vert \vec{p}_2\vert +\vert
\vec{p}_3\vert=E_1+E_2+E_3=Q$. Thus, using (\ref{vert_pi}), we have
$T=x_i$. For a {\em massive} gluon, however, the denominator is
$\frac{Q}{2}(x_1+x_2+\sqrt{x_3^2-4\epsilon})\,\neq\,Q$. 
On the other hand, since the ``massive gluon'' dissociates into 
massless partons one should actually calculate the thrust taking into account
the final partons. In this case the denominator is $Q$. Note that 
the numerator remains the same whether or not one takes into account
the gluon dissociation, provided that all the partons produced end 
up in the same hemisphere as the parent gluon -- an assumption to which
we return below.
In conclusion we find that, due to the change in the denominator of 
(\ref{T_def}), by giving mass to the gluon (in order to represent 
its dissociation through the dispersive approach) 
we unwillingly change the calculated value of the thrust.
To solve this difficulty it has been suggested \cite{DMW} to modify the 
definition of the thrust such that the normalization will be with 
respect to the sum of energies ($Q$),
\beq
T=\frac{\sum_i \left\vert \vec{p}_i \cdot \vec{n}_T\right\vert}
{\sum _i E_i }=\frac{\sum_i \left\vert \vec{p}_i \cdot \vec{n}_T\right\vert}
{Q}.
\label{T_def2}
\eeq

There is no difference between (\ref{T_def2}) and (\ref{T_def}) so
long as only massless particles are produced. However, for the
theoretical computation with a ``massive gluon'' it is
important to use (\ref{T_def2}) which guarantees that the same 
value of the thrust is obtained with a ``massive gluon'' 
as with the massless products of its dissociation\footnote{The
original definition of the 
thrust, eq.~(\ref{T_def}), with a ``massive gluon'' does not comply
with this requirement. It would lead to a different result 
for ${\cal F}(\epsilon)$ and thus to a different normalization of the power 
corrections (see \cite{BB_DY,DMW}).}. 
The final result for the thrust, using (\ref{T_def2}), is
\beq
t(x_1,x_2,\epsilon)={\rm min}\left\{ 1-x_1\,, \,1-x_2\,
,\,1-\sqrt{(2-x_1-x_2)^2-4\epsilon}
\right\}.
\label{thrust}
\eeq
Fig.~\ref{phase_space} shows the separation of phase space
to regions where each of the three particles has the largest momentum and
thus determines the thrust axis.

Let us now return to the more delicate problem, namely the assumption
we made concerning the decay products of the gluon. 
This assumption is absolutely necessary in order to keep 
the same value of the thrust when referring to the ``massive gluon''
itself as when referring to the massless products of its dissociation.
In the first case the relevant term in the numerator of (\ref{T_def2}) is 
$\vert \vec{p}_3\cdot \vec{n}_T\vert$ while in the second it is larger:
\hbox{$\vert \vec{p_L} \cdot \vec{n}_T\vert+\vert \vec{p_R} \cdot 
\vec{n}_T\vert$}, 
where \hbox{$\vec{p}_3=\vec{p}_L+\vec{p}_R$ and $\vec{p}_{L}$}
($\vec{p}_{R}$) corresponds to the
sum of momenta of the particles that originate in the gluon and 
end up in the left (right) hemisphere. 
It is a priori not clear whether this assumption is
justified: kinematic considerations alone do
not exclude the possibility of dissociation into opposite
hemispheres and in fact when the gluon is close to the transverse
direction, it is quite plausible that it would dissociate this way.
The ``massive gluon'' approach is justified \cite{BB,BBB,DMW} for completely
inclusive quantities: then dressing the gluon or taking into account
its dissociation amounts exactly to building up the running
coupling. We see that the thrust is not inclusive enough.
We can still ask, however, how large is the error we introduce
using the inclusive ``massive gluon'' approach instead of taking into
account separately the contribution of the decay products of the
gluon.
The problem of non-inclusiveness was first raised in 
\cite{Nason_Seymour} in the framework 
of renormalon resummation in the large $N_f$ limit where the terms
which prohibit an inclusive treatment were identified
and evaluated. We shall return to this issue in the next section.

\begin{figure}[htb]
\begin{center}
\mbox{\kern-0.5cm
\epsfig{file=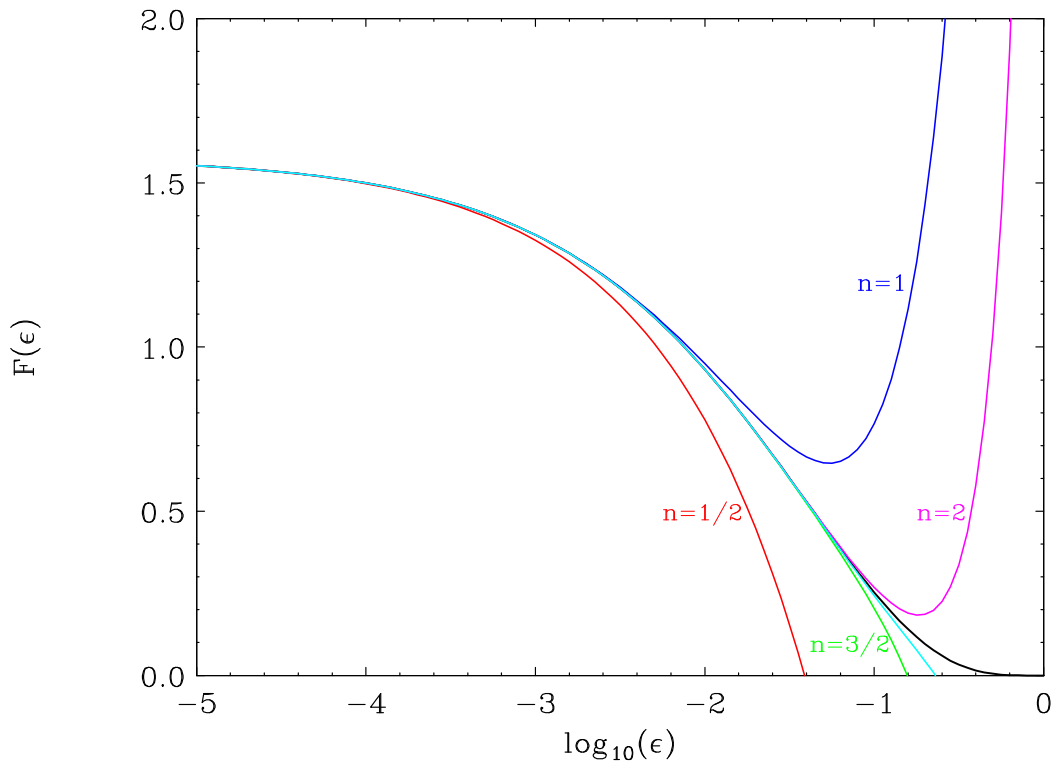,width=10.0truecm,angle=90}
}
\end{center}
\caption{The characteristic function ${\cal F}(\epsilon)$  
for the average thrust (\ref{F_def}) 
as a function of $\log_{10}(\epsilon)$,
where $\mu^2=\epsilon Q^2$ is the ``gluon mass''.
${\cal F}(\epsilon)$ is represented by a thick black line reaching
zero in the limit $\epsilon\longrightarrow 1$. It is 
compared with its asymptotic expansion at small $\epsilon$ (\ref{F_ana}),
for $n=\frac12,1,\frac32,2,\frac52$, where $n$ is the highest order
term ${\cal O}(\epsilon^n)$ taken into account in each approximating curve.
 }
\label{F_analytic}
\end{figure}

Let us now proceed with the calculation of ${\cal F}(\epsilon)$ 
for the thrust, based
on the definition (\ref{F_def}), the squared matrix element
(\ref{M}) the expression for the thrust (\ref{thrust}) and the phase 
space limitations (\ref{small_epsilon}) and (\ref{large_epsilon})
which determine the integration range.
In order to evaluate the integral (\ref{F_def}) one has to treat
separately each of the three regions of phase space (see 
fig.~\ref{phase_space}): the integrand in
each of them is different, as implied by (\ref{thrust}). 
To perform the integrals it is useful to
change the integration variables to the following:
\hbox{$z_1=x_1+x_2+\epsilon-1$} and \hbox{$z_2=x_1-x_2$}, which 
fit better the phase space
limitations. Most eventual integrals can be performed analytically
but the resulting functions are complicated. 
Instead, we calculated the integral numerically for any $0<\epsilon<1$
and in addition obtained an asymptotic expansion of ${\cal
F}(\epsilon)$ for small $\epsilon$,
\begin{eqnarray}
\label{F_ana}
{\cal F}(\epsilon)  &=&  - { \frac {1}{18}}  + \pi ^{2} + 8\,
\rm{ln}(3)\,\rm{ln}(2) - { \frac {3}{4}} \,
\rm{ln}(3) + 4\,\rm{dilog}(4) + 4\,\rm{dilog}(3) \nonumber  \\
&-& 8 \,\epsilon^{\frac{1}{2}} 
+ \left[4 + 12\,\rm{ln}(3)\right]\,\epsilon       
- { \frac {160}{9}} \,\epsilon ^{\frac{3}{2}}\nonumber \\ 
&+& \left[ - 3\,\rm{ln}\left({ \frac {1}{
\epsilon }}\right) + { \frac {17}{6}}  - \pi ^{2} - 8
\,\rm{ln}(3)\,\rm{ln}(2) - 4\,\rm{dilog}(4) +
{ \frac {56}{3}} \,\rm{ln}(2) - 4\,\rm{dilog
}(3)\right]\,\epsilon ^{2}\nonumber \\ 
&-& 8\,\epsilon ^{\frac{5}{2}}
+ \left[{ \frac {28}{15}} \,\rm{ln}(2) -
{ \frac {8}{5}}  + { \frac {16}{3}} \,
\rm{ln}\left({ \frac {1}{\epsilon }} \right)\right]\,
\epsilon ^{3} \nonumber \\ 
&+& \left[ - 11\,\rm{ln}\left({ \frac {1}{
\epsilon }} \right) + { \frac {2719}{630}}  -
{ \frac {128}{105}} \,\rm{ln}(2)\right]\,\epsilon
^{4} + \, \cdots \\ 
&=&1.5776 - 8\,\epsilon^{\frac{1}{2}} + 17.1833\,\epsilon 
- 17.7778\,\epsilon ^{\frac{3}{2}}
+\left[ - 3\,\rm{ln}\left({ \frac {1}{\epsilon }}\right)
+13.3149\right]\,\epsilon ^{2} \nonumber \\  \nonumber
& -& 8\,\epsilon ^{\frac{5}{2}}
+\left[ - 0.3061 + 5.3333\,\rm{ln}\left({ \frac {1}{\epsilon }}\right)\right]
\,\epsilon ^{3}
+\left[ - 11\,\rm{ln}\left({ \frac {1}{\epsilon }}\right) 
+3.4709\right]\,\epsilon ^{4}+\,\cdots
\end{eqnarray}
The agreement between the asymptotic expansion (\ref{F_ana}) and 
the numerical calculation is shown in fig.~\ref{F_analytic}. 

As we saw in the previous sections, the leading terms in the small
$\epsilon$ expansion of ${\cal F}(\epsilon)$ are required for the calculation
of the difference between different regularizations of the perturbative sum and
eventually for the parametrization of power corrections. 
Each term in eq.~(\ref{F_ana}) has the form of eq.~(\ref{F-low}) and so
the numerical coefficients $B_T^{(n)}$ and $C_T^{(n)}$ can be
immediately identified.  
The first non-analytic term in (\ref{F_ana}) is a square root one,
corresponding to a 1/Q infrared power correction. The coefficient of
this term agrees with previous calculations \cite{DMW}: 
$\dot {\cal F}(\epsilon)=4\sqrt{\epsilon}$. As explained there, this
term arises from the limit of phase space (\ref{large_epsilon})
corresponding to the softest gluons.     

A new observation is that there are no $1/Q^2$ infrared power
corrections in this ``leading skeleton'' approximation. 
The next non-analytic term in (\ref{F_ana})
is $\epsilon^{\frac{3}{2}}$ associated with $1/Q^3$ infrared power
corrections.
As a result the apparent convergence of the discontinuity function 
$\Phi(k^2/Q^2)$ 
\beq
\Phi(k^2/Q^2)\simeq \,\frac1\pi\,\left(\frac{k^2}{Q^2}\right)^\frac12
\left[8-17.7778\left(\frac{k^2}{Q^2}\right)+\cdots\right]
\label{conv_of_Phi}
\eeq
is better than that of ${\cal F}(\epsilon)$. In (\ref{conv_of_Phi}) 
the second term becomes about $20\%$ of the first around 
$k^2\,\sim\,0.1\, Q^2$.
Another observation is that the asymptotic expansion (\ref{F_ana})
does not contain double logarithmic terms up to the order considered.

\begin{figure}[htb]
\begin{center}
\mbox{\kern-0.5cm
\epsfig{file=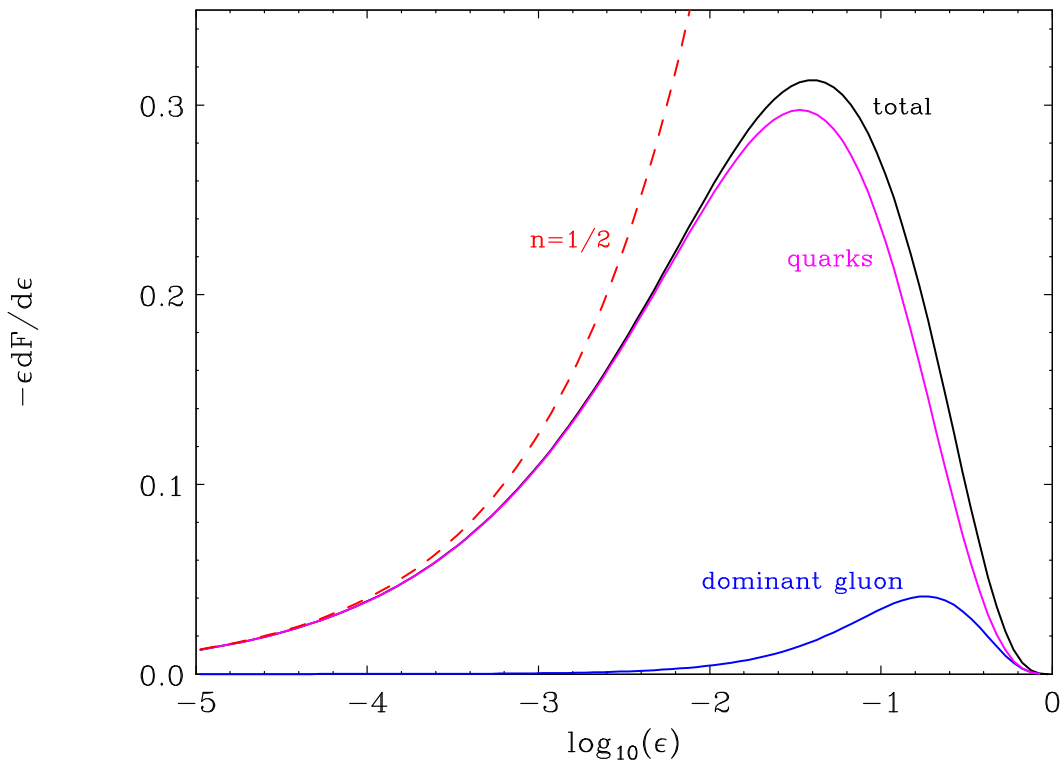,width=10.0truecm,angle=90}
}
\end{center}
\caption{The derivative of the characteristic function 
for the average thrust $\dot{{\cal F}}(\epsilon)$ 
as a function of $\log_{10}(\epsilon)$,
where $\mu^2=\epsilon Q^2$ is the ``gluon mass''.
The separate contributions from
kinematic configurations where one of the quarks or the gluon carries
the largest momentum are shown as well. The dashed line represents the
leading ($n=\frac12$) term in the small $\epsilon$ expansion of this function,
$\dot{{\cal F}}(\epsilon)\simeq 4\sqrt{\epsilon}$.  
 }
\label{F_dot}
\end{figure}

Finally, taking the logarithmic derivative we obtain
$\dot{\cal F}(\epsilon)=-\epsilon\,d{\cal F}/d\epsilon$, 
which agrees with the numerical results of \cite{DMW}.
In fig.~\ref{F_dot} one can see that there is just a small 
contribution to the characteristic function,
and thus also to the value of the average thrust, from the kinematic  
configurations where the gluon momentum is the largest. 
Moreover, this contribution to $\dot{\cal F}(\epsilon)$ 
has some significance only for fairly large ``gluon mass'' 
$\epsilon\gsim 10^{-2}$, and so its
effect on the large order behavior of the perturbative 
series and the related power corrections is negligible.

\subsection{Does the renormalon resummation program apply to
 non-inclusive quantities?}

The ``skeleton expansion'' approach and thus our renormalon 
resummation program are based on the assumption that all the higher
order diagrams related to the running of the coupling contribute
inclusively to the observable. This assumption does not hold in the
case of event shape variables like thrust.
In the previous section we calculated the thrust characteristic
function using a ``massive gluon'', which should represent after
integration with respect to the discontinuity of the running coupling,
the partons to which the gluon dissociates. 
We encountered and eventually ignored the non-inclusive nature of the 
thrust: as explained there, if the decay products of the gluon end up in 
different hemispheres the average thrust value obtained in the
``massive gluon'' calculation is different (and in fact always lower) than it
is in reality. 
Still, we intuitively expect that most decays are roughly
collinear and so for the major part of the phase
space decay into opposite hemispheres is not very likely. 
We therefore conjecture that the inclusive resummation approach can provide a
reasonable approximation to the full higher order corrections.  
In the following we give a quantitative argument at the next-to-leading
order level in favor of this conjecture.

In order to estimate the error we are making by the inclusive
treatment, let us compare the two type of expansions we have:
the ordinary perturbative expansion, e.g. eq.~(\ref{t_MSbar}) in 
the ${\rm \overline{MS}}$ scheme, and the conjectured 
``skeleton expansion'', i.e. the equivalent of (\ref{ske}) and
(\ref{ske_min}) for the average thrust
\beq
\left<t\right>_{\PT}(Q^2)\,
=\,\frac{C_F}{2}\,\left[\left(S_0^{\PT}(Q^2)+ S_1^{\PT}(Q^2)+\cdots\right)
+{\rm Non-Skeleton}\right]
\label{ske_min_NS}
\eeq
where similarly to $R_i^{\PT}$ in eq.~(\ref{ske_min}) $S_0^{\PT}$ is 
the leading  ``dressed skeleton'' term, normalized as $S_0^{\PT}=t_0\, a+{\cal
O}(a^2)$, $S_1^{\PT}$ is the second ``dressed skeleton'' term which
starts at order $a^2$ and so on. 
The ${\rm Non-Skeleton}$ piece allows for additional perturbative 
terms of order $a^2$ or higher which do not fit into 
a ``skeleton expansion'' {\em due to 
the non-inclusive nature} of the observable.
The ``leading skeleton'' term $S_0^{\PT}$ is given by $R_{\APT}$ 
of eq.~(\ref{Rapt}), up to
a regularization related power term which we now ignore.
Since both $S_1^{\PT}$ and the ${\rm Non-Skeleton}$ terms start
at order $a^2$, we can approximate $\left<t\right>_{\PT}(Q^2)$ by
\beq
\left<t\right>_{\PT}(Q^2)\,
\simeq\,\frac{C_F}{2}\,\left[
S_0^{\PT}(Q^2)+\delta_{\NLO}\right]
\label{ske_min_NLO}
\eeq
with
\beq
\delta_{\NLO}= \tilde{t}_1 a^2\,+\,\cdots
\label{S1_est}
\eeq 
where in general the coefficient $\tilde{t}_1$ contains contributions
from both $S_1^{\PT}$ and the ${\rm Non-Skeleton}$ terms.
To compare the two expansions, let us expand the time-like coupling 
within the first ``skeleton'' term $R_{\APT}$ (\ref{Rapt}) in terms of
$a_{\MSbar}$. Note that at a difference with sec.~2, it is the
Minkowskian representation that we start with. 
Taking the one-loop form (\ref{a_oneloop}) of the coupling we
obtain from (\ref{a_eff_oneloop}) the following expansion in 
$a_{\MSbar}(Q^2)$ 
\begin{eqnarray}
\label{abar_pert_expansion}
\lefteqn{\bar{a}^{\PT}_{\eff}(\mu^2)= a_{\MSbar}(Q^2) 
\,+\, \left\{d_1-
\left(c_1+\log\frac{\mu^2}{Q^2}\right)\beta_0\right\}
\,a_{\MSbar}^{2}(Q^2) \,\,+} \\
&&\left\{{d_1}^{2} \,-\, 2{d_1}\,\left(
{c_1}+\log\frac{\mu^2}{Q^2}\right)\beta_0\,+\, \left[
\left({c_1}+\log\frac{\mu^2}{Q^2}\right)^{2}- { \frac
{1}{3}} \,\pi ^{2} \right]\beta_0^2 \right\} a_{\MSbar}^{3}(Q^2)
+\cdots\nonumber  
\end{eqnarray}
where the first two terms are actually valid beyond the one-loop
approximation of the coupling and also coincide with the corresponding 
terms in the expansion of the space-like coupling (\ref{bara}). 
Inserting (\ref{abar_pert_expansion}) into (\ref{Rapt}) yields
\beq
\label{Rapt_pert_expansion}
S_0^{\PT}(Q^2) = t_0\,a_{\MSbar}(Q^2) 
\,+\,t_1^0 \,a_{\MSbar}^{2}(Q^2) \,\,+t_2^0 \,a_{\MSbar}^{3}(Q^2)\,+\,
\cdots\nonumber  
\eeq
with
\begin{eqnarray}
\label{Rapt_pert_coef}
t_0&=&f_{0}\\
t_1^0&=&{d_1}\,{f_{0}} - ( {c_1}\,
{f_{0}} + {f_{1}})\,\beta_0 \nonumber\\ 
t_2^0&=&{d_1}^{2}\,{f_{0}} \,-\, 2\,{d_1}\,(
{c_1}\,{f_{0}} +{f_{1}})\,\beta_0 + \left[\left(
{c_1}^{2} - { \frac {1}{3}} \,\pi ^{2}\right)\,{f_{0
}} + 2\,{c_1}\,{f_{1}} + {f_{2}}\right]\,\beta_0^{2}\nonumber
\end{eqnarray}
and so on.
Note that at the next-to-next-to-leading order ($t_2^0$) we recover 
the characteristic $\pi^2$ terms which appear in perturbative expansions of 
Minkowskian observables.
The constants $f_i$ are the log-moments of the characteristic function
(compare with (\ref{phi_i}))
\begin{equation}
f_i\equiv\int_{0}^\infty{d\mu^2
\over \mu^2}\ \left(\log{\mu^2\over Q^2}\right)^i\
\dot{\cal F}(\mu^2/Q^2). 
\label{f_i}
\end{equation}
Using the numerical result for the characteristic function of the
average thrust one can obtain $f_i$ to any arbitrary order. 
The first values are
\begin{eqnarray}
\label{f_i_thrust}
f_0 &=& 1.577602558 \nonumber \\
f_1 &=& -7.176762311 \nonumber \\
f_2 &=& 42.11235577 \nonumber \\
f_3 &=& -307.991760 \nonumber \\
f_4 &=& 2736.14010 \nonumber \\
f_5 &=& -28923.5429 \nonumber \\
f_6 &=& 357358.9993
\end{eqnarray}
With these coefficients at hand one can construct a power series
approximation\footnote{This expansion is an asymptotic one, badly affected
by infrared renormalons. Note that the explicit sign oscillation in
$f_i$ cancels against the sign oscillation in eq.
(\ref{Rapt_pert_coef}). Note also that the fast growth, that eventually
becomes factorial, is already
apparent in (\ref{f_i_thrust}). This expansion will be discussed 
in sec.~5.4.} (\ref{Rapt_pert_expansion}) to the first ``skeleton'' 
term $S_0^{\PT}$ of eq.~(\ref{Rapt}) to any arbitrary
order, provided one specifies the ``skeleton coupling'' 
$\overline{a}_{\PT}$, namely the parameters $d_1$ and $c_1$. 
As discussed at the end of sec.~2, we use in this work
several different schemes for the ``skeleton coupling'' $\overline{a}_{\PT}$. 
In the Abelian limit, $\overline{a}_{\PT}$
should coincide with the V-scheme coupling and so $c_1=-\frac{5}{3}$.
We can thus determine the $\beta_0$ dependent piece in 
$t_1^0$ in eq.~(\ref{Rapt_pert_coef}) 
\beq
-(c_1f_0+f_1)\, \beta_0 = 9.8061 \, \beta_0.
\label{linear_beta0}
\eeq

We now have all the ingredients for the comparison up to
next-to-leading order between the ``skeleton expansion''
(\ref{ske_min_NLO}) and the standard expansion (\ref{t_MSbar}).
For the first we use (\ref{Rapt_pert_expansion}) and obtain
\begin{eqnarray}
\label{exp_skeleton}
\left< t \right>_{\PT}&=& \,\frac{C_F}{2}\,\left\{ 
t_0\,a_{\MSbar}(Q^2)+ 
\left(\tilde{t}_1+t_1^0\right)\,a_{\MSbar}^{2}(Q^2)
+{\cal O}\left(a_{\MSbar}^{3}(Q^2)\right)
\right\}\\ \nonumber
&=& \,\frac{C_F}{2}\,\left\{{f_{0}\,a_{\MSbar}(Q^2) 
+ \left[\tilde{t}_1+ {d_1}\,{f_{0}} - ( {c_1}\,
{f_{0}} + {f_{1}})\,\beta_0\right]\,a_{\MSbar}^{2}(Q^2)}+{\cal
O}\left(a_{\MSbar}^{3}(Q^2)\right) \right\}
\end{eqnarray}
By construction the leading order is the same.
The comparison at the next-to-leading order gives
\beq
\tilde{t}_1+ {d_1}\,{f_{0}}+ 9.8061\,\beta_0\,=\, -4.128 + 10.134 \, \beta_0
\label{comparison_NLO}
\eeq 
where the l.h.s. corresponds to the ``skeleton expansion''
coefficients $\tilde{t}_1+t_1^0$ of eq.~(\ref{exp_skeleton}) 
and the r.h.s. to the standard expansion
coefficient $t_1$ of eq.~(\ref{t_0_and_t1}), 
with the $N_f$ dependence expressed in terms of $\beta_0$ (\ref{beta_0}).

For an inclusive quantity, where we assume that the ``skeleton 
expansion'' exists (i.e. the ${\rm Non-Skeleton}$ terms in 
(\ref{ske_min_NS}) are absent),
the $N_f$ dependence at the next-to-leading
order comes only from diagrams that
are related to the running of the coupling.  In such a case the entire 
$\beta_0$ dependent piece in the next-to-leading order coefficient 
is accounted for by the leading term $S_0^{\PT}$ in the 
``skeleton expansion'', and the remaining coefficient $\tilde{t}_1$ in
(\ref{ske_min_NLO}),
which coincides now with the normalization of the sub-leading 
``skeleton'' $S_1^{\PT}=\tilde{t}_1a^2+\cdots$, should be free of $\beta_0$.
For the thrust, which is non-inclusive with respect to the
decay products of the gluon, this does not hold.
However, we find that the difference between the term linear
in $\beta_0$ in (\ref{Rapt_pert_expansion}) $9.8061 \, \beta_0$
(l.h.s. in (\ref{comparison_NLO})) and in full next-to-leading QCD 
coefficient $10.134 \,\beta_0$ (r.h.s. in (\ref{comparison_NLO})) 
is quite small: it is about 3 percent.
This finding gives place to hope that the ${\rm Non-Skeleton}$ terms
in (\ref{ske_min_NS}) are small and thus the inclusive treatment is after
all a good approximation for the resummation of a certain class of
diagrams. 

The observation that for a non-inclusive quantity the $N_f$ dependence
of the next-to-leading order coefficient cannot be explained in terms
of the running coupling also implies that the usual motivation for 
BLM scale fixing and the Naive Non-Abelianization procedure\footnote{
The ambiguity of the Naive Non-Abelianization procedure for 
non-inclusive quantities was pointed out in \cite{Beneke-Braun-Magnea}.}
does not hold. As opposed to the inclusive case discussed in sec.~2, 
the BLM scale computed from the full next-to-leading order coefficient
does not coincide with the one of the ``leading skeleton''. 
As a result, the identification made following eq.~(\ref{nlocoef}), 
between the remaining next-to-leading order 
coefficient\footnote{There $r_1-r_1^0$ 
was also identified with the normalization of the ``sub-leading skeleton''.}  
and the BLM coefficient in the ``skeleton scheme'' fails: 
the former still contains some $N_f$ dependence while the latter does not. 
Like our resummation program, the BLM procedure becomes
relevant once the observation is made that the $\beta_0$ dependent
terms in the two sides of eq.~(\ref{comparison_NLO}) are numerically
very close.

Finally, using (\ref{comparison_NLO}) we evaluate $\tilde{t}_1$. 
If the coupling $\overline{a}_{\eff}^{\PT}$ in the ``leading skeleton'' term
(\ref{Rapt}) is the time-like coupling associated (\ref{aeffpt}) 
with the ``gluon bremsstrahlung'' coupling \cite{CMW}  
(see the end of sec.~2), where $d_1^{\brem}=1-{\pi^2\over 4}$, then 
\beq
\tilde{t}_1^{\brem}=-1.813+0.328\,\beta_0.
\label{tilde_f1_CMW}
\eeq
For the pinch technique coupling \cite{Watson} $d_1^{\pinch}=1$, and so
\beq
\tilde{t}_1^{\pinch}=-5.706+0.328\,\beta_0.
\label{tilde_f1_pinch}
\eeq
For the V-scheme coupling \cite{BLM}, $d_1^{\V}=-2$, and then  
\beq
\tilde{t}_1^{\V}=-0.973+0.328\,\beta_0.
\label{tilde_f1_V}
\eeq
For $N_f=5$ the coefficients are $\tilde{t}_1^{\brem}=-1.184$,
$\tilde{t}_1^{\pinch}=-5.077$ and $\tilde{t}_1^{\V}=-0.344$.
These coefficients can be compared with the standard 
next-to-leading coefficient in ${\rm {\overline{MS}}}$ 
(\ref{t_0_and_t1}) which equals $t_1\simeq 15.296$. We conclude
that at least the apparent convergence of the suggested expansion is
better than that of the standard one. The coefficient is extremely
small in the ``gluon bremsstrahlung'' and V schemes. 
We shall thus choose for our phenomenological analysis two couplings:
\begin{description}
\item{a) } the ``gluon bremsstrahlung'' scheme which was used before 
in the analysis of
power corrections to the thrust \cite{DW,DMW}. This coupling
will be quite convenient in practice: having a small next-to-leading 
coefficient (\ref{tilde_f1_CMW}), 
the full result should be close to the first term in the 
``skeleton expansion''.
\item{b) }  the pinch technique coupling which may be the correct
physical ``skeleton coupling'' $\bar{a}$. 
Using this coupling, with its relatively large
next-to-leading coefficient (\ref{tilde_f1_pinch}), 
in addition to the ``gluon bremsstrahlung'' coupling, 
will be useful to measure the sensitivity of our procedure to the
value of $d_1$.
\end{description}

\subsection{The perturbative sum vs. experimental data}

The most convenient way to calculate the perturbative sum 
is to use the APT formula (\ref{Rapt}). 
As explained in sec.~3, other regularizations of the perturbative sum,
such as the principal value Borel sum can then be obtained from
$R_{\APT}$.

The ingredients required for the calculation of $R_{\APT}$ are the
numerical function ${\cal F}(\mu^2/Q^2)-{\cal F}(0)$ we obtained in
sec.~5.1 and the discontinuity of the perturbative coupling on the 
time-like axis, $\overline{\rho}_{\PT}(\mu^2)$.
For the latter, we shall use here the one and two loop
couplings.
In the one-loop case (\ref{a_oneloop}), the expression obtained from 
(\ref{discpt}) is simply
\beq
\overline{\rho}_{\1loop}(\mu^2)=-\frac{1}{\beta_0}\,\frac{1}{\pi^2+\log{\,^2}
\left(\frac{\mu^2}{\Lambda^2}\right)}.
\eeq
In the two-loop case we use the Lambert W function representation 
of the coupling \cite{FP,GGK}
\beq
\begin{array}{c}
\displaystyle
\bar{a}_{\2loop}(k^2)=-\frac{\beta_0}{\beta_1}\,\,\frac{1}{1+W_{-1}(z)}
\nonumber\\
\phantom{a}\\
\displaystyle
z = -\frac{1}{e}
\left(\frac{k^2}{\Lambda^2}\right)^{-\beta_0^2/\beta_1}
\end{array}
 \label{W_sol_2loop}
\eeq
where $W(z)$ is the Lambert W function defined by 
$W(z) \exp\left[W(z)\right]=z$ and the particular branch 
$W_{-1}(z)$ is implied by asymptotic 
freedom: in the ultraviolet $z\longrightarrow 0^-$ 
and $W_{-1}(z)\longrightarrow -\infty$ \cite{GGK}.
Note that in (\ref{W_sol_2loop}) the explicit Landau pole and the 
tip of the branch cut coincide ($W_{-1}(-1/e)=-1$) and so there is
only one singularity in the complex momentum plane, at
$k^2=\Lambda^2$, with a cut $-\infty<k^2<\Lambda^2$.  
Using the computer algebra program Maple, $W_{-1}(z)$ is readily
available at any given complex $z$. It is then straightforward to obtain the
time-like discontinuity $\overline{\rho}_{\2loop}(\mu^2)$ 
(\ref{discpt}) corresponding to $\bar{a}_{\2loop}(k^2)$. 

As a first trial, let us calculate $R_{\APT}$ based on the world
average value of $\alpha_s$, $\alpha_s^{\MSbar}({\rm M_Z})=0.117$.
We choose $\bar{a}_{\PT}$ in the ``gluon bremsstrahlung'' scheme. 
Taking\footnote{This point will be discussed in sec.~5.6.} $N_f=5$
we find $\Lambda_{\brem}^{\1loop}=0.130\, {\rm GeV}$ 
and $\Lambda_{\brem}^{\2loop}=0.361\, {\rm GeV}$.
We evaluate the perturbative sum $R_{\APT}(Q^2)$ in 
(\ref{Rapt}) by a
numerical integration of \hbox{${\cal F}(\mu^2/Q^2)-{\cal F}(0)$} times either
$\overline{\rho}_{\1loop}(\mu^2)$ or $\overline{\rho}_{\2loop}(\mu^2)$. 

Next, we use $R_{\APT}$ to calculate the principal value Borel sum,
according to 
$R_{\PT\vert\PV}=R_{\APT}-{\rm Re}\left\{ \delta R_{\APT}\right\}$, 
where ${\rm Re}\left\{ \delta R_{\APT}\right\}$ is evaluated for a
generic term in the small $\epsilon$ expansion of ${\cal F}(\epsilon)$
(\ref{F_ana}) by eq.~(\ref{delta_R_n_1loop}) and (\ref{rebn-apt2}) in the
one and two loop cases, respectively. 
At order $n$ in the expansion, the contribution to ${\rm Re}\left\{
\delta R_{\APT}\right\}$ is \hbox{${\cal O}(1/Q^{2n})$.}
Since ${\rm Re} \left\{\delta R^{\APT}_{\frac12}\right\}$ in
eq.~(\ref{delta_R_n_1loop}) and (\ref{rebn-apt2}) vanishes identically,
the leading contribution is at order $n=1$. 
We note that the $n=1$ term in ${\cal F}(\epsilon)$ is
analytic, and so this contribution is {\em not} of infrared origin.
We obtain
\begin{eqnarray}
\label{delta_R_APT_thrust}
\left.{\rm Re}\left\{ \delta R_{\APT}\right\}\right\vert_{\1loop}&=&
\frac{\Lambda^2_{\1loop}}{Q^2}\,C_T^{(1)}\frac{1}{\beta_0}=8.96\, 
\frac{\Lambda^2_{\1loop}}{Q^2} \nonumber \\
\left.{\rm Re} \left\{\delta R_{\APT}\right\}\right\vert_{\2loop}&=&
\frac{\Lambda^2_{\2loop}}{Q^2}\,C_T^{(1)}\frac{1}{\beta_0}\,\Gamma(1+\delta_1)
\,\delta_1^{\delta_1}\, e^{-\delta_1} =3.18\,\frac{\Lambda^2_{\2loop}}{Q^2}
\end{eqnarray}
where the numerical values were obtained using 
$\delta_1=\beta_1/\beta_0^2$ and $C_T^{(1)}=17.1833$.
We find that ${\rm Re}\left\{\delta R_{\APT}\right\}$ 
is absolutely negligible, already at the
lowest relevant experimental value $Q=12\, {\rm GeV}$, where it is
less than one percent of $R_{\APT}$ (see table~\ref{Q12} in sec.~5.5).  

\begin{figure}[htb]
\begin{center}
\mbox{\kern-0.5cm
\epsfig{file=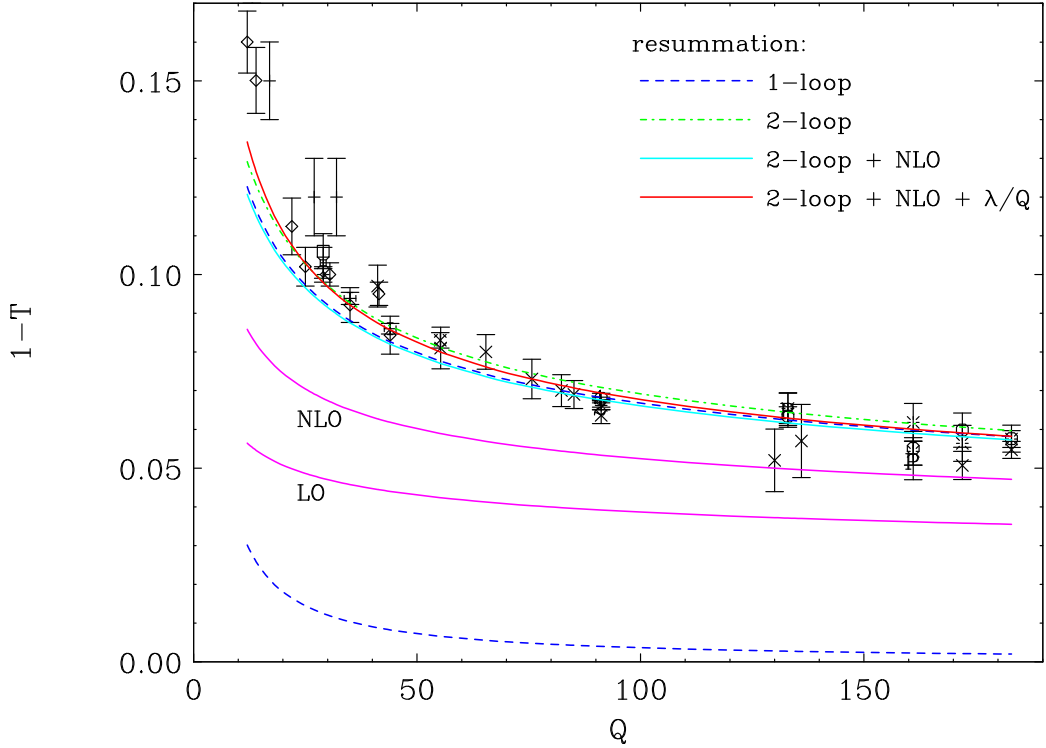,width=10.0truecm,angle=90}
}
\end{center}
\caption{Average of $1-{\rm thrust}$ as a function of Q:
 experimental data is compared with naive perturbative QCD results
 (LO and NLO in $\overline {\rm MS}$ with $\mu_R=Q$) and with the
 resummed perturbative series in the APT or principal value Borel sum
regularizations (the two coincide). 
All the theoretical calculations are based on
 $\alpha_s^{\MSbar}({\rm M_Z})=0.117$.
Both one-loop (upper dashed) and two-loop
 (dot-dash) resummation results ($R_{\APT}$) are presented, where the running
 coupling $\bar{a}_{\PT}$ is in the ``gluon bremsstrahlung'' scheme.  
In the two-loop case we also show eq.~(\ref{inc_NLO})
 (lower continuous line; just below the upper dashed line) which
 includes the full ${\cal O}(\alpha_s^2)$ term and
 eq.~(\ref{inc_lambda}) (upper continuous line) which includes in
 addition a fitted $\lambda/Q$
 term.  The lower
 dashed line is the absolute value of the imaginary part of the
 one-loop Borel sum, which reflects the magnitude of the renormalon 
ambiguity.
 }
\label{K_117_PV}
\end{figure}

The results for the
average thrust $\left<t\right>_{\PT}(Q^2)\simeq \frac{C_F}{2}
R_{\APT}\simeq \frac{C_F}{2} R_{\PT\vert\PV}$ are
shown in fig.~\ref{K_117_PV} (within the resolution of the 
figure $R_{\APT}$ and $R_{\PT\vert\PV}$ could hardly be distinguished). 
The first observation is that the difference between the one-loop and
two-loop resummation results is quite small. This stability, which
shall be discussed further in sec.~5.6, is reassuring 
since in our approach $\bar{a}_{\PT}$ should actually be an all-order running
coupling, and so its replacement by the one-loop coupling at all
scales is not obviously justified, as already noted in sec.~3.3.
 
\setcounter{footnote}{0}
In order to compare our results with experimental data, we should 
include the estimated contribution of the terms not included in the 
``leading skeleton'' according to eqs.~(\ref{ske_min_NLO}) 
and (\ref{S1_est}).
For a concrete estimate of $\delta_{\NLO}$ we replace
the arbitrary scheme coupling $a$ 
by\footnote{An alternative choice (that would be
numerically close) is to replace $a$ by the ``skeleton coupling''
$\bar{a}_{\PT}$ at the BLM scale (\ref{mu_BLM_skeleton}).} 
the natural effective charge at hand, namely the
value of the ``leading skeleton'' with the appropriate 
normalization 
\beq
\delta_{\NLO}= \tilde{t}_1 a^2\simeq \tilde{t}_1
\left(\frac{R_{\APT}}{f_0}\right)^2
\label{S1_est_mod}
\eeq
and thus 
\beq
\left<t\right>_{\PT}=\frac{C_F}{2}\left[R_{\APT}+\tilde{t}_1
\left(\frac{R_{\APT}}{f_0}\right)^2\right].
\label{inc_NLO}
\eeq
This expression exhausts our knowledge concerning the perturbative
contribution to the average thrust, as it includes, in addition to the
resummation of the first ``skeleton'', the full next-to-leading 
order coefficient.
As seen in the figure, the line representing (\ref{inc_NLO}) does not
deviate much from the ``leading skeleton'' results. This is due, of
course, to the small $\tilde{t}_1$ coefficient (\ref{tilde_f1_CMW}). 

The next crucial observation in fig.~\ref{K_117_PV} is that the
resummed results turn out to be quite
close to the experimental data, significantly closer than the 
next-to-leading order result in ${\rm {\overline{MS}}}$ with $\mu_R=Q$
given in eq.~(\ref{t_MSbar}). 
As implied by the $1/Q$ nature of the leading renormalon term in the
expansion (\ref{F_ana}) we introduce a non-perturbative parameter
$\lambda$ and add an explicit power correction of the form $\lambda/Q$ to the 
perturbative prediction (\ref{inc_NLO})
\beq
\left<t\right>=\frac{C_F}{2}\left[R_{\APT}+\delta_{\NLO}\right]
+\frac{\lambda}{Q}.
\label{inc_lambda}
\eeq
Being unable to compute $\lambda$ from the theory, we determine
it by performing a $\chi^2$ fit of (\ref{inc_lambda}) to the data. 
The results of such a fit are summarized in table~\ref{117_fit}, where
$R_{\APT}$ is calculated with the ``skeleton coupling'' (assumed to be
the ``gluon bremsstrahlung'' coupling) at one or two loops.
\begin{table}[H]
\[
\begin{array}{|c|c|c|}
\hline
\,\bar{a}_{\eff}^{\PT}\,& \,\lambda\, ({\rm GeV})\,&\,\chi^2/{\rm point}\,\\
\hline
\hline
{\rm one-loop}&0.36&2.43\\
\hline
{\rm two-loop}&0.16&3.46\\
\hline
\end{array}
\]
\caption{Power term ($\lambda$) fit results with 
$\alpha_s^{\MSbar}({\rm  M_z})=0.117$ based on (\ref{inc_lambda}).}
\label{117_fit}
\end{table}
The fit results in the two-loop case\footnote{The fit results in the
one-loop case (not shown in the plot) are very close to those of 
the two-loop case. The difference between the best fits in the two
cases is of some significance only for  $Q\lsim 30\, {\rm GeV}$ and it
reaches $6\%$ at the lowest data point $Q=12\,{\rm GeV}$. The
difference at low $Q$ explains the variation in $\chi^2$ in 
table~\ref{117_fit}.
We further comment on the comparison between the one and two loop 
resummation results in sec.~5.6.}  
are presented together with the perturbative sum 
in fig.~\ref{K_117_PV}: they turn out to be quite close.
We conclude that a major part of the discrepancy between the 
next-to-leading order result and the data is due to neglecting higher
order perturbative corrections that can be resummed using the
suggested program.
This finding will be discussed in more detail in the next sections.

\begin{figure}[htb]
\begin{center}
\mbox{\kern-0.5cm
\epsfig{file=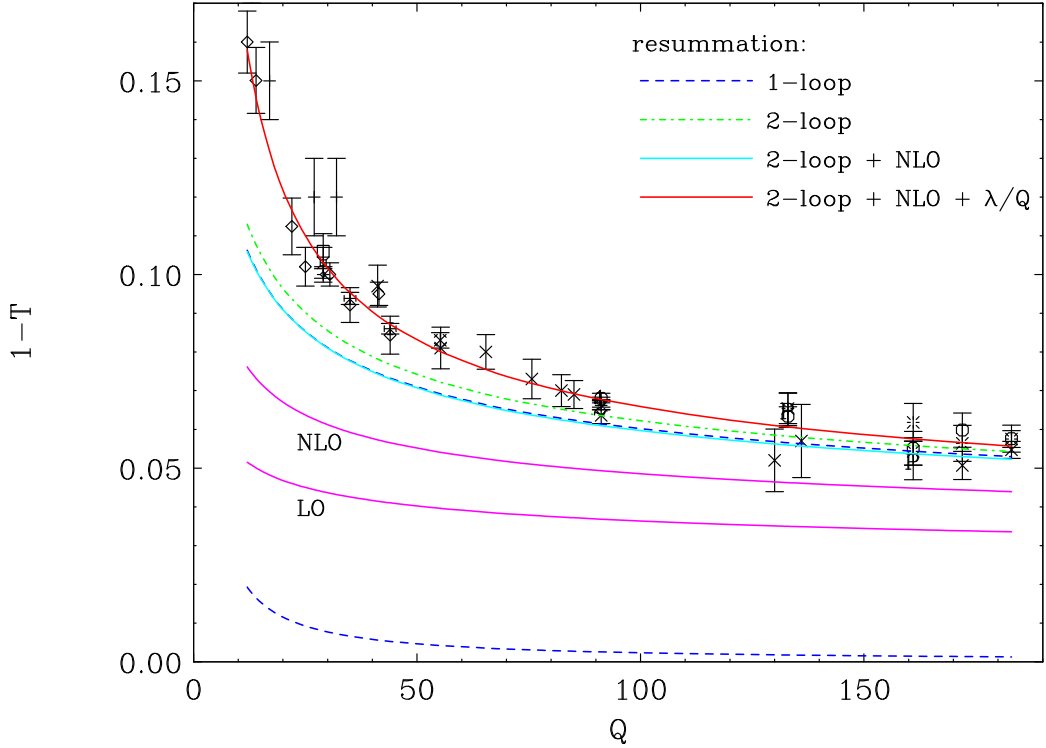,width=10.0truecm,angle=90}
}
\end{center}
\caption{Average of $1-{\rm thrust}$: experimental data vs. naive
 and resummed QCD predictions, given $\alpha_s^{\MSbar}({\rm
 M_Z})=0.110$ -- the value which yields the best fit in
 eq.~(\ref{inc_lambda}). The resummation is performed with $\bar{a}_{\PT}$
 in the ``gluon bremsstrahlung'' scheme. 
See the caption of fig.~\ref{K_117_PV} for further details.  
 }
\label{K_110_PV}
\end{figure}

A closer look at fig.~\ref{K_117_PV} reveals that
our theoretical results undershoot the low $Q$ data points,
while they overshoots at least part of the high $Q$ data points.
Indeed performing a two parameter fit where both the coupling and the
power term are free can lead to better agreement with the data. 
This is shown in table~\ref{best_fit_tab}. 
\begin{table}[H]
\[
\begin{array}{|c|c|c|c|}
\hline
\,\bar{a}_{\eff}^{\PT}\,&\,\alpha_s^{\MSbar}({\rm M_Z})\,& \,
\lambda\, ({\rm GeV})\,&\,\chi^2/{\rm point}\,\\
\hline
\hline
{\rm one-loop}&0.111&0.73&1.33\\
\hline
{\rm two-loop}&0.110&0.62&1.35\\
\hline
\end{array}
\]
\caption{Fit results for $\alpha_s$ and $\lambda$
based on (\ref{inc_lambda}).}
\label{best_fit_tab}
\end{table}
Fig.~\ref{K_110_PV} shows the perturbative summation results for 
$\alpha_s^{\MSbar}({\rm M_Z})=0.110$ 
together with the best fit line (\ref{inc_lambda}) in the two-loop case. 
While in fig.~\ref{K_110_PV} the perturbative sum (\ref{inc_NLO})
by itself is not as close to the data, it is clear that the main
conclusion we drew from fig.~\ref{K_117_PV} concerning the
significance of the resummation holds.

The curvature of $\chi^2$ as a function of $\alpha_s$ and
$\lambda$ around the minimum reflects the spread of the experimental
results. The two parameter fit of (\ref{inc_lambda}), calculated at two-loop, 
to the entire set of data points yields an experimental error of
\beq
\alpha_s^{\MSbar}({\rm M_Z})=0.110\pm 0.0017, 
\label{bf_alpha}
\eeq
and
\beq
\lambda=0.62\pm 0.12 \,{\rm GeV}
\label{bf_lambda}
\eeq
for a confidence level of $95\%$.

Further statistical analysis shows that the various experiments are
quite consistent and that there is no significant difference between
small $Q$ and large $Q$ data points as far as our fit is concerned. 
For instance, if we exclude the lowest data points $Q<22\, {\rm GeV}$,
which seem quite spread, we find an improvement in the fit with a
minimal $\chi^2/{\rm point}=1.15$ but the corresponding values of
$\alpha_s$ and $\lambda$ change just a little, namely:
$\alpha_s^{\MSbar}({\rm M_Z})=0.111$ and 
$\lambda=0.54 \,{\rm GeV}$\footnote{The
latter central values are obtained also if we exclude the 4 data points of the
Mark J experiment which are higher than the rest. Indeed here the fit
is better: $\chi^2/{\rm point}=1.01$.}. If we exclude
the highest data points $Q\geq 172\,{\rm GeV}$ we find $\chi^2/{\rm
point}=1.44$ with the same central values as in (\ref{bf_alpha}) and
(\ref{bf_lambda}). The most striking evidence that there is no systematic 
trend in the data which is missed by our fit is that even if we exclude
{\em all} the data points above or below ${\rm M_Z}=91.2\,{\rm GeV}$, the 
best fit values are hardly affected: in the former case, having 29 data
points, we get $\chi^2/{\rm point}=1.61$ with the same central 
values as in  (\ref{bf_alpha}) and (\ref{bf_lambda}) and in the
latter, having 20 data points, we get $\chi^2/{\rm point}=1.17$
with $\alpha_s^{\MSbar}({\rm M_Z})=0.111$ and $\lambda=0.53 \,{\rm
GeV}$. Note, however, that the effective experimental error 
changes significantly, 
e.g. in the latter case, the error in a two parameter fit with
$95\%$ confidence level on the extracted value of 
$\alpha_s$ becomes $\pm 0.011$.

\subsection{Truncation of the perturbative expansion}

Let us consider now the expansion of the renormalon integral 
(\ref{Rapt}) in some renormalization scheme, e.g. the expansion 
(\ref{Rapt_pert_expansion}) in $a_{\MSbar}(Q^2)$. As explained in
sec.~5.2 this series diverges at large orders due to the factorial
increase of the coefficients induced by (infrared)
renormalons. A standard procedure dealing with asymptotic
expansions is to sum the series up to the minimal term.
This can be regarded as an effective regularization of the all order
sum.

\begin{figure}[htb]
\begin{center}
\mbox{\kern-0.5cm
\epsfig{file=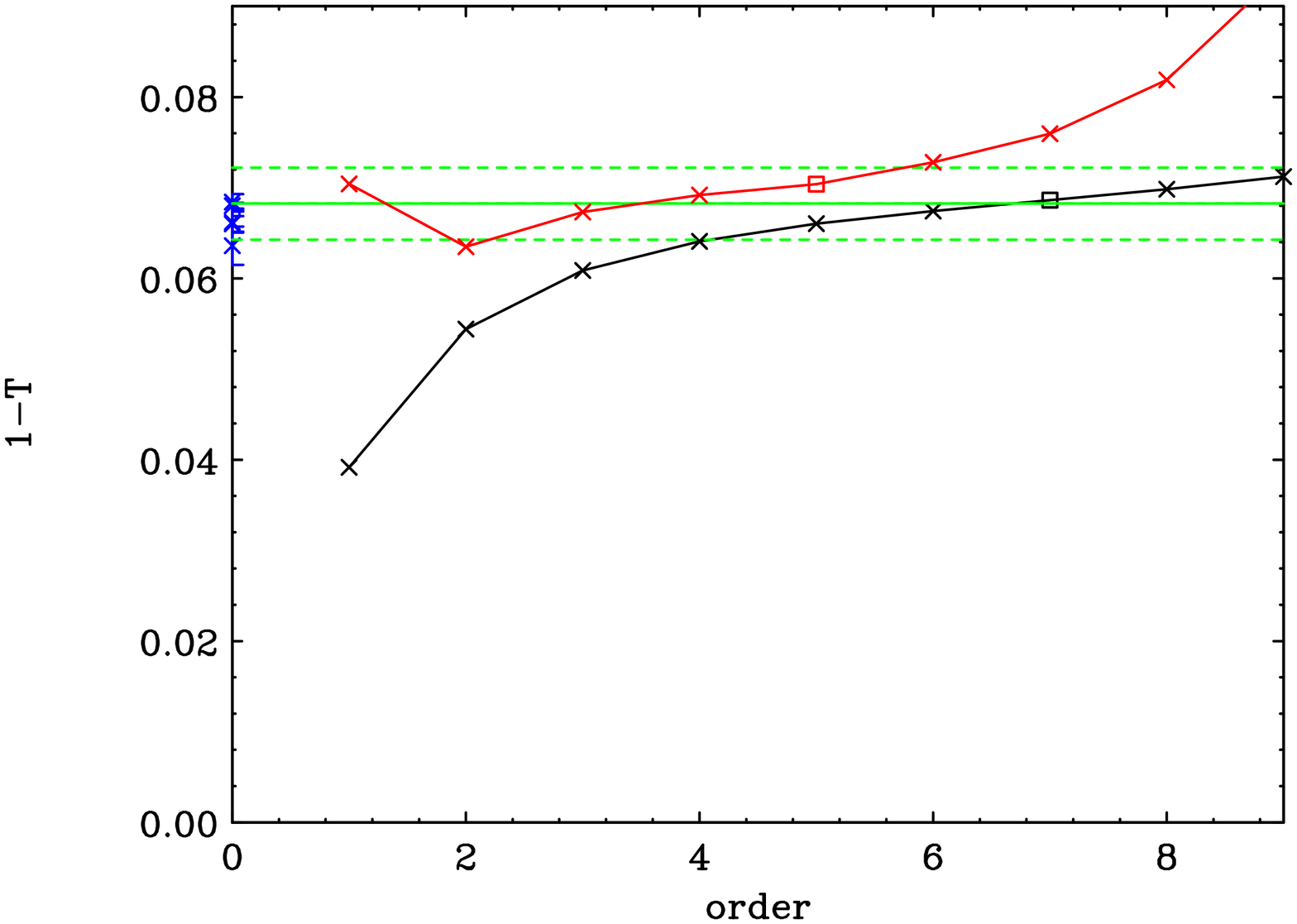,
width=10.0truecm,angle=90}
}
\end{center}
\caption{Partial sums corresponding to the expansion of the one-loop
 renormalon integral (\ref{Rapt}), with $\bar{a}_{\PT}$ in the ``gluon
 bremsstrahlung'' scheme, in terms of $a_{\MSbar}(\mu_R^2)$
as a function of the truncation order, for $Q={\rm M_Z}=91.2\,{\rm GeV}$
with $\alpha_s^{\MSbar}({\rm M_Z})=0.117$.
The lower line corresponds to $\mu_R^2=Q^2$, eq.~(\ref{Rapt_pert_expansion}), 
and the upper line to $\mu_R^2=\mu_{\BLM}^2$, eq.~(\ref{BLM_expansion}).
The square symbol is the minimal term in each expansion.
The horizontal band represents Borel summation, where the middle line
is the principal value regularization and the two dashed lines show
the estimated renormalon ambiguity based on the imaginary part of the
Borel sum. The symbols on the left show the experimental data points.
Note that the small next-to-leading order correction of (\ref{S1_est})
is not included in the theoretical results shown here.
 }
\label{BLM}
\end{figure}

\begin{figure}[htb]
\begin{center}
\mbox{\kern-0.5cm
\epsfig{file=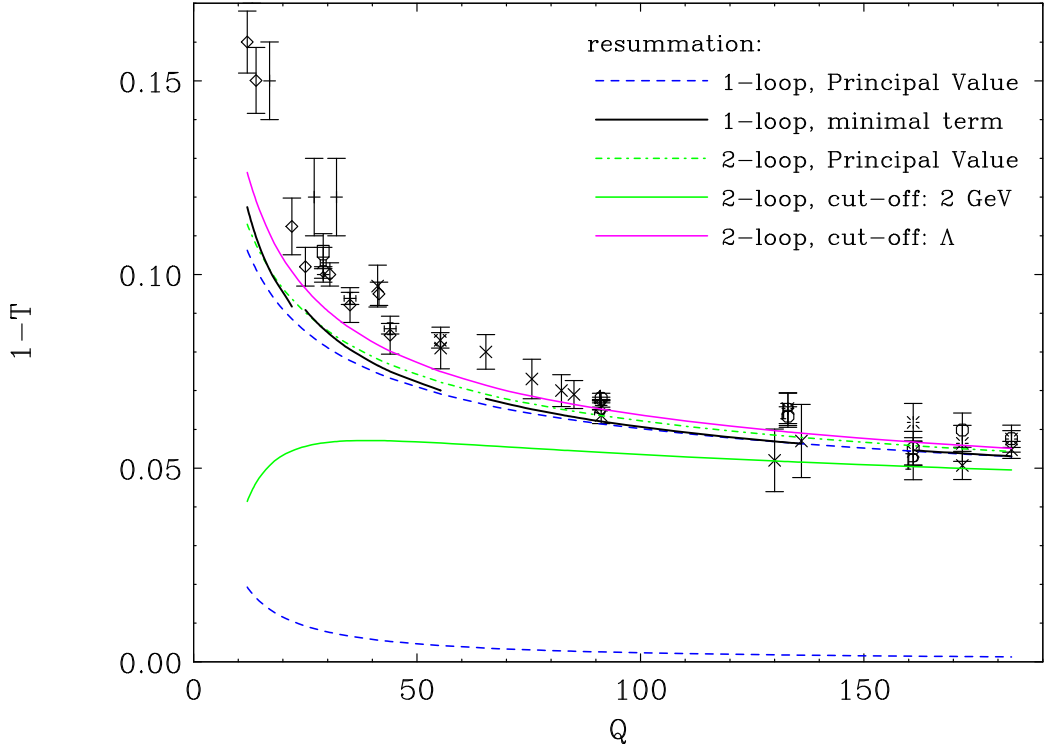,width=10.0truecm,angle=90}
}
\end{center}
\caption{Comparison between different regularizations of the perturbative
 sum with $\bar{a}_{\PT}$ in the ``gluon bremsstrahlung'' scheme 
for $\alpha_s^{\MSbar}({\rm
 M_Z})=0.110$ as a function of the center of mass energy $Q$.
The regularizations shown are: one-loop (dashed) and two-loop
 (dot-dash) principal value Borel sum (or $R_{\APT}$), truncation of
 the $\overline{\rm MS}$ expansion (\ref{Rapt_pert_expansion}) at
 the minimal term 
and the two-loop $R_{\UV}^{\PT}$ with the cutoff scales 
$\mu_I=2 \,{\rm GeV}$
 and \hbox{$\mu_I=\Lambda_{\brem}^{\2loop}=0.361 \,{\rm GeV}$}.
Note that the small next-to-leading order correction of (\ref{S1_est})
is not included in the perturbative sum in this plot.
 }
\label{K_110_reg}
\end{figure}

As an example we analyze here the expansion
(\ref{Rapt_pert_expansion}) in some detail. 
Fig.~\ref{BLM} shows the increasing
order partial sums at a given center of mass energy $Q={\rm M_Z}$
while fig.~\ref{K_110_reg} shows the results obtained when truncating
the series at the minimal term as a function of $Q$.
In the latter, the relevant curve is made of four distinct pieces
according to the number of terms included in the sum: the minimal term
is reached between the sixth and the ninth term depending on $Q$. 
Both figures show that truncation at the minimal term
is quite close to the principal value Borel sum regularization.

It is clear from fig.~\ref{BLM} that the contribution of the 
${\cal O}(a_{\MSbar}^3)$ and ${\cal
O}(a_{\MSbar}^4)$ terms is significant, as one could already
guess based on the largeness of the ${\cal O}(a_{\MSbar}^2)$ term
in (\ref{t_MSbar}). The minimal term, which turns out to be 
close to the principal value Borel sum, 
is obtained for this center of mass energy at order ${\cal O}(a_{\MSbar}^7)$.
In (\ref{Rapt_pert_expansion}), a new log-moment $f_i$ enters at each 
order in the perturbative expansion, such that the 
term ${\cal O}(a_{\MSbar}^i)$
depends on all $f_j$ with $j<i$. From the definition of the log-moments 
(\ref{f_i}) it follows that the larger the order $i$, 
the more sensitive $f_i$ is to the small $\epsilon$ behavior of 
${\cal F}(\epsilon)$. Eventually, when the asymptotic regime is reached, 
the added terms reflect just the leading term in the small $\mu^2$ 
expansion of the 
characteristic function, namely the leading renormalon, which is related in 
our case to a $1/Q$ power correction. Since the value of the added 
non-perturbative term $\lambda/Q$ is determined by a fit, it may not
be important at which order exactly the series is truncated, provided 
it is close enough to the asymptotic regime. 
On the other hand, a qualitative difference 
exists between truncation of the series at the minimal term and
truncation much before the asymptotic regime, say at the 
next-to-leading order. The latter would not differ from a generic 
regularization of the perturbative sum just by power terms. 
This also implies that the value of $\alpha_s$ obtained by fitting 
the data with a next-to-leading order 
series plus a power term would be different than in the current
approach.

Next consider, as a pedagogical exercise, a fit to experimental 
data based on the truncated 
expansion (\ref{Rapt_pert_expansion}) in the $\overline{\rm MS}$ scheme, namely
\begin{eqnarray}
\label{trunc_fit}
\left<t\right>&=&\frac{C_F}{2}\left[\left. S_0^{\PT}\right\vert_{{\cal
O}(a^k)} +\delta_{\NLO}\right]\,+\,\frac{\lambda_{\rm pert}}{Q}\\ \nonumber
&=&\frac{C_F}{2}\left[t_0 \,a_{\MSbar}(Q^2) \,+\, t_1
\,a_{\MSbar}^2(Q^2) \,+\, \sum_{i=3}^{k}\,t_{i-1}^0\,a_{\MSbar}^{i}(Q^2)
\,\right]\,+\,\frac{\lambda_{\rm pert}}{Q}
\end{eqnarray}
where $k$ is the order of truncation. The fit results are listed in
table~\ref{trunc_fit_tab} for $k=2$ through $6$. 
For these values of $k$ the series is still 
convergent: the diverging part of the expansion 
is not reached for any $Q$, since the minimal term is between the 
sixth and the ninth term. 
The coupling $a_{\MSbar}(Q^2)$ in (\ref{trunc_fit}) is assumed
to obey the two-loop renormalization group equation
with $N_f=5$.
\begin{table}[H]
\[
\begin{array}{|c|c|c|}
\hline
\,k\,&\,\alpha_s^{\MSbar}({\rm M_Z})\,& \,\lambda_{\rm pert}\, ({\rm GeV})\,\\
\hline
\hline
2&0.128&0.72\\
\hline
3&0.118&0.65\\
\hline
4&0.115&0.58\\
\hline
5&0.114&0.50\\
\hline
6&0.114&0.40\\
\hline
\end{array}
\]
\caption{Fit results based on (\ref{trunc_fit}) using the expansion
(\ref{Rapt_pert_expansion}) up to order $k$.}
\label{trunc_fit_tab}
\end{table}
Note that nothing can be learned from the quality of the fit: it is roughly 
the same in all cases, $\chi^2/{\rm point}\simeq 1.3$, and it is also 
very close to the resummation based fit of the previous section,
$\chi^2/{\rm point}\simeq 1.33$. The corresponding 
experimental error in the extracted value of $\alpha_s$ in
table~\ref{trunc_fit_tab} based on a two parameter fit with $95\%$
confidence level is $\pm 0.0024$.  

Care should be taken comparing the results for $k=2$
in table~\ref{trunc_fit_tab} with the fit in \cite{DW}
as well as with recent experimental fits \cite{Moriond_Stenzel}. 
In the latter, the finite infrared coupling formula of 
refs.~\cite{DW,DMW,Mil} is used, namely the coefficient in front of the
$1/Q$ term is modified, $\lambda \longrightarrow
\lambda-\lambda_{\PT}$ to avoid double counting in the perturbative
and power correction pieces. Since $\lambda_{\PT}$ depends 
on the coupling this change has an effect on the central
values of the fit. For example, using the formula of
\cite{DW,DMW,Mil} with the current data set one obtains at the
next-to-leading order ($k=2$) a central value of 
$\alpha_s^{\MSbar}({\rm M_Z})=0.124$ (the Milan
factor \cite{Mil} is not included) rather than $\alpha_s^{\MSbar}({\rm
M_Z})=0.128$. 

Comparing table~\ref{trunc_fit_tab} with the 
resummation\footnote{We recall that the coefficients 
in (\ref{Rapt_pert_expansion}) are
computed based on the one-loop $\bar{a}_{\eff}^{\PT}$ and so the 
relevant comparison is with the one-loop resummation fit in table 
~\ref{best_fit_tab}.} results of table~\ref{best_fit_tab} we find that the
truncation leads to an overestimated value of $\alpha_s$.
This comparison invalidates the next-to-leading order procedure of
ref.~\cite{DW,DMW,Mil}.
As more terms are included, the value of $\alpha_s$ becomes closer 
to the resummation result, $\alpha_s^{\MSbar}({\rm M_Z})=0.111$.  
This value is not reached even for $k=6$: indeed there is a
difference between a fixed order calculation and a regularized sum,
e.g. in the principal value regularization. The latter is close,
as we saw, to truncation of the series at the minimal term,
but then the order of truncation is not fixed but rather depends on $Q$.
 
The most drastic change in table~\ref{trunc_fit_tab} is, 
of course, between the next-to-leading order
based fit ($k=2$) and the $k=3$ fit which includes an estimated
(\ref{Rapt_pert_coef})
next-to-next-to-leading order contribution from the ``leading
skeleton'', $t_2^0 \,a_{\MSbar}^3$ with $t_2^0\simeq 188$. 
There is some uncertainty in the estimated coefficient
$t_2^0$ so long as the identity of the ``skeleton coupling'' is not
known. Here we assumed that $\bar{a}_{\PT}$ is in the ``gluon bremsstrahlung''
scheme, and so we used $d_1=1-\frac{\pi^2}{4}$ in
eq.~(\ref{Rapt_pert_coef}). If we assume instead\footnote{The 
dependence of the resummation based fit on the ``skeleton scheme'' is
discussed in sec.~5.6.} the pinch technique
scheme, with $d_1=1$, we obtain $t_2^0\simeq 279$, which yields a best
fit for $k=3$ at $\alpha_s^{\MSbar}({\rm M_Z})=0.114$ with
$\lambda_{\rm pert}=0.66$ (cf. $k=3$ in table~\ref{trunc_fit_tab}). 
Note that if $d_1$ (which characterizes the relation between the
``skeleton scheme'' and $\overline{\rm MS}$) is not large, it is the
``large $\beta_0$'' term (which is proportional to $\beta_0^2$ and  
independent of $d_1$) that dominates 
$t_2^0$ in eq.~(\ref{Rapt_pert_coef}): 
\hbox{$t_2^0\,=\,1.5776\, d_1^2\,+\,37.59\,d_1\,+\,239.62$}.

We stress that the choice we made in (\ref{Rapt_pert_expansion}) to expand
$\overline{a}_{\eff}^{\PT}(\mu^2)$ in terms of
$a_{\MSbar}(Q^2)$ is arbitrary. We could in principle pick any
renormalization scale and scheme. A particularly good choice, 
using still the ${\overline {\rm MS}}$ scheme, is to set the scale equal
to the BLM scale, namely to eliminate the term proportional to
$\beta_0$ from the next-to-leading order coefficient
\beq
\mu^2_{\BLM}=Q^2\exp\left(\frac{f_1}{f_0}+c_1 \right)
\eeq
yielding $\mu_{\BLM}^{\MSbar}\simeq 0.0447 \, Q$.
We then obtain from (\ref{Rapt_pert_expansion}) the following 
series\footnote{Since $c_1$ can be swallowed into the definition of the 
scale, it disappears from the BLM series (\ref{BLM_expansion}) completely. 
Note also that in this expansion all the higher order terms linear 
in $\beta_0$ vanish.}  
\begin{eqnarray}
\label{BLM_expansion}
S_0^{\PT} ={f_{0}}\,a_{\MSbar}(\mu^2_{\BLM}) &+&
{d_1}\,{f_{0}}\,a_{\MSbar}^{2}(\mu^2_{\BLM})\\ 
&+& \left\{{d_1}^{2}\,{f_{0}} - { \frac {1}{3}} \,{
\frac {3\,{f_{1}}^{2} + \pi ^{2}\,{f_{0}}^{2} - 3\,{f_{2}}\,{f_{0
}}}{{f_{0}}}} \,\beta_0^{2}\right\}\,a_{\MSbar}^{3}(\mu^2_{\BLM})+\cdots
\nonumber
\end{eqnarray}
which provides a good approximation to $R_{\APT}\simeq
R_{\PT\vert\PV}$ already at the
leading order, as shown in fig.~\ref{BLM}. 
We mention that a particularly low renormalization scale 
was suggested for this observable in \cite{Beneke,CGM}. 
Such a choice can now be justified from another view point,
noting that the BLM scale approximates well the resummed perturbative 
series. 

The proximity of the leading order BLM result to the Borel 
sum in fig.~\ref{BLM} suggests that performing the fit based on a 
next-to-leading order partial sum in $\overline{\rm MS}$  with 
$\mu_R=\mu_{\BLM}$ would be much better than with $\mu_R=Q$ 
corresponding to $k=2$ in table~\ref{trunc_fit_tab}. 
Indeed, performing such a fit (with a power term of the form
$\lambda_{\rm pert}/Q$), we find a significant change in the 
extracted parameters. The central value is
$\alpha_s^{\MSbar}({\rm M_Z})=0.116$ 
(with $\lambda_{\rm pert}=0.55\, {\rm GeV}$),
which is much closer to the best fit result of our
resummation, $\alpha_s^{\MSbar}({\rm M_Z})=0.111$. It should be noted
that the results do not coincide: leading order BLM scale-setting is
not a substitute to actually performing the resummation (see related
observations in \cite{BBB}).   

Note that the success of BLM in ${\overline {\rm MS}}$ is not
guaranteed a priori. It is the smallness of $d_1$ in the relation (\ref{bara})
between the scheme coupling (chosen as ${\overline {\rm MS}}$) and the
``skeleton coupling'' (assumed to be the ``gluon bremsstrahlung''coupling) 
which plays a role here.
The most natural scheme to apply BLM is the ``skeleton scheme'' 
$\bar{a}_{\PT}$, where $c_1=d_1=0$. 
Then, similarly to the Euclidean case (\ref{BLM_skeleton}), 
there is a scheme invariant interpretation to the BLM scale
\beq
\mu^2_{\BLM}=Q^2\exp\left(\frac{f_1}{f_0}\right)
\label{BLM_skeleton_F}
\eeq
as the center of $\dot{\cal F}(\epsilon)$.
In the case of the average thrust it is 
\beq
\mu_{\BLM}= 0.1028\, Q,
\label{mu_BLM_skeleton}
\eeq
as can be verified directly in fig.~\ref{F_dot}.
Now the leading term in the BLM expansion (\ref{BLM_expansion}) 
is $f_0\,\bar{a}_{\PT}(\mu_{\BLM}^2)$ and the next-to-leading order correction
vanishes. Higher order corrections have a simple
interpretation \cite{Neu,mom} in terms of the properties for the 
function $\dot{\cal F}(\epsilon)$. For instance, the next-to-next-to-leading 
order is related to the width of $\dot{\cal F}(\epsilon)$ through its
second moment $f_2$.

\subsection{Cutoff regularization and the infrared finite coupling approach}

As explained in sec.~3.3 the cutoff regularized perturbative sum 
$R_{\UV}^{\PT}$ of eq.~(\ref{Rptuv}) is of special interest because
it is fully under control in perturbation theory. In particular, in
this regularization the replacement of the all order 
coupling $\bar{a}_{\PT}$ by a one-loop or two-loop coupling is 
justified provided $\mu_I/\Lambda$ is large enough.
In addition, as we saw in sec.~4, the infrared cutoff regularization appears
naturally in the framework of the infrared finite coupling \cite{DW,DMW}.
Let us therefore repeat the analysis of the average thrust in terms of
this regularization.

In general, $R_{\UV}^{\PT}$ can be obtained from $R_{\APT}$ by
$R_{\UV}^{\PT}=R_{\APT}-\Delta R$.
Since in our case ${\rm Re} \left\{\delta R_{\APT}\right\}$ 
is negligibly small, we simply have 
$\Delta R\simeq {\rm Re} \left\{R^{\PT}_{\IR}\right\}$. The ${\cal
O}(1/Q^{2n})$ contribution corresponding to the $n$-th order term 
in (\ref{F_ana}) is given by eq.~(\ref{qn_form}) and
eq.~(\ref{reRirpt-low1}) with~(\ref{In-2loop1}) (or equivalently
eq.~(\ref{reRirpt-low2})) in the one and two loop cases, respectively.
The leading $n=\frac{1}{2}$ term is given in the one-loop case by
\beq
\left. {\rm
Re}\left\{R^{\PT}_{\IR,\frac12}\right\}\right\vert_{\1loop}=\left(\frac{\mu_I}{Q}\right)
\frac{-C_T^{\left(\frac12\right)}}{\beta_0\pi}\,e^{-\frac{t_I}{2}}\,{\rm
Ei}\left(\frac{t_I}{2}\right)
\label{R_PT_IR_half}
\eeq
with $t_I \equiv \ln\left(\mu_I^2/\Lambda^2\right)$ and in the
two-loop case by
\beq
\left. {\rm Re}\left\{R^{\PT}_{\IR,\frac12}\right\}\right\vert_{\2loop}
=\left(\frac{\mu_I}{Q}\right)
\frac{-C_T^{\left(\frac12\right)}}
{\beta_0\pi}\,e^{-\frac{\tilde{t}_I}{2}}\,{\rm
Re}\left\{-\left(-\frac{\tilde{t}_I}{2}\right)^{\delta}\,
\Gamma\left(-\delta,-\frac{\tilde{t}_I}{2}\right)\right\}
\label{R_PT_IR_half_2loop}
\eeq
with $\tilde{t}_I\equiv1/(\beta_0\tilde{a}_I)$ where $\tilde{a}_I$ is
defined in (\ref{atilde-I}) and
$\delta=\delta_{\frac12}=\frac12\left(\beta_1/\beta_0^2\right)$
according to eq.~(\ref{delta_n}).

\begin{figure}[htb]
\begin{center}
\mbox{\kern-0.5cm
\epsfig{file=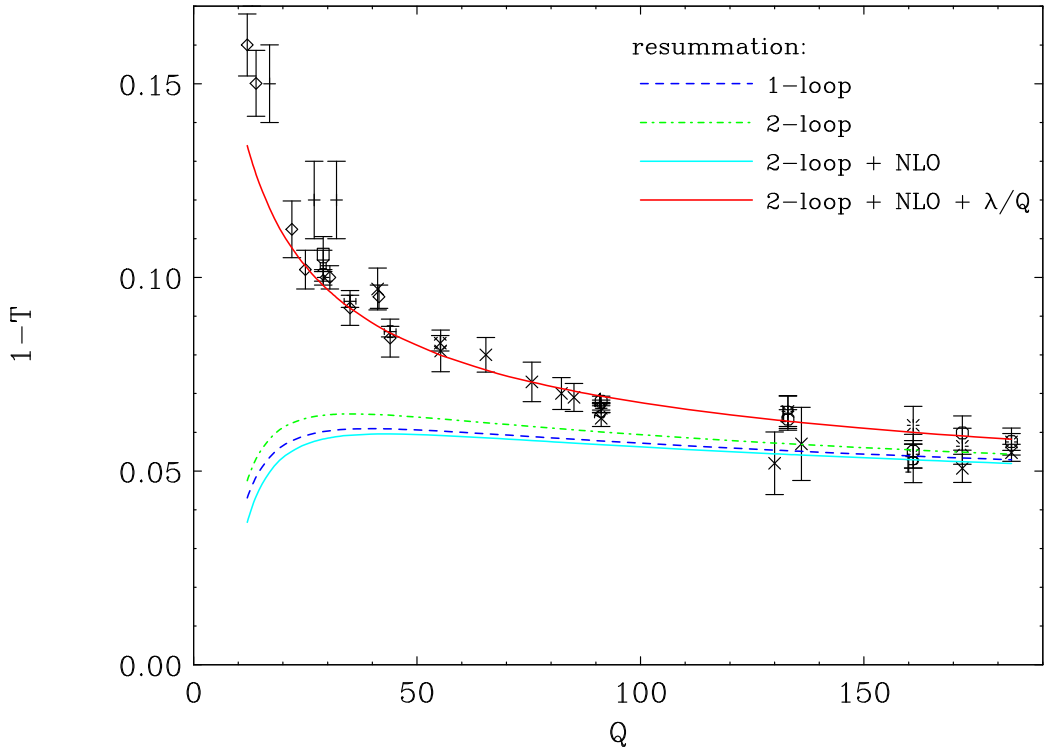,width=10.0truecm,angle=90}
}
\end{center}
\caption{Cutoff regularized perturbative sum  with 
$\mu_I=2\,{\rm GeV}$, given $\alpha_s^{\MSbar}({\rm M_Z})=0.117$. As
in fig.~\ref{K_117_PV}, the resummation is calculated with $\bar{a}_{\PT}$
as a one or two loop coupling in the ``gluon bremsstrahlung'' 
scheme. In the
two-loop case we show also a curve that includes the full ${\cal
O}(\alpha_s^2)$ term, as well as a best fit curve that includes in
addition a fitted $\lambda/Q$ correction (\ref{inc_lambda_cutoff}).  
 }
\label{K_117_cutoff}
\end{figure}

Choosing $\mu_I=2\,{\rm GeV}$, with the same value of $\alpha_s$ as in
fig.~\ref{K_117_PV}, we obtain the cutoff regularized sum 
\hbox{$\left<t\right>_{\PT}=\frac{Cf}{2}
R_{\UV}^{\PT}$} which is presented in fig.~\ref{K_117_cutoff}.
Comparing fig.~\ref{K_117_cutoff} with fig.~\ref{K_117_PV}, the first
observation is that $R_{\UV}^{\PT}(Q^2)$ is significantly lower than
$R_{\APT}(Q^2)$ (or $R_{\PT\vert\PV}(Q^2)$), 
especially\footnote{Fig.~\ref{K_117_cutoff} shows 
that $R_{\UV}^{\PT}(Q^2)$ is not a
monotonous function of $Q$. This is a unique feature of this
regularization (with a reasonably high $\mu_I$) which distinguishes 
it from both $R_{\PT\vert\PV}(Q^2)$ and typical truncated perturbative
series, which are monotonic decreasing functions of $Q$.}
for low $Q$. This means that a major contribution to the resummation 
(\ref{Rapt}) is from space-like momentum scales below the cutoff
$\mu_I=2\, {\rm GeV}$, i.e. from large distance scales in which
the perturbative treatment does not apply. This point will be
elaborated in the conclusion section.

As discussed in sec.~3.3, the replacement of the all order coupling by
some approximation like the one or two loop coupling is best
justified in terms of the cutoff regularization. It is thus important
to verify that $R^{\PT}_{\UV}$ remains stable going from one to two
loops. Fig.~\ref{K_117_cutoff} shows that indeed even at 
low $Q$ the difference between the one and two loop 
$R_{\UV}^{\PT}(Q^2)$ is rather small. 
We conclude that $R_{\UV}^{\PT}(Q^2)$ is under perturbative 
control for $\mu_I=2 \,{\rm GeV}$.
To improve further the accuracy at which $R_{\UV}^{\PT}(Q^2)$ is determined
one can, in principle, either go to higher orders in the $\beta$
function of $\bar{a}_{\PT}$ or increase $\mu_I$. 
In the latter case, however, one should be careful not to 
invalidate the expansion of $\dot{\cal F}(\mu^2/Q^2)$ in 
$\Delta R$ (or $\Phi(k^2/Q^2)$ in $R^{\PT}_{\IR}$)
which holds only for small enough $\mu_I^2/Q^2$. The physical value of
$\mu_I$, separating at once short distance physics from large distance
physics and the perturbative domain from the non-perturbative one,
should be set satisfying these two constraints.

Next, we fit the data with
\beq
\left<t\right>=\frac{C_F}{2}\left[R_{\UV}^{\PT}+\delta_{\NLO}\right]
+\frac{\lambda_{\mu_I}}{Q}.
\label{inc_lambda_cutoff}
\eeq
The best fit values of $\lambda_{2\,{\rm GeV}}$ for 
$a_{\MSbar}({\rm M_Z})=0.117$ (fig.~\ref{K_117_cutoff}) are listed in 
table~\ref{117_cutoff_fit} below.
Note that the fit (\ref{inc_lambda_cutoff})
in fig.~\ref{K_117_cutoff}, and the fit (\ref{inc_lambda})
in fig.~\ref{K_117_PV} are almost identical. This is reflected also in the
similar $\chi^2$ values quoted in tables~\ref{117_cutoff_fit} and
\ref{117_fit}.
The reason is that the two regularizations, 
$R_{\APT}$ (or $R_{\PT\vert\PV}$) and 
$R_{\UV}^{\PT}(Q^2)$ differ, to a very good approximation, just by the
$1/Q$ power correction of the type we add as a free parameter in the 
fit (see sec.~4). 
One can thus check that the difference between the fitted
parameters in the two regularizations 
is as implied by the calculation. For instance, in the one-loop case,
the difference is calculable from eq.~(\ref{R_PT_IR_half}), namely
\beq
\lambda_{\mu_I}-\lambda\simeq \mu_I\,\frac{C_F}{2}\,
\frac{-C_T^{\left(\frac12\right)}}{\beta_0\pi} e^{-\frac{t_I}{2}}{\rm
Ei}\left(\frac{t_I}{2}\right).
\eeq
\begin{table}[H]
\[
\begin{array}{|c|c|c|}
\hline
\,\bar{a}_{\eff}^{\PT}\,& \,\lambda_{2\,{\rm GeV}}\, 
({\rm GeV})\,&\,\chi^2/{\rm point}\,\\
\hline
\hline
{\rm one-loop}&1.32&2.43\\
\hline
{\rm two-loop}&1.14&3.46\\
\hline
\end{array}
\]
\caption{Power term ($\lambda_{2\,{\rm GeV}}$) fit results with 
$\alpha_s^{\MSbar}({\rm  M_z})=0.117$ based on (\ref{inc_lambda_cutoff}).}
\label{117_cutoff_fit}
\end{table}

The sub-leading power term making $R_{\PT\vert\PV}$
and $R_{\UV}^{\PT}$ different is related to the next non-analytic
term in the expansion of ${\cal F}(\epsilon)$,
$n=\frac32$, which leads to negligibly small $1/Q^3$ power
corrections. 
As mentioned above, the latter are calculable using eq.~(\ref{qn_form}) and 
(\ref{reRirpt-low2}) in the one and two loop cases, respectively. 

The effect of leading and sub-leading power terms
in the expansion of ${\cal F}(\epsilon)$ in (\ref{F_ana}) 
is summarized in table \ref{Q12} for the lowest
experimentally relevant energy $Q=12 \,{\rm GeV}$.
\begin{table}[H]
\[
\begin{array}{|c||c|c||c||c|c|}
\hline
&&&&&\\
\,n^{\,} & {{C_F}\over{2}}{\rm Re} \left\{R_{\IR}^{\PT}\right\} &
\frac{C_F}{2}{\rm Re}\left\{\delta R_{\APT}\right\} & \frac{C_F}{2}\Delta R &
\frac{C_F}{2}\left(R_{\IR}^{\PT}-R_<^{\PT}\right) &\frac{C_F}{2} R_<^{\APT}\\
\hline
\,\frac12_{\, }&0.0796          &0               &0.0796         &-0.0401&0.1197 \\
\hline
\,1_{\, }      &0               &\,0.701\,\,10^{-3}  &0.701\,\,10^{-3} &0.0341 &-0.0334  \\
\hline
\,\frac32_{\, } &-0.131\,\,10^{-2}&0&-0.131\,\,10^{-2}&-0.666\,\,10^{-2}&\,0.535\,\, 10^{-2} \\
\hline
\,2_ {\, } &-0.822\,\,10^{-4}&\,0.663\,\,10^{-7} &-0.821\,\,10^{-4}&\,0.353\,\,10^{-4}&-0.117\,\,10^{-3}\\
\hline
\end{array}
\]
\caption{Contributions ${\cal O}(1/Q^{2n})$ from increasing order terms
in the expansion of ${\cal F}(\epsilon)$ to the difference between different
regularizations of the perturbative sum with the one-loop running
coupling in the ``gluon bremsstrahlung'' scheme, at $Q=12\, {\rm
GeV}$. The infrared cutoff is set
to $\mu_I=2\,{\rm GeV}$. For comparison, at this energy
$\left<t\right>\simeq\frac{C_F}{2}R_{\APT}=0.1233$. 
 }
\label{Q12}
\end{table}
Table \ref{Q12} contains contributions of both analytic and
non-analytic terms in ${\cal F}(\epsilon)$. It
presents the two possible separations of $\Delta R$
discussed in sec.~3.3 -- the separation into ${\rm Re}
\left\{R_{\IR}^{\PT}\right\}$ and ${\rm Re}\left\{\delta
R_{\APT}\right\}$, corresponding to eq.~(\ref{dR-1}), on the left side of the
table, and the separation into $R_{\IR}^{\PT}-R_<^{\PT}$ and
$R_<^{\APT}$, corresponding to eq.~(\ref{delta_R_sep}) on the right.
We find that in the separate pieces on the right 
the sub-leading terms (e.g. $n=1$) are relatively
important, while not so on the left, where the total $\Delta R$ can be
approximated by the leading ${\rm Re} \left\{ R_{\IR}^{\PT}\right\}$ piece
\beq
\Delta R\simeq {\rm Re} \left\{R_{\IR,\frac{1}{2}}^{\PT}\right\}.
\eeq

Next, consider the dependence of the cutoff regulated perturbative
sum $R^{\PT}_{\UV}(Q^2)$ on the cutoff scale $\mu_I$.
Any two cutoff regularizations differ, in general, by renormalon related
infrared power corrections. In our case, they differ to a very good
approximation by a $1/Q$ term which can be calculated using
eq.~(\ref{R_PT_IR_half}) and (\ref{R_PT_IR_half_2loop}) 
in the one and two loop cases, respectively.
As a result, one would practically obtain the same best fit in 
(\ref{inc_lambda_cutoff}) for any arbitrary $\Lambda \leq\mu_I\ll Q$.
In the two-loop case, $R^{\PT}_{\UV}(Q^2)$ is well
defined even for the extreme choice $\mu_I=\Lambda$. The resulting
regularization is shown in fig.~\ref{K_110_reg} together with the
standard $\mu_I=2\,{\rm GeV}$ choice.
The $\mu_I=\Lambda$ regularization turns out to be close to the
best fit in this case. This regularization is, of course, unacceptable 
in the infrared finite coupling approach \cite{DW,DMW}, where one requires
that the non-perturbative coupling $\bar{a}$ is well approximated by
$\bar{a}_{\PT}$ above $\mu_I$. As explained in sec.~4, within the
infrared finite coupling approach, the non-perturbative
parameter $\lambda_{\mu_I}$ acquires a physical interpretation as the
small gluon virtuality moment of the ``skeleton coupling''. 
Using (\ref{Rir_n}) with (\ref{F_ana}), we have
\beq
R_{\IR}=\int_0^{\mu_I^2}\bar{a}(k^2)\,\Phi\left(k^2/Q^2\right)\,
\frac{dk^2}{k^2}\simeq -\left(\frac{\mu_I}{Q}\right)
C_T^{\left(\frac12\right)}\,\frac{2}{\pi}\,\frac{1}{\mu_I}\int_0^{\mu_I}
\bar{a}(k) dk
\label{R_IR_half}
\eeq 
which implies
\beq
\lambda_{\mu_I}= - \mu_I \frac{C_F}{2} C_T^{\left(\frac12\right)}\,
\frac{2}{\pi}\,\frac{1}{\mu_I}\int_0^{\mu_I}
\bar{a}(k) dk.
\label{lambda_ident}
\eeq
We stress that the identification of the fit parameter
$\lambda_{\mu_I}$ with a small virtuality moment of the coupling is
meaningful only under 
the strong assumption of universality: it is the same
``skeleton coupling'' for any observable. Eq.~(\ref{lambda_ident}) as
written implies that the coefficient in front of the ``skeleton
coupling'' average is simply related to the small
$\mu^2$ behavior of the characteristic function, in spite of the
non-inclusive nature of the thrust. In ref.~\cite{Mil}
it is argued that this coefficient is in fact modified by the so called
``Milan factor'' arising at the next-to-leading order level, taking 
into account the emission of two gluons. We ignore these corrections
here (see sec.~5.6).
 
\begin{figure}[H]
\begin{center}
\mbox{\kern-0.5cm
\epsfig{file=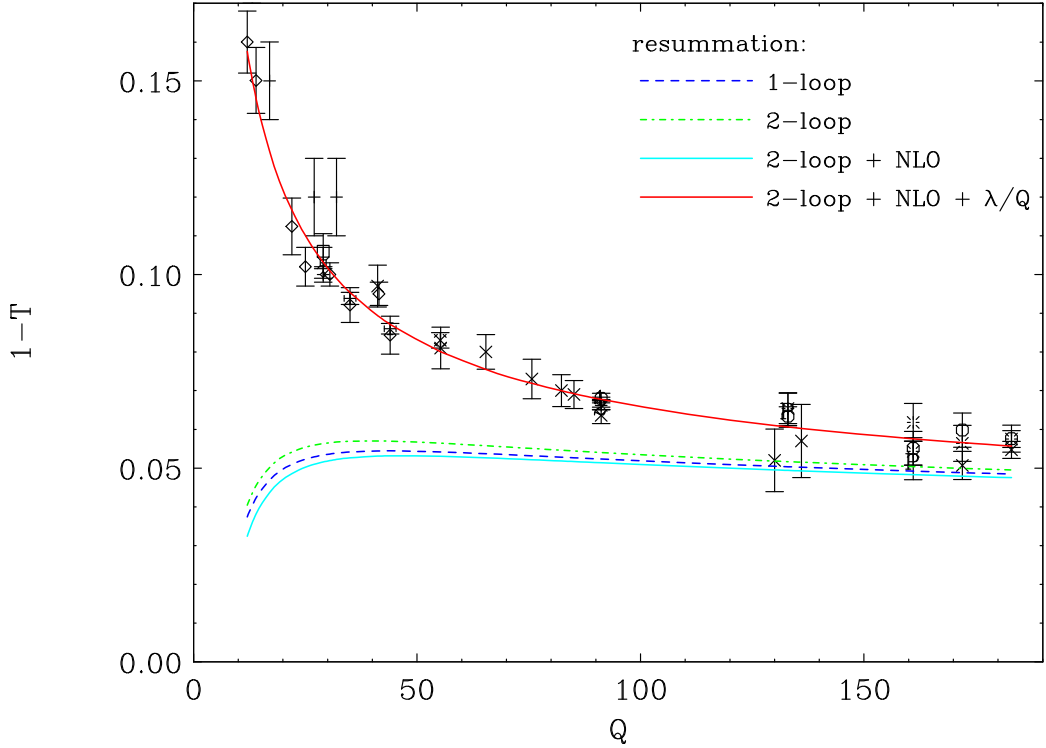,width=10.0truecm,angle=90}
}
\end{center}
\caption{Cutoff regularized perturbative sum  with 
$\mu_I=2\,{\rm GeV}$, for the best fit value 
\hbox{$\alpha_s^{\MSbar}({\rm M_Z})=0.110$}. As
in fig.~\ref{K_110_PV}, the resummation is calculated with $\bar{a}_{\PT}$
as a one or two loop coupling in the ``gluon bremsstrahlung'' 
scheme. In the
two-loop case we show also a curve that includes the full ${\cal
O}(\alpha_s^2)$ term, as well as a best fit curve that includes in
addition a fitted $\lambda/Q$ correction (\ref{inc_lambda_cutoff}).  
 }
\label{K_110_cutoff}
\end{figure}

Let us now extract $\lambda_{2\,{\rm GeV}}$ based on the best fit of 
(\ref{inc_lambda_cutoff}). As we know already from
sec.~5.3 (see table~\ref{best_fit_tab}), the best fit is obtained 
in the one-loop case at \hbox{$\alpha_s^{\MSbar}({\rm M_Z})=0.111$}. 
The corresponding power correction coefficient in the cutoff
regularization is 
\hbox{$\lambda_{2\,{\rm GeV}}=1.57\pm 0.12\, {\rm GeV}$}, where the
error corresponds to the combined systematic and statistical
experimental uncertainties. In the two-loop case, the
best fit is at \hbox{$\alpha_s^{\MSbar}({\rm M_Z})=0.110$} with
\hbox{$\lambda_{2\,{\rm GeV}}=1.49\pm 0.12\, {\rm GeV}$}. 
In the latter case the cutoff regularized sum and the best fit 
are shown in fig.~\ref{K_110_cutoff}.

According to eq.~(\ref{lambda_ident}) we have for $\mu_I=2\,{\rm GeV}$
\beq
\frac{1}{\mu_I}\int_0^{\mu_I}
\bar{a}(k) dk=-\frac{\lambda_{\mu_I}}{\mu_I} \frac{\pi}{C_F}
\frac{1}{C_T^{\left(\frac12\right)}}=\left\{\begin{array}{ll}
0.231\pm 0.017&\,\,\,\,\,\,\,\,\,\,\,{\rm one-loop}\\
0.219\pm 0.017&\,\,\,\,\,\,\,\,\,\,\, {\rm two-loop}\\ 
\end{array}
\right.
\label{first_moment}
\eeq
where $\bar{a}$ is the ``gluon bremsstrahlung'' coupling.

\subsection{Sources of theoretical uncertainty}

The experimental error on the value of $\alpha_s$ and $\lambda$ was
discussed in sec.~5.3. The theoretical error, which is harder to
determine, will be briefly discussed here.

\setcounter{footnote}{0}
A primary source of uncertainty is the unknown identity of the running
coupling used in the calculation of $R_{\APT}$. In sec.~5.3 we
approximated this coupling by the one or two loop coupling
in the ``gluon bremsstrahlung'' scheme with $\beta_0$ and $\beta_1$ 
calculated for $N_f=5$. 
In principle, this coupling should be an ``all-order'' coupling
in the ``skeleton scheme'' (yet unknown), where the number of active
flavors depends on the scale. Let us try to quantify the errors.
\begin{description}
\item{\bf Five active flavors:}
Considering the center of mass energy of most relevant 
experiments, using\footnote{Another source of error is the
assumption, taken in the calculation of ${\cal F}$ and the
perturbative coefficients (\ref{t_0_and_t1}), that the {\em primary} quark
and anti-quark are massless. This problem is common to the standard
perturbative approach and to ours and it certainly deserves
attention. We do not attempt to estimate this error
here.} the running coupling with $N_f=5$ is known to 
be a reasonable approximation in the standard perturbative approach. 
However, in the resummation performed here the coupling runs over 
small gluon virtualities as well, and then there are only four or 
three active flavors. 
The significance of the $N_f=5$ approximation in
our approach can be checked by calculating the difference between the 
contribution to perturbative sum $R_{\APT}$ from small space-like 
momentum scales with $N_f=3$ or $4$ and the corresponding contribution
with $N_f=5$. 
We use for this calculation\footnote{One could evaluate also ${\rm
Re}\left\{R_{\IR,\frac12}^{\PT}\right\}$ or $\Delta R_{\frac12}$ which
are all roughly the same in this case.} the formula for 
$R_{\IR,\frac12}^{\APT}$ at the one-loop order, 
which is obtained in the Appendix, eq.~(\ref{Rirptn_APT}) and (\ref{Jn}).
The relative difference between $R_{\IR,\frac12}^{\APT}(N_f=3)$ and 
$R_{\IR,\frac12}^{\APT}(N_f=5)$ for the lowest experimental data point
$Q=12\,{\rm GeV}$, where the effect is the largest, is just $1.9$
percent. This is negligible compared to other sources of error.
\item{\bf Two-loop coupling:}
In the phenomenological analysis we used the one or two loop coupling
to evaluate the perturbative sum, instead of the ``all 
order skeleton coupling''. 
As explained in sec.~3.3 this replacement is justified in the cutoff
regularization provided $\mu_I/\Lambda$ is large enough. Indeed we
saw in sec.~5.5 (see figs.~\ref{K_117_cutoff} and \ref{K_110_cutoff})
that the cutoff regularized sum $R^{\PT}_{\UV}$ with $\mu_I=2\,{\rm
GeV}$ does not change much going from one to two loops. This
stability suggests that the replacement of the all order coupling by
some low order coupling is a reasonable approximation for this value
of the cutoff.
To go further one would like to examine the effect of a three-loop correction
$\beta_2$ in the $\beta$ function of the ``skeleton coupling'' on
$R^{\PT}_{\UV}$. However, technically, this is a complicated
task. Below, we shall examine instead the effect on $R_{\APT}$.   
At the one and two loop level $R_{\APT}$ is rather stable, as shown in 
figs.~\ref{K_117_PV} and~\ref{K_110_PV}. A priori, the stability 
of $R_{\APT}$ or the principal value Borel sum may be less intuitive
because these regularizations do involve the coupling in the infrared. 
The stability of $R_{\APT}$ may be explained in terms of the infrared 
stability of the APT coupling itself, which is discussed in 
\cite{SS,G5,GGK}, and goes beyond the two-loop level.
Returning to our case, the effect of a three-loop correction in 
$\beta(\bar{a})$ on $R_{\APT}$ can be examined explicitly. 
Since we do not know the identity of the ``skeleton coupling'', 
we just try various choices of $\beta_2$ where for
simplicity we use the Pad\'e improved form of the three-loop $\beta$
function which can be solved analytically using the Lambert W function
\cite{GGK}. As an example consider the case where the
$\beta$ function is modified to include a {\em large}  
three-loop term, $\beta_2/\beta_0=-45$ (both positive and negative
values were studied). Taking the world average value
of $\alpha_s$ ($\alpha_s^{\MSbar}({\rm M}_Z)=0.117$) we obtain at the
lowest experimental data point $Q=12\,{\rm GeV}$, where the effect 
is the largest, 
the following results for $\left<t\right>\simeq\frac{C_F}{2}R_{\APT}$:
at one-loop
$\left<t\right>\simeq 0.123$, at two-loop
$\left<t\right>\simeq 0.131$, and at three-loop
$\left<t\right>\simeq 0.116$. These differences are small compared to
the experimental error bars.
The conclusion remains the same: the perturbative sum is 
quite stable with respect to including higher orders in the $\beta$ function.
Whichever regularization is chosen, what counts at the end is to what
extent the extracted parameters depend on the approximation used for
the perturbative coupling. 
It is clear from table~\ref{best_fit_tab} that the resummation
procedure is quite stable going from one to two loops concerning 
the determination of $\alpha_s$.
In particular, the resulting uncertainty in the extracted value of $\alpha_s$ 
is smaller than the experimental error (\ref{bf_alpha}). 
Moreover, as we saw, one can use either $R_{\APT}$ or $R^{\PT}_{\UV}$ 
obtaining the same value of $\alpha_s$. 
Thus, based on the stability of $R_{\APT}$ at three-loop we conclude 
that the extracted value of $\alpha_s$ will remain stable, despite our lack of
knowledge concerning the stability of $R^{\PT}_{\UV}$ at this level.  
On the other hand, for the determination of the power correction 
or the low momentum average of the coupling, 
the relevant theoretical uncertainty is comparable to the experimental error.
This is reflected for example in the difference between the one and two loop
results in tables~\ref{117_fit} and~\ref{117_cutoff_fit}. 
Similarly we have for\footnote{Note that this value 
of $\alpha_s$ is close but not equal to the best fit value in 
the one-loop case, table~\ref{best_fit_tab}.} $\alpha_s=0.110$, 
\begin{table}[H]
\[
\begin{array}{|c|c|c|}
\hline
\,&{\rm APT\,/\,PV} & {\rm cutoff}\\
\,\bar{a}_{\eff}^{\PT}\,&\,\,\,\,\,\,\lambda\, ({\rm GeV})
 \,\,\,\,\,\,&\,\lambda_{2\,{\rm GeV}}\, ({\rm GeV})\,\\
\hline
\hline
{\rm one-loop}&0.79& 1.62 \\
\hline
{\rm two-loop}&0.62& 1.49\\
\hline
\end{array}
\]
\caption{Power term fit results with $\alpha_s^{\MSbar}({\rm M_z})=0.110$. }
\label{110_fit}
\end{table}
Here it is important, in principle, which regularization is used.
In our case, as shown in table~\ref{110_fit}, the cutoff
regularization with $\mu_I=2\,{\rm GeV}$ yields slightly more 
stable values of the power correction coefficient than the 
principal value Borel sum (or APT) regularization.
Nevertheless, this uncertainty in $\lambda_{\mu_I}$ is significant and
should be taken into account when discussing the universality of 
the infrared finite coupling.
\item{\bf ``gluon bremsstrahlung'' scheme:}
In the phenomenological analysis above we used the ``gluon
bremsstrahlung'' renormalization 
scheme as the scheme of the ``skeleton coupling'' $\bar{a}_{\PT}$. As
explained in sec.~2, the ``skeleton coupling'' should actually be
defined diagrammatically in a unique way and the only thing we know 
about it, is that it coincides with the V-scheme in the Abelian limit.
Alternative schemes were discussed in sec.~2 and 5.2. 
Of particular interest is the pinch technique coupling which has
some systematic justification \cite{Watson} 
and which differs from the ``gluon bremsstrahlung'' scheme by
having a relatively large next-to-leading order correction
(\ref{tilde_f1_pinch}) on top 
of the terms included in the resummation.
The leading ${\cal O}(a^2)$ ``scheme dependence'' effects
corresponding to taking $\bar{a}_{\PT}$ in different schemes cancel 
between $R_{\APT}$ and remaining next-to-leading order
correction in eq.~(\ref{inc_NLO}). The residual dependence 
is thus formally of order $a^3$. 
In order to check the sensitivity of our procedure to the choice of
$\bar{a}_{\PT}$, we
repeated the whole analysis with $\bar{a}_{\PT}$ as $a_{\pinch}$.
Starting as in sec.~5.3 with the world average value of $\alpha_s$, 
$\alpha_s^{\MSbar}({\rm M_Z})=0.117$ which corresponds to
$\Lambda_{\pinch}^{\1loop}=0.274\, {\rm GeV}$ and
$\Lambda_{\pinch}^{\2loop}=0.686\, {\rm GeV}$, we obtain the results shown in
fig.~\ref{watson_117_PV}.
\begin{figure}[htb]
\begin{center}
\mbox{\kern-0.5cm
\epsfig{file=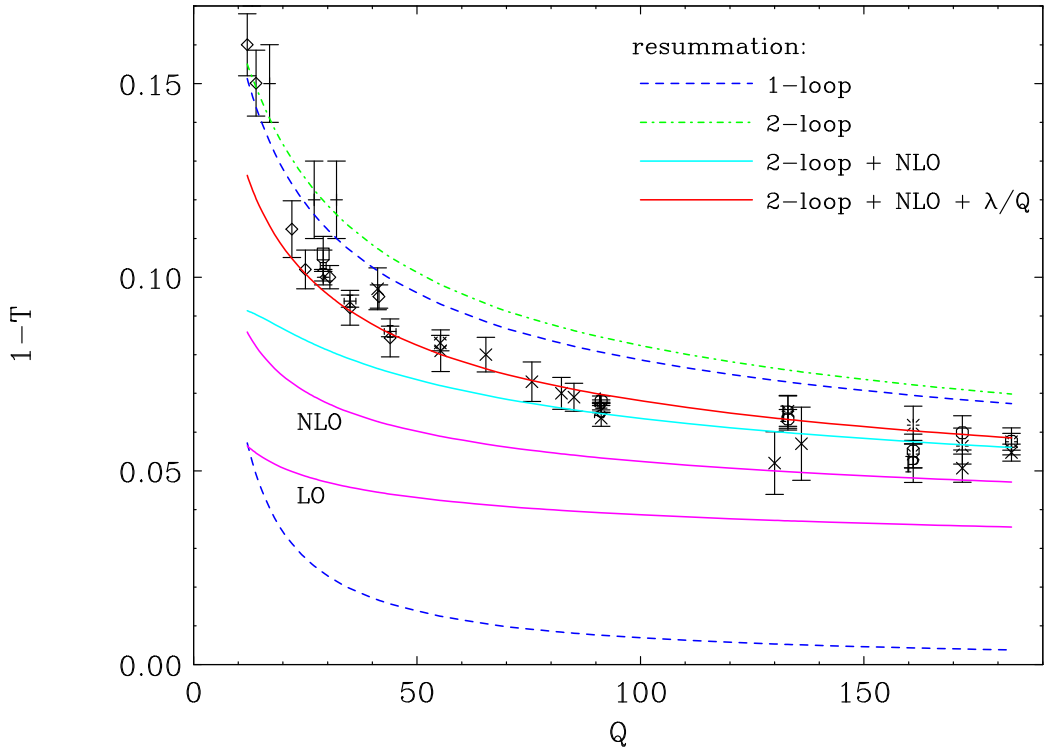,width=10.0truecm,angle=90}
}
\end{center}
\caption{Principal value Borel sum (or $R_{\APT}$) regularization of
 the perturbative sum (\ref{Rapt}) with the ``skeleton coupling''
 $\bar{a}_{\PT}$ in
 the pinch technique scheme \cite{Watson}, given
 $\alpha_s^{\MSbar}({\rm M_Z})=0.117$. For more details, see the
 caption in fig.~\ref{K_117_PV}. 
 }
\label{watson_117_PV}
\end{figure}
Here the resummation result, namely $R_{\APT}$ or principal value Borel sum 
with $\bar{a}_{\PT}$ at one-loop or two-loops, overshoots most data points and
especially the large $Q$ ones. However, after including the
next-to-leading order correction (\ref{inc_NLO}) with the 
appropriate negative coefficient 
(\ref{tilde_f1_pinch}), the perturbative result becomes again quite 
close to the large $Q$ data points. Adding a power correction
(\ref{inc_lambda}) the best fit (with $\bar{a}_{\PT}$ at two-loop) 
is obtained for $\lambda=0.44 \,{\rm GeV}$,
with $\chi^2/{\rm point}=4.35$. In the infrared cutoff
regularization with $\mu_I=2\,{\rm GeV}$ the same fit is obtained with
$\lambda_{2\,{\rm GeV}}=1.53\,{\rm GeV}$. Finally, we fit also $\alpha_s$ using
$\bar{a}_{\PT}$ in the pinch scheme. The results of the best fit are shown
in fig.~\ref{watson_109_PV}. 
\begin{figure}[htb]
\begin{center}
\mbox{\kern-0.5cm
\epsfig{file=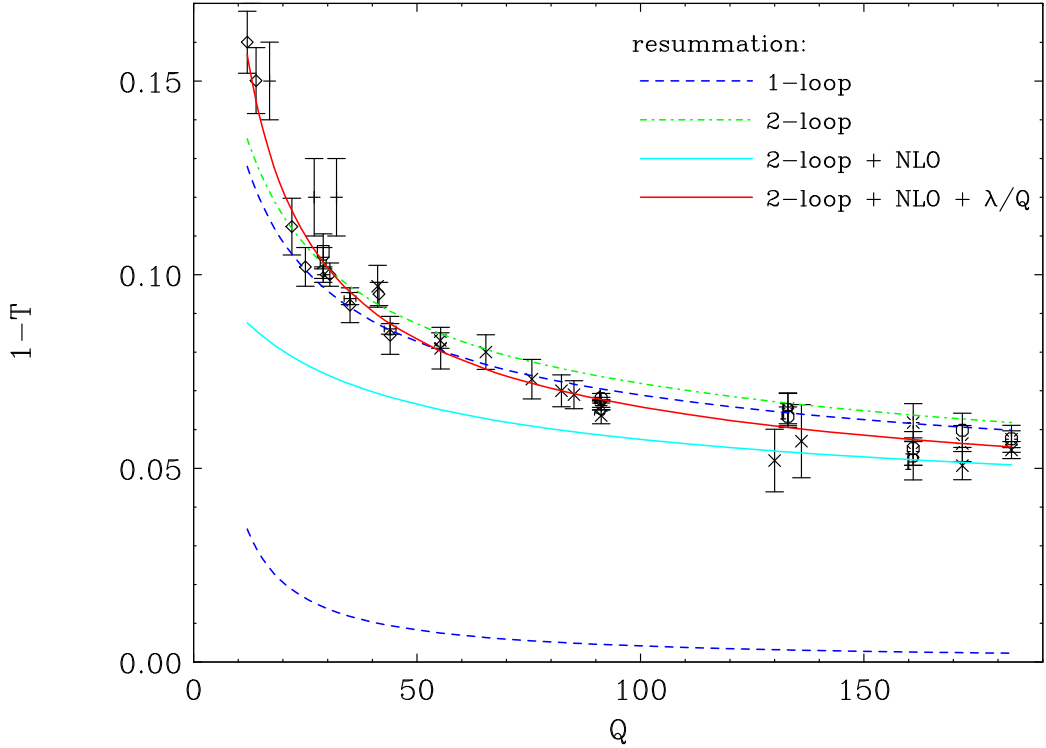,width=10.0truecm,angle=90}
}
\end{center}
\caption{Principal value Borel sum (or $R_{\APT}$) regularization of
 the perturbative sum (\ref{Rapt}) with the ``skeleton coupling''
 $\bar{a}_{\PT}$ in
 the pinch technique scheme \cite{Watson}, for the best fit value of  
 $\alpha_s^{\MSbar}({\rm M_Z})=0.109$. For more details, see the
 caption in fig.~\ref{K_117_PV}. 
 }
\label{watson_109_PV}
\end{figure}
The best fit with the two-loop running coupling in this scheme is obtained for 
$\alpha_s^{\MSbar}({\rm M_Z})=0.109$,
with a power correction $\lambda=0.83 \,{\rm GeV}$ in the principal
value regularization and $\lambda_{2\,{\rm GeV}}=1.85\, {\rm GeV}$ in
the cutoff regularization. In this fit $\chi^2/{\rm point}=1.37$.
The comparison of these results with eqs.~(\ref{bf_alpha}) 
and (\ref{bf_lambda}) suggests that precise identification of 
the ``skeleton coupling'' $\bar{a}_{\PT}$ is not crucial in this case for the 
determination of $\alpha_s$. 
\end{description}

Another possible source of uncertainty are ${\cal O}(a^3)$ terms
which are not related to the ``leading skeleton''.
At the single dressed gluon approximation level, such effects can be taken
into account by calculating 
\beq
\left<t\right>_{\PT}(Q^2)\,
\simeq\,\frac{C_F}{2}\,\left[
S_0^{\PT}(Q^2)\,+\,\tilde{t}_1\, a^2\,+\,\tilde{t}_2\, a^3\right]
\label{ske_min_NNLO}
\eeq   
instead of (\ref{ske_min_NLO}) where only the next-to-leading term was
included. The coefficient $\tilde{t}_2$ can be determined
from the full next-to-next-to-leading order perturbative coefficient
$t_2$, $\tilde{t}_2=t_2-t_2^0$, 
once the time-like ``skeleton coupling'' $\bar{a}_{\eff}^{\PT}$ 
in (\ref{Rapt}) and the scheme coupling $a$ are specified.
Unfortunately, $t_2$ is not known yet.
Since the running coupling effects we resum here are understood to be
the major source of growth of the standard coefficients $t_i$, 
the general expectation is that the remaining coefficients
$\tilde{t}_i$ are small. Nevertheless, let us see what are the
consequences if this assumption does not hold.
We proceed as before using the coupling $a=R_{\APT}(Q^2)/f_0$
\beq
\left<t\right>_{\PT}(Q^2)\,
\simeq\,\frac{C_F}{2}\,
R_{\APT}\left[1\,+\,\frac{\tilde{t}_1}{f_0}\, a\,+\,\frac{\tilde{t}_2}{f_0}\,
a^2\right].
\label{ske_min_NNLO_mod}
\eeq 
For the perturbative expansion to make sense one should require that the   
next-to-next-to-leading order term would be smaller than the leading
term, i.e. \hbox{$\tilde{t}_2/f_0<1/a^2$} or 
\hbox{$\tilde{t}_2/f_0<f_0^2/R_{\APT}^2$}.
For the lowest data points $Q=12\,{\rm GeV}$ the r.h.s. equals roughly
$64$ so we shall consider possible next-to-next-to-leading corrections
with $\left\vert \tilde{t}_2/f_0\right\vert \lsim 64$.
Once $\tilde{t}_2$ is fixed we can fit 
the thrust data with (\ref{ske_min_NNLO_mod}) plus a $\lambda/Q$ power
term. The results of such a fit turn out to depend significantly on
the value of $\tilde{t}_2$. For instance, with $\tilde{t}_2/f_0=-63$,
the best fit is obtained for $\alpha_s^{\MSbar}({\rm M_Z})=0.117$ with
$\lambda=2.4 \,{\rm GeV}$ and $\chi^2/{\rm point}=1.478$ while for 
$\tilde{t}_2/f_0=56$ the best fit is $\alpha_s^{\MSbar}({\rm M_Z})=0.104$ with
$\lambda=0.96 \,{\rm GeV}$ and $\chi^2/{\rm point}=1.313$.
Note that one could gain nothing by performing 
a three parameter fit, varying $\tilde{t}_2$ in addition
to $\alpha_s$ and $\lambda$, since the $\chi^2$ curve becomes almost flat.
The conclusion is that existence of a {\em large} next-to-next-to-leading
coefficient on top of the resummation performed would indeed modify 
the fit results significantly.

Finally, we relied heavily on the assumption that the {\em inclusive}
resummation program based on the ``skeleton expansion'' applies to
the case of the average thrust. 
The physical intuition behind this assumption is that
most gluon splittings are roughly collinear, and thus one can ignore the cases
where the split is to opposite hemispheres. This was further justified
in sec.~5.2 by showing that the magnitude of the non-inclusive correction 
at the next-to-leading order is small, at least as far as the ``Abelian'' 
($\beta_0$ dependent) part is concerned.
On these grounds we assumed that the non-inclusive effect is small also at 
higher orders.
In the large $N_f$ calculation \cite{Nason_Seymour} it was found that
non-inclusive corrections do affect the asymptotic factorial growth of the
perturbative coefficients, and therefore also the $1/Q$ power
correction. However, if such corrections are small before the asymptotic
regime it reached (like they are at the next-to-leading order) they would not
change the extracted value of $\alpha_s$.
Further, concentrating on the {\em single} gluon emission approximation, we
did not take into account perturbative and non-perturbative 
effects which are related to emission of more than one gluon.
Such effects were analyzed in \cite{Mil} at the next-to-leading order level, 
and were argued to give a significant contribution to the $1/Q$ power term. 
To understand how to incorporate this type of corrections into
the present approach requires further work.

\section{Conclusions}

In this paper we suggest a calculation method for inclusive enough 
observables in QCD, which is based on the conjecture that the Abelian
skeleton expansion can be extended to the non-Abelian case.
We did not make here an attempt to establish the existence of such an
expansion in QCD, but rather took a practical approach to use it
at the single dressed gluon level based on the analogy with the Abelian case. 
Clearly, to go beyond this approximation, a more rigorous treatment is
required, which would hopefully establish a diagrammatic interpretation
to the sub-leading terms in the ``skeleton expansion'' and an all
order identification of the ``skeleton coupling''.
Such a program was initiated by Watson in \cite{Watson} based on the 
pinch technique. It certainly deserves more attention.

The assumption of a ``skeleton expansion'' leads to the following crucial
consequences already at the single dressed gluon approximation:
\begin{description}
\item{a) } resummation of all order {\em perturbative} corrections 
that are related to the running coupling (renormalons) and
parametrization of {\em power corrections} must be performed together.
\item{b) } the resummation yields a renormalization scheme invariant result.
\end{description}

The idea that a certain type of infrared power corrections can be 
related to a universal {\em infrared finite} coupling \cite{DW,DMW} 
fits very well into the ``skeleton expansion'' approach. If the
``skeleton coupling'' is well defined on the non-perturbative level 
down to the infrared, there is no renormalon ambiguity and the
perturbative sum plus the power corrections are obtained at once by
performing the renormalon integral.

In this work we considered the specific example of the average thrust
in $e^+e^-$ annihilation, which is just appropriate to demonstrate the 
advantages of the suggested procedure over the standard perturbative
treatment. 
For this observable the available next-to-leading order
perturbative calculation with a standard choice of renormalization
scheme and scale
turns out to be far from the experimental data (fig.~\ref{pert}).
The perturbative series suffers from large renormalization scale
dependence. We assume that the most important higher order corrections
are related to running coupling effects which have a precise meaning
and can be resummed in the ``dressed skeleton expansion'' framework.
Assuming a ``skeleton expansion'' with a uniqely
identified ``skeleton coupling'', once the resummation 
is performed there is no more scale or scheme dependence.
The resummation itself, however, is ambiguous due to infrared renormalons. 
The ambiguity is resolved only when the regularized perturbative sum is 
combined with the appropriate power correction. Since for this observable
the leading renormalon is so pronounced ($n=\frac12$) one
cannot ignore the resummation ambiguity and the related $1/Q$ 
power corrections.

At the first stage we perform the ``leading skeleton'' resummation 
in a standard regularization such as the principal value Borel sum
(sec.~5.3) or truncation of the series at the minimal term (sec.~5.4).
We find (figs.~\ref{K_117_PV} and \ref{K_110_PV}) that a major part of the 
discrepancy between the next-to-leading order result and experimental 
data can be explained in terms of the resummation. 
The most important consequence is that the extracted value of
$\alpha_s$ is {\em different} than in a standard next-to-leading order based
fit. We obtain $\alpha_s^{\MSbar}({\rm M_Z})=0.110\pm 0.002$, where
the error reflects the combined statistical and systematic
experimental uncertainties. The discrepancy between this result and 
the world average value of $\alpha_s$ is interesting and calls for
further study.
It is necessary to apply a similar resummation program to other
QCD observables that are prone to having large higher order corrections of the
same type, e.g. other average event shapes\footnote{The analysis of
event shape distributions is more involved since there one has to
perform multiple soft gluon resummation in addition to the renormalon
resummation performed here.}, in order to extract a 
reliable value for $\alpha_s$ from experimental data.

At the next stage (sec.~5.5) we specify to the infrared finite coupling 
approach and reinterpret the resummation plus power corrections as
originating together in the renormalon integral performed down to zero
momentum. The regularization we use in this case is an infrared
cutoff $\mu_I=2\, {\rm GeV}$ on the space-like momentum. 
We assume that the full ``non-perturbative'' coupling coincides 
with the perturbative one above this scale. 
Thus the cutoff separates short and long distance physics 
as well as perturbative and ``non-perturbative'' physics.
Using the cutoff regularization, the perturbative
sum ($R_{\UV}^{\PT}$) includes only short distance contributions
related to scales where the ``skeleton coupling'' is controlled 
by the leading terms in the perturbative $\beta$ function.
The ``non-perturbative'' long distance contribution is expressed in 
terms of moments of the assumed universal ``skeleton coupling'' 
$\bar {a}(k^2)$ in the infrared region: $0<k^2<\mu_I^2$.    
Comparing fig.~\ref{K_117_cutoff} and \ref{K_110_cutoff} with 
fig.~\ref{K_117_PV} and \ref{K_110_PV}, we find that the gap between 
the cutoff regularized sum $R^{\PT}_{\UV}$ and the standard
regularizations discussed previously (e.g. principal value Borel sum) 
is quite large, especially for low $Q$ values (see also fig.~\ref{K_110_reg}). 
This means that an important part of
the contribution to what is called perturbative sum\footnote{Note that
the prescription dependence of the separate pieces, namely the regularized
perturbative sum and the power correction, does not
appear in the standard treatment when a power correction is added to the
truncated perturbative sum, e.g. at next-to-leading order.}
in the standard
regularization actually originates in small momentum
scales where the coupling is non-perturbative. 
On the other hand, this does not imply that the resummation is 
entirely related to long distance contribution: 
at large $Q$, $R^{\PT}_{\UV}$ becomes quite close to the standard
principal value Borel sum regularization.  

It is useful to examine the physical scales which contribute to the
perturbative sum by applying the BLM scale-setting method to 
the ``leading skeleton'' term.   
For Euclidean quantities, the momentum scales from which the major 
contribution to the perturbative sum (say, in the principal
value regularization) originates, is 
the BLM scale in the ``skeleton scheme'' (\ref{BLM_skeleton}).
This characteristic scale corresponds to the average virtuality
of the exchanged gluon. It is natural to compare it with a typical hadronic
scale, e.g. in our context with the cutoff $\mu_I$. 
If $\mu_{\BLM}\gg \mu_I$ the
resummation program is purely perturbative in the sense that it does
not depend on the coupling at infrared scales. If $\mu_{\BLM}\lsim
\mu_I$, the most important contribution to the ``perturbative sum'' 
comes from long distances and then the role of power corrections 
will be crucial.
Applying the same considerations to the Minkowskian representation
(\ref{Rapt}) we find that it is the center 
of $\dot{\cal F}(\mu^2/Q^2)$ which determines the BLM scale 
in the ``skeleton scheme'' (\ref{BLM_skeleton_F}),
the scale where the most significant contribution to the perturbative
sum comes from\footnote{We note, however, that the integral 
(\ref{Rapt}) runs over {\em time-like} 
momentum, and so the BLM scale in (\ref{BLM_skeleton_F}) cannot be 
identified immediately as related to some characteristic distance scale.}. 
Comparing $\mu_{\BLM}$ of
(\ref{BLM_skeleton_F}) with the (space-like) cutoff $\mu_I$ we can
learn to what extent the corrections are controlled by perturbation theory.
In the case of thrust, $\mu_{\BLM}$ in the ``skeleton scheme'' 
(\ref{mu_BLM_skeleton}) is $0.1028\, Q$, 
which is above the $\mu_I=2\, {\rm GeV}$ cutoff
for most of the relevant experimental data. For the low $Q$ data
points, the proximity of the two scales implies that an important
part of the contribution comes from the non-perturbative large
distance regime, as we learned already based on
fig.~\ref{K_117_cutoff} and \ref{K_110_cutoff}.

A crucial assumption we made in this work is that the inclusive ``skeleton
expansion'' approach is useful also for not-completely-inclusive
quantities like the average thrust.
As explained in sec.~5.1, the non-inclusive nature of the thrust
appears first in perturbation theory due to a split of a gluon into 
the two opposite hemispheres defined by the thrust axis. 
This possibility is ignored in the resummation we perform: 
non-inclusive corrections do not fit into a ``skeleton expansion''.
Physically, the assumption that the non-inclusive corrections are
small seems reasonable since in QCD one expects
most parton splittings to be roughly collinear. We further justified it
in sec.~5.2 by showing that the ``Abelian'' ($\beta_0$ dependent)
part of the next-to-leading order term that emerges from the inclusive
gluon dressing almost coincides with the corresponding term in the full
next-to-leading QCD calculation -- the two would coincide exactly for an
inclusive quantity.  
It would be interesting to examine how well this approximation works
at higher order, i.e. what is the error due to neglecting the 
``${\rm Non-Skeleton}$'' terms in eq.~(\ref{ske_min_NS}) beyond the
next-to-leading order. 
The Abelian case can be studied to all orders in perturbation theory 
comparing the present inclusive ``massive gluon'' calculation with that of
the non-inclusive large $N_f$ renormalon calculation \cite{Nason_Seymour}, 
similarly to the comparison made in \cite{Beneke-Braun-Magnea} for the
longitudinal cross section.
Further contributions relevant to the $1/Q$ power term 
which are specific to the non-Abelian theory were analyzed in
ref.~\cite{Mil} at the next-to-leading order level.
The connection of this type of corrections with the 
``skeleton expansion'' and renormalons should be clarified. 
 
It is tempting to speculate about the perturbative nature of the
``non perturbative'' infrared ``skeleton coupling'', namely
that the infrared part of the ``skeleton coupling'' could
itself be determined from information contained in perturbation theory. It
could be, for instance, that the {\em all order Borel resummed} ``skeleton
coupling'' is infrared regular, i.e. free of 
Landau singularities\footnote{For a
realistic number of flavors, the first two universal coefficients 
of the $\beta$ function are negative. It is therefore quite clear that 
without resummation a perturbative fixed point cannot appear in any 
coupling, including the ``skeleton coupling''. See the related discussion
in \cite{FP,GGK,conformal}.}. 
This last possibility seems natural as a continuation of the
perturbative fixed point scenario which presumably holds
within the conformal window \cite{conformal}:  
for $10\leq N_f\leq 16$ there is a fixed point in the perturbative
$\beta$ function and the perturbative coupling appears to be causal. 
One might then speculate that analytic continuation in $N_f$ below the
conformal window\footnote{This does {\em not} imply that more 
genuine non-perturbative phenomena like spontaneous chiral symmetry 
breaking could be studied by analytic continuation in
$N_f$ from the conformal window, since there these phenomena are 
probably absent.} might yield the correct infrared finite 
``skeleton coupling''. 
In this sense the physics of the universal infrared finite ``skeleton
coupling'', and in particular that of the gluon condensate, would be 
entirely perturbative. 

Finally we recall that the BLM scale-setting procedure is aimed to 
approximate the ``leading skeleton'' integral. In the case of the
average thrust we saw that choosing the BLM scale leads to a
significant improvement of the next-to-leading order partial sum
compared to the naive choice of scale $\mu_R=Q$. However, a fit based
on the next-to-leading order series with $\mu_R=\mu_{\BLM}$, gives
$\alpha_s^{\MSbar}({\rm M_Z})=0.116$, which is still 
appreciably different from the resummation based fit. 

\acknowledgments{We wish to thank S.J. Brodsky, Yu.L. Dokshitzer, L. Frankfurt,
G.P. Korchemsky, G. Marchesini, A.H. Mueller, M. Karliner, G.P. Salam and 
A.I. Vainshtein for very interesting discussions. 
We appreciate a lot the help we got from O. Biebel concerning  
the experimental data and ref.~\cite{EVENT}. We are also grateful to
the referee for very useful comments.}

\newpage
\appendix
\section{Cutoff regularization in terms of APT coupling}

One can give a useful alternative expression for $\Delta R$, based on the
general formula eq.~(\ref{R-split}) in sec.~4 specialized to the APT coupling
\begin{equation}
\Delta R=R_{\IR}^{\APT}+\delta
R_{\UV}^{\APT}
\label{DR-2}
\end{equation}
where each of the terms is separately
unambiguous\footnote{Eq.~(\ref{DR-2}) actually defines $\delta 
R_{\UV}^{\APT}$.}.

The first term
\begin{equation}
R^{\APT}_{\IR}(Q^2)\equiv \int_0^{\mu_I^2}{dk^2\over k^2}
\ \bar{a}_{\APT}(k^2)\ \Phi_R(k^2/Q^2)
\label{Rir_APT}
\end{equation}
can be evaluated explicitly in the one-loop case (\ref{d-apt1})
\beq
\bar{a}_{\APT}=\frac{1}{\beta_0}\left(\frac{1}{\log\frac{k^2}{\Lambda^2}}+
\frac{1}{1-\frac{k^2}{\Lambda^2}}\right).
\eeq
Consider a generic term in the small $\epsilon$ expansion of ${\cal
F}(\epsilon)$, as in (\ref{F-low}). Similarly to eq.~(\ref{Rirptn}) we
have
\begin{eqnarray}
\label{Rirptn_APT}
&R&_{\IR,n}^{\APT}(Q^2)= \nonumber \\
&-&\left({\mu_I^2\over Q^2}\right)^n
\left[\left( B^{(n)}_R \log{Q^2\over \mu_I^2}+ C^{(n)}_R\right)  \,
{\rm sinc}\ \pi n - {B^{(n)}_R\over n} \cos \,\pi n\right]\,
J_n\left(\mu_I^2/\Lambda^2\right)\nonumber\\
&-&\left({\mu_I^2\over Q^2}\right)^n {B^{(n)}_R\over n}\ {\rm sinc}\ \pi n\,\,
J^1_n\left(\mu_I^2/\Lambda^2\right)
\end{eqnarray} 
with 
\begin{equation}
J_n\left(\mu_I^2/\Lambda^2\right)\equiv\int_0^{\mu_I^2}\,n\,{dk^2\over k^2}
\left({k^2\over \mu_I^2}\right)^n \bar{a}_{\APT}(k^2)
\label{Jn}
\end{equation}
and
\begin{equation}
J^1_n\left(\mu_I^2/\Lambda^2\right)
\equiv\int_0^{\mu_I^2}\,n^2\,{dk^2\over k^2}
\left({k^2\over \mu_I^2}\right)^n \log{\mu_I^2\over k^2
}\,\, \bar{a}_{\APT}(k^2).
\label{barJn}
\end{equation}
In the one-loop case these integrals give
\begin{equation}
J_n\vert_{\1loop}=\,-\,{n\over \beta_0}\left[ 
e^{-nt_I}\, {\rm Ei}_1\left(-n t_I\right) -{\rm
Lerch}\Phi\left(e^{t_I},1,n\right)\right]
\label{Jn_lerch}
\end{equation}
and
\begin{equation}
\left. J^1_n\right\vert_{\1loop}=\,-\,{n\over \beta_0} 
\left[n t_I\,e^{-nt_I}\, {\rm Ei}_1\left(-nt_I\right)+1 -n\,{\rm
Lerch}\Phi(e^{t_I},2,n)  \right] 
\label{barJn_lerchlog}
\end{equation}
where as before $t_I\equiv\log\left( \mu_I^2/\Lambda^2\right)$,  
${\rm Ei}_k$ is defined in (\ref{Ei_def}) and ${\rm Lerch}\Phi$
is defined by analytic continuation of
\beq
{\rm Lerch}\Phi(z,k,n)\equiv\sum_{i=0}^{\infty}\frac{z^i}{(n+i)^k}
\label{lerch_def}
\eeq
from $\vert z\vert<1$ to the whole complex $z$ plane.
We have used the property that for a positive $n$,
\beq
\int_0^{z_I}\frac{z^n}{1-z}\frac{dz}{z}=
z_I^n\,{\rm Lerch}\Phi(z_I,1,n)
\label{larech_def_2}
\eeq
(here $z_I=\mu_I^2/\Lambda^2$) and taking the derivative with respect to $n$,
\beq
\int_0^{z_I}\log(z)
\frac{z^n}{1-z}\frac{dz}{z}=
\log(z_I)\,z_I^n\, {\rm Lerch}\Phi(z_I,1,n)
-z_I^n \,{\rm Lerch}\Phi(z_I,2,n).
\label{larech_def_3}
\eeq 
In (\ref{Jn_lerch}) and (\ref{barJn_lerchlog}), as opposed 
to $I_n$ and $I_n^1$ in (\ref{In_Ei}) and (\ref{barIn_Eilog}), 
there is no cut for
$\mu_I>\Lambda$ since the imaginary part from the ${\rm Ei}$ function is
cancelled by the one from the ${\rm Lerch}\Phi$ function. 
This is easy to understand, since the ${\rm Lerch}\Phi$ part
originates in the integration over $\delta {\bar a}_{\APT}$  of 
eq.~(\ref{d-apt1}). The simple pole in $\delta {\bar a}_{\APT}$ cancels by
construction the simple Landau pole in the perturbative one-loop
coupling, and thus also the related ambiguity in $I_n$ and $I_n^1$ 
should cancel.

\setcounter{footnote}{0}
One can write explicit formulae for the integral of
eq.~(\ref{larech_def_2}) in terms of elementary functions for simple
cases such as integer $n$ values
\beq
 {\rm Lerch}\Phi(z_I,1,n)=-   { \sum _{j=1}^{n - 1}} \,{ \frac
{{z_I}^{- j}}{n - j}}  - {z_I}^{ - n}\,\ln(1 - {z_I})
\eeq
or $n=\frac12$
\beq
 {\rm Lerch}\Phi(z_I,1,\small{\frac12})
=- \frac 1{\sqrt{z_I}}\, \ln{ \frac {1 - \sqrt{z_I} }{1 + \sqrt{z_I}}}.
\eeq
An interesting empirical observation is that 
$J_n\simeq {\rm Re}\left\{I_n\right\}$
and $J_n^1\simeq {\rm Re}\left\{I_n^1\right\}$.
The ratios $J_n/{\rm Re}\left\{I_n\right\}$ and $J_n^1/{\rm
Re}\left\{I_n^1\right\}$ approach $1$ as $\mu_I/\Lambda$
increases, due to the smallness of the ${\rm Lerch}\Phi$ piece in
$J_n$ and $J_n^1$. Already\footnote{An exception is the log term 
$J_n^1/{\rm Re}\left\{I_n^1\right\}$
for $n=\frac12$.} at $\mu_I/\Lambda\lsim 10$ these ratios
are not too far from $1$, as shown in table \ref{table_APT_IR}.  
\begin{table}[H]
\[
\begin{array}{|c|c|c|c|}
\hline
n&\mu_I/\Lambda& J_n/{\rm Re}\left\{I_n\right\}&J_n^1/{\rm
Re}\left\{I_n^1\right\}\\
\hline
\hline
\frac12&5&1.11&6.33\\
\hline
\frac12&8&1.05&2.60\\
\hline
\hline
1&5&0.72&0.84\\
\hline
1&8&0.81&0.80\\
\hline
\hline
\frac32&5&0.73&0.64\\
\hline
\frac32&8&0.85&0.75\\
\hline
\hline
2&5&0.77&0.66\\
\hline
2&8&0.88&0.80\\
\hline
\end{array}
\]
\caption{Comparison between $J_n$ and ${\rm Re}\left\{I_n\right\}$
and between $J_n^1$ and ${\rm Re}\left\{I_n^1\right\}$ for $\mu_I/\Lambda<10$.}
\label{table_APT_IR}
\end{table}
This implies that even for not too large $\mu_I/\Lambda$ the contribution of
a generic {\em non-analytic} term in the small $\epsilon$ expansion 
of ${\cal F}_R$ admits 
\begin{equation}
{\rm Re}\left\{R^{\PT}_{\IR,n} \right\}\simeq R_{\IR,n}^{\APT}.
\label{DR-APT_IR}
\end{equation}
If we assume in addition that (\ref{delta_R_APT_small}) holds,
i.e. that  ${\rm Re}\left\{\delta R^{\APT}_{n} \right\}$ is negligible
compared to  ${\rm Re}\left\{R^{\PT}_{\IR,n} \right\}$ 
then in eq.~(\ref{dR-1}) 
$\Delta R_n \simeq {\rm Re}\left\{R^{\PT}_{\IR,n} \right\}$, which implies  
\begin{equation}
\Delta R_n\simeq R_{\IR,n}^{\APT}.
\label{DR-3}
\end{equation}
By comparison with eq.~(\ref{DR-2}) it follows that already for not
too large $\mu_I/\Lambda$, the corresponding ultraviolet contribution 
in $\delta R_{\UV,n}^{\APT}$ is small.
For analytic terms, the infrared contributions 
${\rm Re}\left\{R^{\PT}_{\IR,n} \right\}$ and  $R_{\IR,n}^{\APT}$ vanishes
identically. 

Note that eq.~(\ref{DR-3}) is equivalent to the statement that the
$\mu_I$ cutoffs in the Euclidean and in the Minkowskian
representations are practically equivalent for the APT coupling, in the
sense that
\begin{equation}
R_{\IR,n}^{\APT}-R_{\IR,n}^{\PT}\simeq
R_{<,n}^{\APT}-R_{<,n}^{\PT}\label{cutoff}
\end{equation}

One can also compute the Minkowskian ``ultraviolet correction'' $\delta
R_{\UV}^{\APT}$ in terms of the associated 
Euclidean one $\delta D_{\UV}^{\APT}$ using the Minkowskian-Euclidean
connection pointed out in \cite{Grunberg-pow}.
We have
\begin{equation}
\delta R_{\UV}^{\APT}(Q^2)={1\over 2\pi i} 
\oint_{|Q'^2|=Q^2}{dQ'^2\over Q'^2}\
\delta D_{\UV}^{\APT}(Q'^2)
\label{dRuv-dDuv}
\end{equation}
with
\begin{equation}
\delta D_{\UV}^{\APT}(Q^2)=\int_{\mu_I^2}^\infty{dk^2\over k^2}\ 
\delta\bar{a}_{\APT}(k^2)\
\dot{{\cal F}}_R(k^2/Q^2)
\label{dDuv}
\end{equation}

This observation allows to express $\delta R_{\UV}^{\APT}$ in terms of
integrals over $\delta {\bar a}_{\APT}$. 
Indeed  one can derive \cite{Grunberg-pow} from eq.~(\ref{dDuv})
the large $Q^2$ behavior of $\delta D_{\UV}^{\APT}(Q^2)$. Assuming e.g. the
leading term in the small gluon mass expansion of ${\cal F}_R$ is a 
$n=\frac12$ pure power
 term (this example is relevant to the thrust case), one finds
\begin{equation}
\delta D_{\UV}^{\APT}(Q^2)\simeq -{1\over 2} C^{\left(\frac12\right)}_R
K(\mu_I){\Lambda\over Q}+\cdots
\label{dDuv-high}
\end{equation}
where
\begin{equation}
K(\mu_I)\equiv\int_{\mu_I^2}^\infty{dk^2\over k^2}\ 
\delta\bar{a}_{\APT}(k^2)\
\left({k^2\over \Lambda^2}\right)^{1/2}\label{K}
\end{equation}
is ultraviolet convergent, since $\delta\bar{a}_{\APT}(k^2)$ is ${\cal
O}(1/k^2)$ at large $k^2$, and
may eventually be computed analytically for the one and two loop couplings.
Eq.~(\ref{dRuv-dDuv}) then gives
\begin{equation}
\delta R_{\UV}^{\APT}(Q^2)\simeq -{1\over 2} C^{\left(\frac12\right)}_R
  K(\mu_I) \, \frac{2}{\pi} \ {\Lambda\over Q}+\cdots
\label{dRuv-high}
\end{equation}

\end{document}